\documentclass[preprints,article,accept,pdftex,moreauthors]{Definitions/mdpi} 

\firstpage{1} 
\pubvolume{1}
\issuenum{1}
\articlenumber{0}
\pubyear{2023}
\copyrightyear{2022}
\externaleditor{Academic Editor: {Jaziel Goulart Coelho, Rita C. Anjos}}
\datereceived{29 November 2022} 
\daterevised{22 December 2022 }
\dateaccepted{28 December 2022 } 
\datepublished{} 
\hreflink{https://doi.org/} 
\pdfoutput=1

\usepackage{shortcuts}

\Title {An {Overview} of Compact Star Populations and Some of Its Open Problems}
\TitleCitation{An Overview of Compact Star Populations and Some of Its Open Problems}

\Author{{Lucas M. de}  S\'a $^{1}$\orcidA, {Ant\^onio Bernardo} $^{1}$\orcidB, {Riis R. A. }Bachega $^{1}$\orcidC, {Livia S.} Rocha$^{1}$\orcidD, {Pedro H. R. S. }Moraes $^{2}$\orcidE and{ Jorge E. }Horvath $^{1}$*\orcidF}

\AuthorNames{ Lucas M. de S\'a, Ant\^onio Bernardo, Riis R. A. Bachega, Lívia S. Rocha, Pedro H. R. S. Moraes and Jorge E. Horvath}

\AuthorCitation{de S\'a, Lucas M.; Bernardo, Ant\^onio; Bachega, Riis R.A.; Rocha, Livia S.; Moraes, Pedro H.R.S.; Horvath, Jorge E.}

\address{$^{1}$ \quad {Universidade} de S\~ao Paulo (USP). Instituto de Astronomia, Geof\'isica e Ci\^encias Atmosf\'ericas (IAG), Departamento de Astronomia, R. do Mat\~ao 1226, Cidade Universit\'aria, 05508-090, S\~ao {Paulo, SP}, Brazil; lucasmdesa@usp.br (L.M.S.); alc.bernardo@gmail.com (A.B.); rrhavia@gmail.com (R.R.A.B.); livia.silva.rocha@usp.br (L.S.R.) \\
$^{2}$ \quad Universidade Federal do ABC ({UFABC}), Centro de Ci\^encias Naturais e Humanas (CCNH), Avenida dos Estados 5001, 09210-580, Santo Andr\'e, {SP}, Brazil; moraes.phrs@gmail.com}

\corres{Correspondence: foton@iag.usp.br; Tel.: 55-11-30912710}

\abstract{The study of compact object populations has come a long way since the determination of the mass of the Hulse--Taylor pulsar, and we now count on more than 150 known Galactic neutron stars and black hole masses, as well as another 180 objects from binary mergers detected from gravitational-waves by the Ligo--Virgo--KAGRA Collaboration. With a growing understanding of the variety of systems that host these objects, their formation, evolution and frequency, we are now in a position to evaluate the statistical nature of these populations, their properties, parameter correlations and long-standing problems, such as the maximum mass of neutron stars and the black hole lower mass gap, to a reasonable level of statistical significance. Here, we give an overview of the evolution and current state of the field and point to some of its standing issues. We focus on Galactic black holes, and offer an updated catalog of 35 black hole masses and orbital parameters, as well as a standardized procedure for dealing with uncertainties.}


\begin{document}


\section{Introduction}

The recognition that extreme states of matter constitute the endpoints of Stellar Evolution is one of the most important achievements of the 20th century. In fact, all the concepts and developments of Stellar Evolution started as such in the 19th century, and evolved symbiotically with the new exciting ``modern'' Physics, General Relativity, Nuclear Physics, Quantum Mechanics, and Statistical Mechanics. Stellar Evolution is their legitimate daughter and combined many things to create a consistent and predictive picture of how stars evolve.

This happy confluence is particularly important for the compact remnants, leftovers of massive stars in which the final stages prompted matter to show its ultimate nature. This is how the idea of {\it neutron stars} (NS) was raised, and although the {\it black hole} (BH) concept followed a different path, their recognition as stellar remnants unifies the two classes as stellar corpses.  

Neutron stars are now recognized to come in many varieties. In addition to the celebrated pulsar group, dim isolated neutron stars (also known as XDINSs) \cite{Walter1996,Zampieri2001,Turolla2009,Ertan2014}, magnetars \cite{MagnetarReview}, Compact Central Objects in supernova remnants \cite{CCODeLuca,CCOMayer,CCOho2013,CCOho2021}, Rotating Radio Transients \cite{RRATabhishek,RRATkeane} and some amazing new detections tentatively labeled as Ultra-Slow Magnetars \cite{GLEAM} form a full family which we would like to understand as a whole. We shall briefly address this issue below.

Galactic black holes, much as has been the case since the confirmation of the first one in Cygnus X-1, are observed generally as components of X-ray binaries, in most cases with a dwarf companion, although a few systems with giant companions are known. This means that most known BH masses have been determined from dynamical parameters (orbital period, mass function, mass ratio), which have been compiled in current catalogs such as {BlackCAT}\footnote{\url{https://www.astro.puc.cl/BlackCAT/}} \cite{blackcat} and WATCHDOG\footnote{\url{https://sites.ualberta.ca/~btetaren/}} \cite{watchdog}. In recent years, however, novel techniques, such as the study of quasi-periodic oscillations \cite{QPOreview} and of gravitational microlensing \cite{lam_isolated_2022,sahu_isolated_2022}, have allowed for masses to be constrained in different ways. The greatest example of this lies in the case of extragalactic black holes, over 100 of which have had their masses constrained from gravitational-wave (GW) observations of compact object mergers by the {LIGO-Virgo-KAGRA Collaboration (LVK)} \footnote{\url{https://www.ligo.caltech.edu/page/ligo-scientific-collaboration}}. In what follows, we will deal in detail with the current 35 well-constrained Galactic BH masses, and also briefly overview and compare them to the extragalactic BH mass distribution observed so far.

The last twenty years or so produced in fact a large body of evidence in which the simplest theoretical expectations serve as an overall framework only. The process of massive star collapse, for example, has been deeply {explored and there are now different perspectives} on how exactly it happens, and particularly on which outcome can be expected from them \cite{Adam, Ertl}.

 At the same time, the determination of masses of NSs and BHs with good accuracy in larger samples has allowed a better glimpse of the astrophysical processes that lead to their birth. This is indeed a long-term task, since many complicated issues in the evolution of progenitors and explosions themselves are involved. We shall not address these issues here, but rather indicate some recent works that illustrate the state-of-the-art understanding of them. The connection with the compact star features is also dependent on the binarity of forming systems to a high degree, and in fact it is in binaries (in which one of the members is generally non-compact) where most of the masses (and some radii for NSs) have been measured. Last, but not least, the production of NSs by accretion induced collapse \cite{AICtauris,AICWang2020,AICWang2022} is not properly understood but may be important for the whole picture, as we shall see. All these issues are entangled when we have to address, for example, the suggested absence of objects between $2$ and $5\Msun$, a paucity called {\it lower mass gap} in the literature \cite{bailyn1998,ozel_black_2010,farr_mass_2011,fryer2012,belczynski2012}. We shall address the existence of a mass gap below according to analyses of the latest data \cite{farah2022,desa_gap,Ye2022}. Finally, a novel form of ``seeing'' compact objects is the now systematic monitoring of gravitational-wave events. Great insights on NS matter have been gathered from the event GW170817 and a substantial set of compact binary masses has been collected through three LVK runs, and these we also discuss briefly.
 
We start in Section \ref{sec:neutron_stars} by offering a brief overview of current issues under study with regard to NSs, including their demographics, maximum mass, mass distribution and radius measurements. We follow this in Section \ref{sec:blackholes} with a more thorough discussion of the current state of BH observations, in particular that of Galactic objects, where we present, case-by-case, an updated catalog of 35 known masses, calculated in a standardized way from dynamical parameters. We display in Section \ref{sec:full_mass_distr} the resulting full Galactic BH mass distribution, and briefly discuss and compare it to the extragalactic distribution from LVK. Finally, we summarize the evolution and current state of the lower mass gap problem in Section \ref{sec:gap}. Our concluding remarks are presented in Section \ref{sec:conclusions}.

\section{Neutron Stars}
\label{sec:neutron_stars}

Neutron stars were predicted simultaneously with the discovery of the neutron itself \cite{Landau}, and related to collapsing/exploding stars by \citet{BZ} without any real proof of their existence. ``Real'' neutron stars became a reality 30 years later, when the first pulsating radio sources were discovered \cite{Jocelyn} and the present model (or close to it) was put forward by \citet{Franco} and \citet{Gold}. However, it became clear over the years that not all NSs pulse, since for that rapid rotation and intense magnetic fields are necessary, according to the basic model in which the torque is given by the electromagnetic emission and scales $\propto B^{2} \Omega^{3}$. In addition, new classes of NSs have been identified, and other properties such as masses and radii measured with increasing precision. We shall briefly point out the main features of each group, the expected demographics and the inferred physical quantities in the following (for a recent full review, see \cite{WSChapter}).

\subsection{Neutron Star Demographics}
\label{sec:NSdemographics}

It is commonplace understanding that NSs are born in massive star supernova explosions. However, it is almost certain that contributions from the explosions of ``low-mass'' massive stars, in the range of 8--$10\Msun$, which are thought to undergo electron capture onto an O-Ne-Mg core and leave ``light'' NSs is an important channel, since progenitors in this mass range are abundant. In addition, the rate of accretion induced collapses (AICs), either in their single-degenerate or double-degenerate (merger) versions, is uncertain but must be added to the total if some NSs are formed by them \citep{AICtauris,AICWang2020,AICWang2022}. Simply multiplying by the Galactic lifetime the proper core--collapse supernova rate extrapolated from  observations in galaxies similar to the Milky Way gives a number of NSs births of $\sim$$10^{8}$ over the whole life of the Galaxy \cite{BisnovatyiKogan1992}. The large group of {\it active} pulsars today has been estimated to be around $70,000$ \cite{LMT, dirson}, although this number quite depends on the birth parameters and evolution of magnetic fields. However, it is highly unlikely that the estimate is wrong by an order of magnitude, leading to the conclusion that most of the Galactic NSs are not pulsars, but instead belong to one of several classes that may be termed ``hidden'', which we describe below. These are NSs that may or may not have functioned as pulsars in the past, but the presence of which would be much more difficult to establish today. Attempts to unify all NSs have been presented before \cite{Kaspi}, but as we shall see, new puzzling detections and unresolved problems are still ubiquitous.

Some well-known NSs are natural candidates for the large ``hidden'' group. These include the central compact objects (CCOs) in a few supernova remnants without detected pulsations \cite{CCODeLuca,CCOMayer}. Most of the population is likely to have faded away due to its old age, well beyond the $\sim \text{Myr}$ scale, and therefore to establish their presence and statistics is quite difficult, although progress has been made; see, e.g., \citep{CCOho2013,CCOho2021}.

Detected between 1996, with the observation of RX J1856.4–3754 by \cite{Walter1996}, and 2001, when RBS 1774 was first observed by \citet{Zampieri2001}, the subpopulation termed the {\it Magnificent Seven} or X-ray Dim Isolated Neutron Stars (XDINSs), in addition to the isolated NS {\it Calvera} \cite{CalveraDiscovery}, consists of relatively close, blackbody-like cooling NSs \citep{CoolingNS} that were once considered the tip of the population iceberg. However, pulsations and non-zero period derivatives were detected over time, proving that they are actually middle-aged objects \cite{Turolla2009,Ertan2014}. Nevertheless, the paucity of this type of NS is somewhat unexpected, and no additional candidates (except for {\it Calvera}) were added over many years. 

Another subpopulation which was identified unexpectedly is that of Rotating Radio Transients (RRATs), emitting occasionally isolated pulses of a few $\text{ms}$ and going silent for days or more \citep{RRATkeane}. This narrow duty cycle makes them virtually invisible most of the time, and therefore their number could be very large. A simple estimate from the $\geq$$100$ sources already known is 

\begin{figure}[H]
{\includegraphics[width=0.8\linewidth]{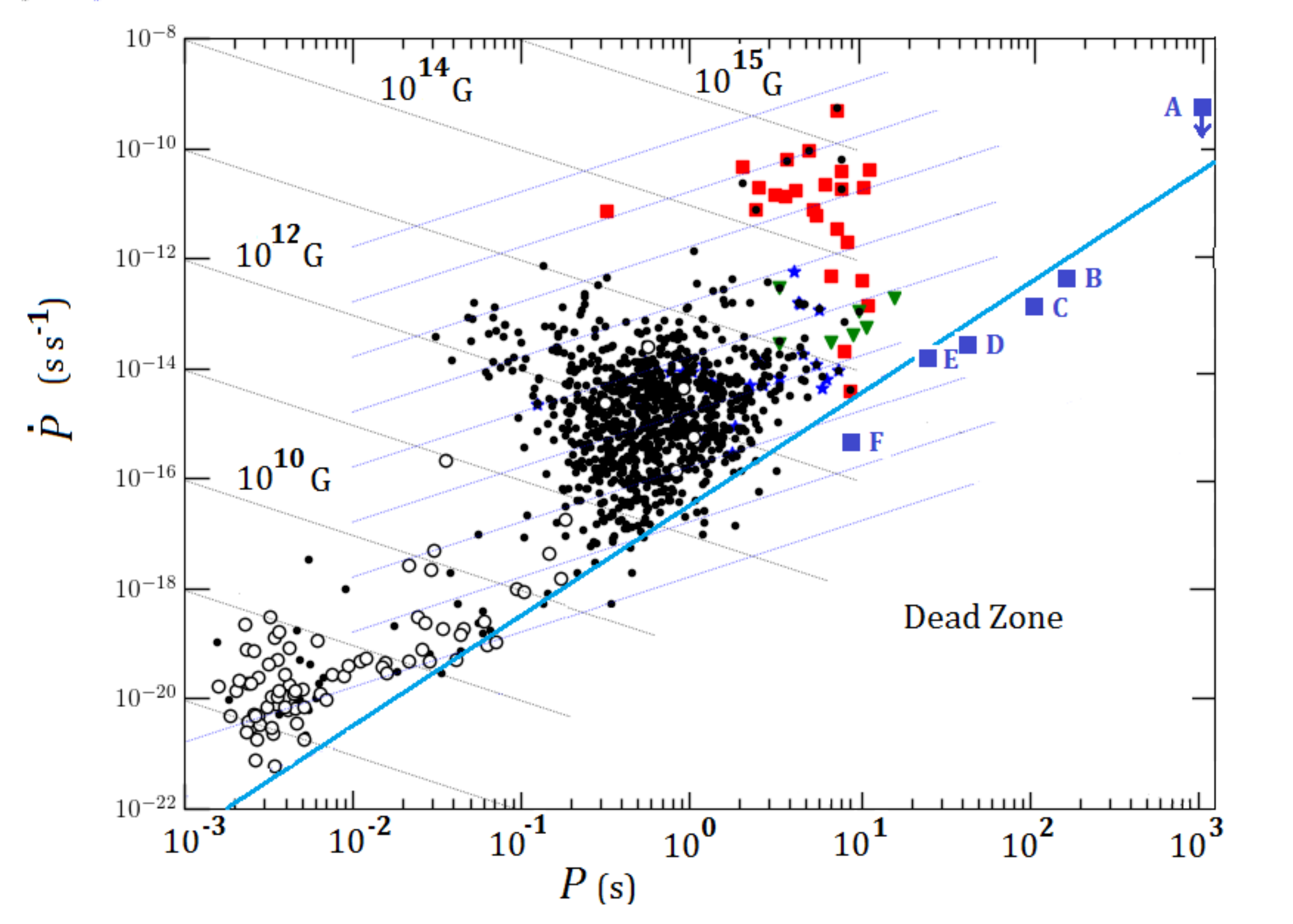}}
\caption{{The} $P-{\dot{P}}$ diagram with the new objects. The diagram had to be extended up to periods of $\sim10^{3}\text{ s}$ to accommodate the ultra-long period magnetar candidates. The symbols are as follows: ordinary pulsars (black dots), millisecond and binary pulsars (open circles), XDINSs (green triangles), RRATs (blue stars), Magnetars (red squares), and the new objects in blue squares. {A)} GLEAM--X 162759.5$-$523504.3 \cite{GLEAM}, {B)} AR Scorpii \cite{WDPulsar}, {C)} J0901$-$4046 \cite{UltraLongPeriodPSR}, D) J0250+5884 \cite{Agar}, E) J2251$-$3711 \cite{Morello}, F) J2144$-$3933 \cite{Young}.}
\label{fig:demo}
\end{figure}

\begin{linenomath}
\begin{equation}
   N_\mathrm{RRATs} = 2 \times 10^{5} {\biggl( {L_\mathrm{min}\over{100\text{ mJy kpc}^{2}}} \biggr)} {\biggl( {0.5 \over{f_\mathrm{on}}} \biggr)} , 
\end{equation}
\end{linenomath}

\noindent where $f_\mathrm{on}$ is the duty fraction and all other uncertainties have been set to unity multiplication factors. We see that their number could be larger than the estimated number of ordinary radio pulsars, but it is unlikely that the bulk of born NSs can be accommodated. In fact, the construction of coherent solutions for the sporadic pulses allowed an estimate of some of the period derivatives ${\dot{P}}$, and thus an estimate of the characteristic spin-down ages. They seem to be not too different from ordinary pulsars in this aspect (see \mbox{Figure \ref{fig:demo}}), and it has been conjectured that for some reason the ``spark'' leading to a pulse is not always operating, like they do in the pulsar case \citep{RRATabhishek}. Of course, we do not know enough about the origin of pulses to discard or prove these ideas.

A third subgroup of importance is that of the so-called {\it magnetars}, objects in which the emission is related to the existence of a large magnetic field, dominating the energetics of the rotation (i.e., satisfying that the X-ray luminosity ${\dot{E}}_{X}$ exceeds the rotational energy $I\Omega{\dot{\Omega}}$); see, for a recent review, \citep{MagnetarReview}. Magnetars seem to possess magnetic fields a few orders of magnitude larger than the ordinary pulsar variety. The natural question is whether there is a continuum of NSs, in the sense that the magnetic fields are generated by a continuous distribution, or if there is some kind of gap in this quantity instead. The presence of transition pulsars, with magnetic fields as high as some identified magnetars seems to argue in favor of the former. The existence of ``low-field'' magnetars \cite{Roberto} is also suggestive of a continuum (Figure \ref{fig:demo}). In this way, the magnetar group was defined as the NSs in which the emission is powered by the magnetic field, irrespective of the precise value of its intensity.

The latest news about the NSs population include the discovery of progressively longer period pulsars \citep{UltraLongPeriodPSR} and magnetars \citep{GLEAM} (see caption of Figure \ref{fig:demo}). These unexpected objects are not only below the ``classical'' {\it death line}, but also challenge a definite division between the two groups. At some point the energy condition is employed to separate them, but otherwise their position in the $P-{\dot{P}}$ plane does not allow one to establish a clear classification. In addition, the first (and unique up to now) pulsar-like white dwarf AR Scorpii (object B in Figure \ref{fig:demo}) \citep{WDPulsar} is included in this group. For now, the latter is just a singularity, but it is likely that a subpopulation could be found, although unrelated to the NSs demographics.

What are the main questions one can ask about these subpopulations and their relationships? The list is quite extensive, and we shall only point out some of them to contribute to this ongoing task.

\begin{itemize}
\item	Do magnetic fields decay?
\end{itemize}

This question is one of the oldest, and has been answered differently over the years. The idea that there must be a decay of the field within $\sim$$10^{6}\text{ yr}$ is quite popular, as discussed and modeled by \citet{MagneticFieldDecay} for a diverse set of 40 sources including, e.g., the already mentioned magnetars, CCOs and XDINSs; see also, for a review, \cite{Igoshev2021}. However, objects of this age still show substantial magnetic fields, and the oldest ones (``black widows'', several $\text{Gyr}$ old according to evolution calculations \cite{BDH}) reinforce this picture. Magnetic fields may decay partially to a ``bottom field'', but not completely and therefore many evolutionary trajectories in the $P-{\dot{P}}$ plane proposed to explain the transit and parenthood of many subpopulations could be misleading. Theoretical calculations support this kind of picture \cite{nos}.

\begin{itemize}
\item	Which is the relationship between all these subpopulations?
\end{itemize}
The idea that some subpopulations are just an evolutionary stage leading to some other type is behind all the attempts of unification \cite{Kaspi}, in one way or another. It is perhaps useful trying to establish which are the {\it youngest} and {\it oldest} objects. Here, we face a very general problem: in many cases, the age of the object is calculated with the characteristic age, and its magnetic field with the inversion of the electromagnetic torque equation (yielding $B \propto {\sqrt {P{\dot{P}}}}$). However, it is clear by several lines of argument that this is probably an oversimplified picture. Braking indexes are not what the ideal electromagnetic torque predicts \cite{braking}, and there is evidence for variations of the torque itself with time \cite{AllenHorvath}. Several ideas have been put forward to link subpopulations; the analysis by \citet{Yone}, for example, supported the idea that XDINSs are old magnetars, not common-type pulsars. Typical ages of magnetars (soft gamma repeaters \cite{SGRrev} and anomalous X-ray pulsars \cite{AXPreview}) are $\leq$$10^{4}\text{ yr}$, and some support for young ages comes from their associations with young supernova remnants. Other kinships have been suggested; for instance, \citet{keane} points out that, from empirical grounds, RRATs and Fast Radio Bursts sources are hardly distinguishable. The clue for this association, and the possible relation with ordinary pulsars which are side by side with RRATs in the ${P-{\dot{P}}}$ diagram, is, of course, a deeper and solid understanding of the short radio pulse emission, and of ordinary pulsar emission in general (\citet{KaspiKramer2016} present an overview of how these different manifestations of NSs relate to pulsars).

\begin{itemize}
\item	Are the new objects old magnetars?
\end{itemize}
The ultra-slow magnetars should be, logically thinking, a latter stage in which they have cooled and braked to very long periods. However, their inferred magnetic fields are very high, and according to conventional wisdom, they should {\it not} be that old indeed. Are they actually related? \citet{Benia} have argued that there must be a large population of ultra-slow magnetars in the galaxy, stressing the resiliency of magnetic fields. Is there a real difference between RRATs and ultra-slow magnetars? In addition, is it a complete coincidence that the ``WD pulsar'' stands nearby other confirmed NSs of this group? It is premature to give definitive answers to these and other important questions

\subsection{Neutron Star Mass Distribution}
\label{sec:NSmasses}

The NS mass sample has 105 members now, with the addition of a few recent observations. An  analysis of a slightly smaller sample (95 objects), presented in \cite{desa_gap}, was discussed in full by \citet{rocha2021}, and we shall highlight the main features for completeness. 

The first important thing is that, as agreed by several groups \cite{PostnovProkhorov2001,valentim2011mass, ozel2016masses, kiziltan2013neutron, zhang1997neutron, alsing2018evidence}, the sample has a multimodal structure, with at least two mass scales. The analysis of \cite{rocha2021} was performed assuming a Gaussian parametrization with $n$ components, of the type

\begin{linenomath}
\begin{equation}
    {\cal L}(m_p | \theta) = \sum_i^n r_i ~{\cal N}(m_p | \mu_i, \sigma_i),
    \label{}
\end{equation}
\end{linenomath}

\noindent where $\mu_i$ and $\sigma_i$ are the mean and standard deviation of the {\it i-th} component ${\cal N}$, and $r_i$ is its relative weight, satisfying the normalization condition $\sum_i^n r_i = 1$.

As discussed in \citet{WSChapter}, the preferred figures for both Anderson--Darling and Kolmogorov--Smirnov frequentist tests are the ones appearing in Table \ref{tests}. 

\begin{table}[H]
\caption{$p$-value of two hypothesis tests for three different models.} \label{tests}
\setlength{\cellWidtha}{\columnwidth/6-2\tabcolsep+0.0in}
\setlength{\cellWidthb}{\columnwidth/6-2\tabcolsep-0.0in}
\setlength{\cellWidthc}{\columnwidth/6-2\tabcolsep-0.0in}
\setlength{\cellWidthd}{\columnwidth/6-2\tabcolsep-0.2in}
\setlength{\cellWidthe}{\columnwidth/6-2\tabcolsep-0.2in}
\setlength{\cellWidthf}{\columnwidth/6-2\tabcolsep+0.4in}
\scalebox{1}[1]{\begin{tabularx}{\columnwidth}{
>{\PreserveBackslash\centering}m{\cellWidtha}
>{\PreserveBackslash\centering}m{\cellWidthb}
>{\PreserveBackslash\centering}m{\cellWidthc}
>{\PreserveBackslash\centering}m{\cellWidthd}
>{\PreserveBackslash\centering}m{\cellWidthe}
>{\PreserveBackslash\centering}m{\cellWidthf}}
\midrule
\textbf{Model} & \boldmath$\mu$ & \boldmath$\sigma$ & \textbf{K--S} &  \textbf{A--D} &  \textbf{Recommendation} \\
\midrule
Unimodal & 1.48 & 0.35 & 0.025 & 0.032 & Reject \\
Bimodal & 1.38, 1.84 & 0.15, 0.35 & 0.971 & 0.990 & Do not reject \\
Trimodal & 1.25, 1.40, 1.89 & 0.09, 0.14, 0.30 & 0.974 & 0.953 & Do not reject \\
 \bottomrule
\end{tabularx}}
\label{Table 1}
\end{table}

The $p$-values indicate a strong rejection of a ``single mass'' hypothesis (labeled as ``Unimodal'') and hence the confirmation of a structured mass distribution. It should be noted that a peak at 1.25$\Msun$ is expected, related to the NSs produced in electron-capture supernovae, but the analysis does not reveal clear evidence for it. The low-mass NSs appear equally likely to be part of the tail of the strong maximum at 1.38$\Msun$, which is narrower than the second one around 1.8$\Msun$. Massive NSs belong to the latter fully.

To visualize the form of the mass distribution, we have drawn 1000 posterior samples from the master sample discussed in \cite{WSChapter}, and taken their mean to compare with the maximum a posteriori distribution obtained. The result confirms the presence of two maxima as explained.

A Bayesian analysis was implemented to cross-check these results. Again, the Bayesian likelihood is much higher for the bimodal distribution, and the mean and standard deviation of the two peaks quite similar, $\mu_1 = 1.351\Msun$, $\mu_2= 1.756\Msun$ and  $\sigma_{1} = 0.087\Msun$, $\sigma_{2} = 0.286\Msun$. This reinforces the conclusion that there are at least two peaks, and probably a third one ``blended'' with the objects at $\sim$1.38$\Msun$, but not a single mass (called sometimes ``canonical'' in the literature, a name that is now not recommended). 

\begin{figure}[H]
{\includegraphics[width=0.8\linewidth]{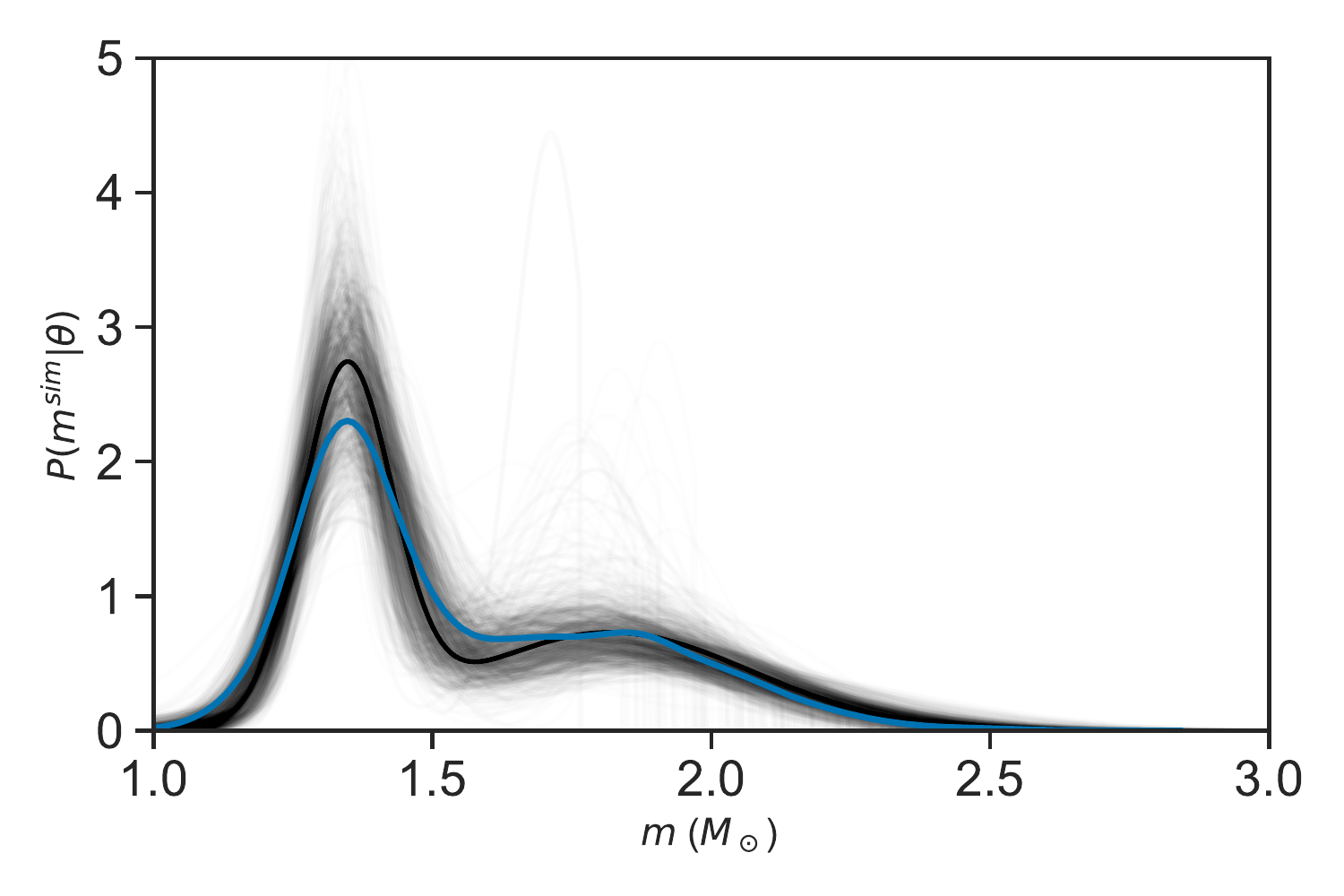}}
\caption{Gray lines represent 1000 posterior samples drawn from the sample in \cite{desa_gap, WSChapter}. The blue curve is the posterior mean of these synthetic samples and the black line is the maximum a posteriori distribution.}
\label{fig:posterior}
\end{figure}

It is much more difficult to attach definite formation events to these Gaussian peaks, although in the long run it will be a rewarding task. The maximum mass achievable by a real NS, $M_\mathrm{max}$, which has as its upper bound the Rhoades--Ruffini limit of $\approx$3.2$\Msun$ \cite{RhoadesRuffini}, should be obtained in a large sample having a statistical upper mass boundary $m_\mathrm{max}$, since in this case, and provided the sample is not biased, $m_\mathrm{max} \to M_\mathrm{max}$. At the present time, we can say that $m_\mathrm{max}$ was studied using the current sample and turns out to be $\sim$$2.5\Msun$ both by a simpler $3\sigma$ estimate of the second peak (Figure \ref{fig:posterior}), and also by a Bayesian approach, both with and without an upper truncation mass introduced as an independent quantity \citep{rocha2021}. 

On the one hand, such a large value of $M_\mathrm{max}$ is at odds with many estimates based on the analysis of the merged object in GW170817, which yield values near 2.1--2.2$\Msun$ \cite{margalit2017,Rezzolla2018,Ruiz2018,Shibata2019}. On the other hand, \citep{Ai2020} have also found from GW170817 that $M_\mathrm{max}$ might be as high as $2.43^{+0.06}_{-0.04}\Msun$, even if the remnant is born as a uniformly-rotating NS. \citet{rocha2021} aim to avoid limitations arising from estimates based on single sources by obtaining their $M_\mathrm{max}$ from a Bayesian analysis of the full sample of Galactic NS masses, which makes their resulting large mass robust. There are also several candidates that would be above the typical $\sim$$2.2\Msun$ inferred value, and one definite quite reliable determination of 2.35 $\pm$ 0.17$\Msun$ \citep{Romani} which challenges a low $M_\mathrm{max}$ limit and has already pushed more recent works to allow for a higher $M_\mathrm{max}$ than previously conducted \cite{RezzollaHighMMax}. A high $M_\mathrm{max}$ makes room for the lighter object in the merger event GW190814 in the NS group, possibly at the highest achievable value of any NS mass. 

It should also be noted that, even if the statistical approach of \citet{rocha2021} is considered more robust than a single-object analysis, the current extragalactic NS mass sample from GW observations is still too small for statistics to be performed confidently upon it, with only two NS-NS and four BH-NS mergers \cite{gwtc3}. Future runs of both current and future GW observatories, with continuously greater sensitivities, should increase the size of this sample, and eventually allow for an analogous procedure to be performed upon it. We point out that an alternate approach is to treat both the extragalactic NS and BH mass sample simultaneously with Bayesian methods. This is the approach of \citet{farah2022}, who take into account NS-NS, NS-BH and BH-BH mergers, finding a break in the compact mass object distribution at $2.4^{+0.5}_{-0.5}\Msun$ ($3\sigma$); and the later work by \citet{Ye2022}, who from the 4 NS-BH mergers find a lower limit of $M_\mathrm{max}>2.53\Msun$ with $90\%$ confidence. The simultaneous study of both NS and BH masses is discussed also with regard to the lower mass gap problem in Section \ref{sec:gap}.

A final remark is related to the recent announcement of a very low mass value for the compact object in the supernova remnant HESS J1731-347, determined using X-ray spectroscopy and GAIA astrometry. The value is just $0.77 {{+0.20}\atop{-0.17}}\Msun$. The reported radius, on the other hand, is small but not overly so, at $10.4 {+0.86\atop{-0.78}}\text{ km}$ \citep{LightNSDoroshenko}. The main problem is that the smallest iron cores, originating the lightest neutron stars, are always heavier than 1--1.1$\Msun$. Therefore, a much smaller mass would be almost impossible to accommodate unless a large fraction of the core is blown away in the very process of the explosion. Alternatively, this could be evidence for a {\it strange star} \citep{DragoSS}, discussed and worked out for almost 40 years \cite{reviewSS}. The difference with the two lowest NSs reliably determined is large, since the latter seem to have a similar mass of 1.17 $M_{\odot}$ \cite{Martinez, Fonsece2016}. In summary, both ends of the NS mass distribution are very interesting and display novelties.

\subsection{Neutron Star Radii}
\label{sec:NSradii}

After dreaming of a simultaneous determination of NSs masses {\it and} radii (see reviews by \cite{ozel2016masses,LattimerRev}), and many works in which this kind of determination was attempted indirectly, with uncertain results, NASA's Neutron Star Interior Composition Explorer {(NICER)}\footnote{\url{https://www.nasa.gov/nicer/}} succeeded in measuring directly the masses and radii of two neutron stars, rendering $2.072 {{+0.067}\atop{-0.066}}\Msun$ and $12.39 {{+1.30}\atop{-0.98}}\text{ km}$ for the massive pulsar J0740+6620 \cite{NICER1}; and $1.34 {{+0.15}\atop{-0.16}}\Msun$ and $12.71 {{+1.14}\atop{-1.19}}\text{ km}$ for the millisecond pulsar J0030+0451 \cite{NICER2}. Even at the $1\sigma$ level, the measurement of essentially the same radius for two neutron stars that are very different in mass means that the $M-R$ sequence varies sharply around this radius, and therefore that the equation of state must be very stiff. The heavier the reported masses (the current record being $2.35 \pm 0.15 \Msun$ reported by \citet{Romani}), the stiffer equations of state need to be. In fact, the consideration of ``exotic'' equations of state complying with high masses and $\sim$$12\text{ km}$ radii is possible and deserves attention \cite{JP}.

\section{Black Holes}
\label{sec:blackholes}

The basic concept of a black hole is older than might be imagined: in the late 18th century, John Michell and Piere-Simon Laplace already, independently, considered the possibility of an object so dense that its escape velocity would exceed that of light. Not surprisingly, at the time, the proposal did not leave the level of pure speculation. Such a step would have to wait for about another 130 years, until Albert Einstein's theory of general relativity \cite{Einstein1915} and Karl Schwarzschild's solution \cite{Schwarzschild1916} for a non-rotating, spherically symmetric, mass, which to this day is the basic description for the spacetime around a non-rotating and electrically uncharged black hole. By 1965, Ezra Newman had arrived at the general solution for a rotating, charged black hole, which we now call the \textit{Kerr--Newman solution} \cite{Kerr1963,Newman1965}. At this time, it was already understood that black holes are relatively simple objects, fully defined by nothing more than their mass, angular momentum and charge, a statement now called the \textit{No-hair theorem}.

Although a very simple kind of object from a physical point of view, the astrophysical nature of a BH is one of the most complex subjects in the area. On the one hand, the fact that a BH can be described by only three numbers also means that we can obtain almost no information on the evolution of its progenitor, as it is forever lost beyond the event horizon \citep{hooft1991,hawking2015information,susskind2006paradox}. On the other hand, the physics behind their disks, jets and magnetic fields are yet on the frontier of our knowledge.

A few years after it was first predicted that collapsing giant stars could be stabilized by degeneracy pressure to form neutron stars (Section \ref{sec:neutron_stars}), Robert Oppenheimer and George Volkoff \cite{OppenheimerVolkoff1939}, in 1939, starting from the work of Richard Tolman \cite{Tolman1939}, determined that even neutron stars have a maximum mass (the Tolman--Oppenheimer--Volkoff, or TOV, mass), beyond which nothing would be able to stop the collapse. Although this was a step in the direction of understanding black holes as one of the endpoints of stellar evolution, at the time, it was posited that yet another mechanism for stopping the collapse should exist. Quasars, which we today know to contain supermassive black holes, were first detected in the 1950s, and, in 1964, Yakov Zeldovich \cite{Zeldovich1964} and Edwin Salpeter \cite{Salpeter1964} independently proposed exactly that they were powered by such objects, but the idea was not taken very seriously then.

\begin{figure}[H]
    \includegraphics[width=0.8\textwidth]{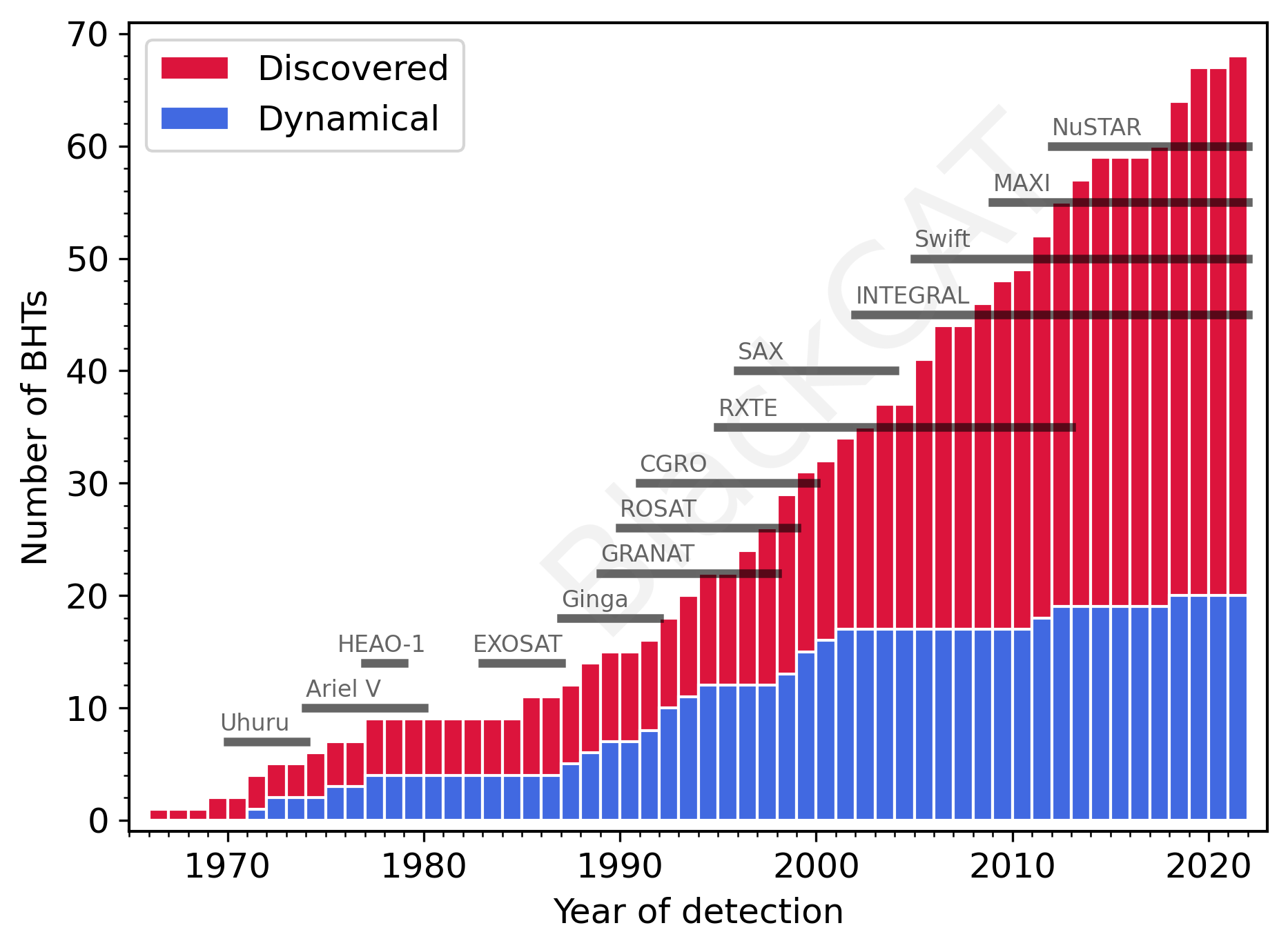}
    \caption{{Cumulative} histogram of discovered and dynamically confirmed BH candidates in X-ray transients, from the BlackCAT catalog \citep{blackcat}, last updated in December 2021. In addition, the durations of X-ray missions that have discovered BH candidates in outburst are indicated.}
    \label{fig:blackcat_history}
\end{figure}

In 1972, it was found independently by Thomas Bolton \cite{BoltonCyg}, and by Louise Webster and Paul Murdin \cite{LouisePaulCyg} that the X-ray source Cygnus X-1, discovered in 1964, had a massive stellar companion, and from its motion a first estimate for the mass of Cygnus X-1 was derived, exceeding the maximum mass of a NS and making it the first stellar black hole candidate. This finding helped to finally convince the scientific community of the existence of black holes in the Universe, and since then the set of known BHs has slowly grown, along with the set of known BH masses, in most cases still determined from the observation of X-ray sources. Along the way, a rich zoo of X-ray binaries has developed ({see Figure} \ref{fig:blackcat_history} {for the evolution of the number of discovered BH X-ray transients}), followed by a diversification of the methods through which BHs can be detected. Besides X-ray binaries, BH masses have now also been constrained in non-interacting binaries, and for the first time from the microlensing of background light by a stellar black hole. Beyond the Galaxy, the GW measurements by LVK have since 2015 built up a catalog of compact object mergers that as of its latest iteration, the third GW Transient Catalog (GWTC-3) \cite{gwtc3} contains 90 different events with well constrained masses, 83 of which are confirmed BH-BH mergers, totaling 166 confirmed extragalactic BH masses. {Five BH-NS merger or merger candidates add another 5 masses to the sample.}

This diversity, however, has also meant that the treatment of mass estimates has not always been consistent across the field, as credibility interval conventions have changed and the advent of Monte Carlo methods for dealing with distributed quantities has brought about an ease-of-use of the entire mass probability distribution, whatever its shape, and not just of central values and credibility intervals. 

In this section, we build a regularized catalog of stellar BH masses and orbital parameters. For this, we have extensively searched the current literature and collected data both already present {and not present} in existing black hole catalogs \citep{blackcat,watchdog}, resulting in a total of 35 objects for which we have recalculated all masses, where possible, based on a standardized method and Monte Carlo computations. In what follows, we first discuss the treatment of uncertainty when it comes to the mass estimates (Section \ref{sec:uncertainties}), before briefly reviewing the nature of the observed systems (Section \ref{sec:nature_obs}), introducing the standard mass computation procedure (Section \ref{sec:mass_comp}) and presenting them system-by-system, along with some outstanding conflicts in their study (Section \ref{sec:catalog}). In Section \ref{sec:full_mass_distr}, we fit simple distributions to the entire mass sample, and make a first, simple, comparison of the resulting Galactic BH mass distribution to the extragalactic distribution from the LVK observations. Finally, in Section \ref{sec:gap}, we present a short overview of the current state of the lower mass gap problem in light of recent observations.

\subsection{Treatment of Uncertainties}
\label{sec:uncertainties}

When probing reality, no measurement of continuous quantities is exact. With some degree of uncertainty always present, the best that can be hoped for is a \textit{probability distribution} (rigorously, a probability density function, or PDF) for the actual value of each observable. More often than not, as a consequence of the \textit{central limit theorem}, the resulting distribution is best described by a Gaussian, which, conveniently, is fully defined by its mean (equal to its mode), given as the observable's nominal value; and its standard deviation $\sigma$, which measures the degree of certainty with which the observable's actual value has been constrained.

Even though it is a very elegant way to describe data, representing the distribution of a physical quantity with only two values can distort the actual measurement, especially when uncertainties are large and the quantity of interest has been derived from other quantities with their own distributions. A problem that one finds when delving into the literature of compact objects is the lack of clarity in the definition of each measurement. Over many different authors and years of publication, definitions of the credibility ranges of reported results have not always been explicitly given, and nominal values as well are not clearly defined to report either the mean or mode of the corresponding distribution, which are the same only for unimodal symmetric distributions.

\begin{figure}[H]
{\includegraphics[width=0.8\linewidth]{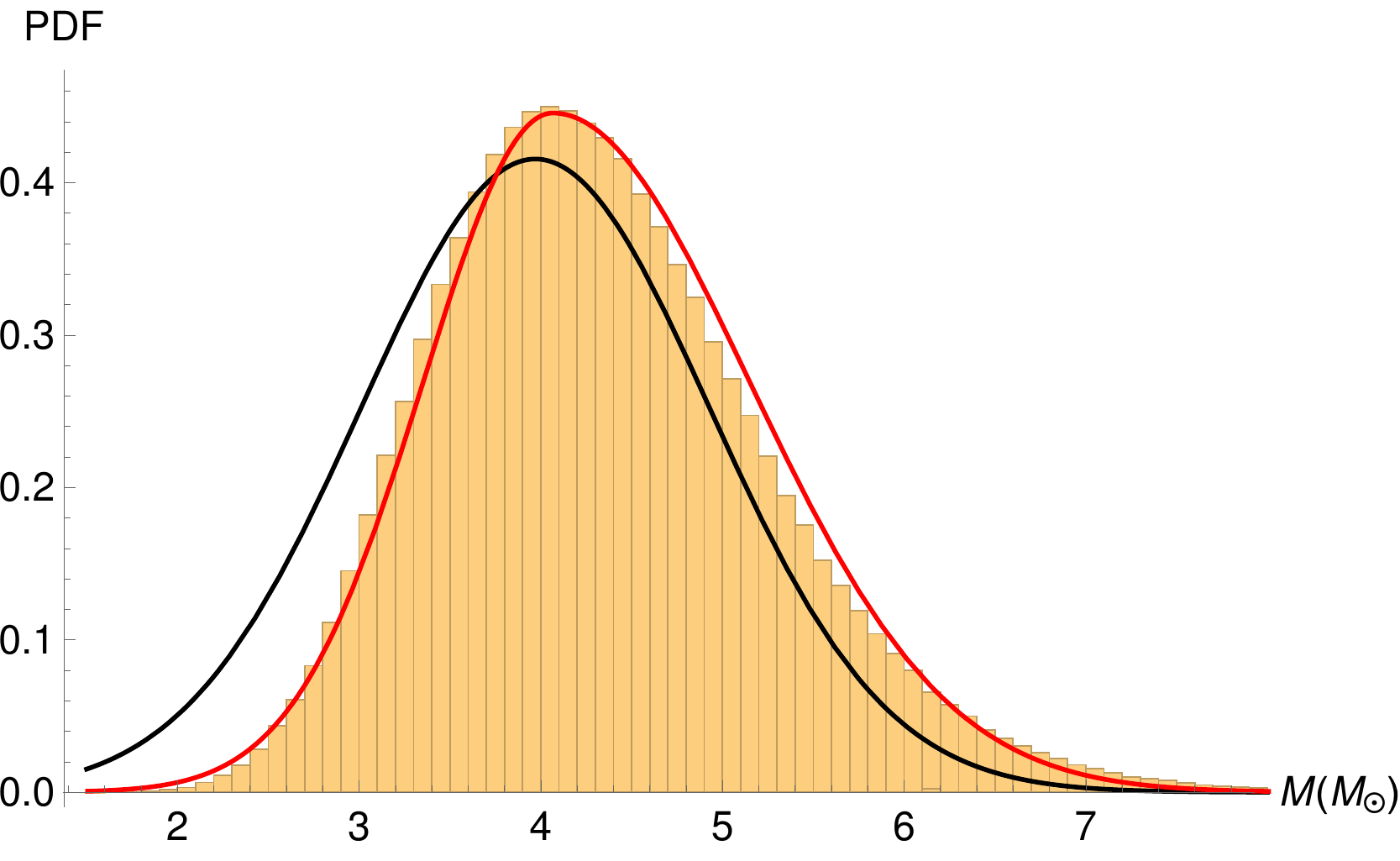}}
\caption{$10^6$ Monte Carlo generated masses from GRO J0422+32 (histogram), along with the Gaussian distribution $N(3.97,0.95)\Msun$ from \cite{gelino_gro_2003} (black line) and our fit of an asymmetric Gaussian $AN(4.06, 1.08, 0.71)\Msun$ (red line). The distributions $N$, $AN$ are defined in the text. }
\label{fig:hmxb_example_mass}
\end{figure}

Dynamic mass measurements of black holes can be particularly vulnerable to this, as they are derived from the measurement of orbital parameters such as the period, velocity semi-amplitude of the companion and, of particular significance, the inclination of the orbital plane, on which mass estimates depend as $\propto\sin i^3$, and which tends to make its uncertainties more asymmetric. Figure \ref{fig:hmxb_example_mass} shows the degree to which BH dynamic masses can deviate from a simple Gaussian due to a broad inclination range. This gives rise to the necessity of a more precise, yet still easily reproducible way of describing such quantities.

We therefore suggest and follow in this paper the practice of always providing any parameters of interest, whether directly observed or calculated, in the form of best-fit probability distributions of their value, so that a minimum of information is lost when using the data elsewhere. For simplicity, we will work with only three types of distribution: a uniform distribution, $U(x_1,x_2)$, between $x_1$ and $x_2$; a simple Gaussian distribution, $N(\mu,\sigma)$, with mean $\mu$ and standard deviation $\sigma$; and what we term an \textit{asymmetric Gaussian}, $AN(m,\sigma_1,\sigma_2)$, defined as 

\begin{equation}\label{AN}
    x\sim AN(m,\sigma_1,\sigma_2)=
    \begin{cases}
       \text{$\sigma_2 N(m,\sigma_2)$,} &\quad\text{if } x\le m, \\
       \text{$\sigma_1 N(m,\sigma_1)$,} &\quad\text{if }x > m, \\    
     \end{cases}
\end{equation}

\noindent as a distribution over a variable $x$, where \textit{m} is the mode of the distribution (not equal to its mean), while $\sigma_1$ and $\sigma_2$ we call the superior and inferior "standard deviations". Formally, the parameter $\sigma_1$ ($\sigma_2$) is defined st. the probability of a random variable $x$ to be drawn from $[m,m+\sigma_1]$ ($[m-\sigma_2,m]$) is $\frac{\mathrm{Erf}(1/\sqrt 2)}{2} \sim 0.34$ times the probability of $x$ to be drawn from $[m,\infty)$ (-$(\infty,m]$).

We highlight that, in any case, the probability distribution for the mass of an individual BH is a reflection solely of current limitations in our knowledge of its related observable properties (orbital parameters, generally). Neither the asymmetric nor the simple Gaussian are taken as definite best-fits to the known distributions; as stated above, a simple Gaussian often results naturally to good approximation as a consequence of the central limit theorem. The proposed asymmetric Gaussian is the simplest extension of the symmetric distribution for observables with asymmetrical confidence intervals.

Equipped with these distributions, we examined all BHs with available mass measurements and carefully re-estimated their masses from known orbital parameters. We also included some well-constrained masses as they are reported, even if orbital parameters were not available, as discussed next.

\subsection{Nature of Observations}
\label{sec:nature_obs}

As dark objects, and except in the case of GW signals from BH--BH mergers, BHs are observed exclusively by their interaction with other luminous sources, which in all but one existing detection means a binary companion. In some of these cases, transversal and radial velocity (RV) measurements from an observed star can be obtained and are enough to constrain the presence of a binary companion in order to explain the motion. If a massive companion is required, but none is observed, then it is most likely to be a BH. There have been so far three cases in which BH masses were measured in this manner for long-period giant star-black hole binaries, where only the giant is directly observed \cite{liu_wide_2019,thompson_noninteracting_2019,el-badry_sun-like_2022}.

For closer binaries, however, another luminous source comes into play: accretion. Either by filling its Roche lobe or from stellar winds, the companion loses mass, part of which is captured by the compact component. As this matter falls toward it, it heats up from the conversion of gravitational potential energy and emits chiefly in the X-rays. Although these X-ray sources are collectively called X-ray Binaries (XRBs), they contain a large "zoo" of  binary classes, starting from whether the compact object is an NS or a BH. In the following, we will briefly review some of the terminology and important observational methods for studying BHs in XRBs, but much of the discussion applies also to NS-hosting XRBs.

A  first important distinction to be made among XRBs is whether accretion occurs through an accretion disk or not; the determining factor is if the specific angular moment $J$ of the infalling matter allows for direct impact onto the compact object. If matter loses energy but no angular momentum, then it would orbit at the \textit{circularization radius}:

\begin{equation}
    \label{eq:circ_radius}
    R_\mathrm{circ} = \frac{J^2}{GM},
\end{equation}

\noindent for an accretor of mass $M$ \cite{king2006accretion}. Typically, disk formation occurs if the accretor's effective radius is smaller than $R_\mathrm{circ}$. For white dwarfs and NSs, the "effective radius" is equal to their radius if there is no significant magnetic field, but is of the order of the magnetospheric radius otherwise. For BHs, it is the radius of the innermost stable circular orbit (ISCO). As matter is accreted, both energy and angular momentum loss occur through viscosity. Once the above condition is fulfilled, disk formation occurs as energy is dissipated by viscosity and radiation faster than angular momentum is redistributed throughout the disk. The loss of energy drives the gas towards the lowest-energy orbit with its slowly-varying angular momentum, which is a circular one; thus, all the accreted gas settles into a series of concentric circular orbits, forming a disk. As gas keeps losing energy, the only way to fall to a lower energy orbit is by losing angular momentum, and in the absence of external torques, this occurs mainly by transferring momentum outward, causing the outer parts of the disk to spiral out \cite{ShakuraDIsk,ShapiroDisk}. While this skips over many nuances of the process, the basic picture of the accretion disk is thus formed, as an efficient machine for lowering highly-rotating material onto the accretor, while converting the orbital energy into radiation, which we can observe.

The conditions above are nearly always fulfilled for Roche lobe overflow (RLOF), as the lost matter carries with it the specific angular momentum of the ``parent'', and thus disk formation follows. Accretion from a stellar wind, on the other hand, becomes relevant for binaries with an O- or B-type companion in a close orbit, which is in fact the case for the first confirmed BH, Cygnus X-1. A simple calculation can tell us that in wind accretion the compact object captures only a fraction $\sim$$10^{-4}$--$10^{-3}$ of the mass lost by the companion, making it much less efficient than RLOF \cite{Frank2002}. It is only because wind loss rates are so large ($10^{-6}$--$10^{-5}\Msun\text{ yr}^{-1}$) that these sources can still be observable. The winds, however, carry much less angular momentum, and thus do not necessarily lead to disk formation; the wind-fed pulsar Vela X-1, for example, has shown evidence of disk formation in the past \cite{VelaX1WindDisk}, but no such evidence has been observed so far in the known wind-fed BHs.

Most XRBs are observed in a \textit{{quiescent}} state, in which their behavior broadly fits within the picture presented above. Occasionally, however, they are observed to undergo \textit{outbursts}, during which their luminosity is greatly increased. For BH-hosting XRBs, the clearest examples are called soft X-ray transients (SXTs, which can also involve a NS instead), in which quiescence usually lasts for $\sim$1--50$\text{ yr}$ with luminosities of $\sim$$10^{32}\text{ erg s}^{-1}$, while outbursts last for $\sim$\text{weeks}--\text{months} with a luminosity increase to $\sim$$10^{34}\text{ erg s}^{-1}$ \cite{tauris2006}. The most common picture for explaining these outbursts is that of a \textit{disk instability} \citep{Frank2002}, based on a distinction between two disk states: hot and high-viscosity (outburst); or cool and low-viscosity (quiescence). The state of the disk is determined essentially by the degree of hydrogen ionization, thus quiescence requires the absence of any ionization zones within the disk. This also defines the condition necessary for suppressing outbursts and making the system \textit{persistent}: that the disk temperature $T$ always exceeds a characteristic hydrogen ionization temperature $T_\mathrm{H}\sim6500\text{ K}$. Naturally, disks where is always true that $T<T_\mathrm{H}$ would also be persistent, but very faint and lacking outbursts entirely; thus, if they exist, there is a strong bias against their detection \cite{king2006accretion}. We refer the reader to the recent review by \citet{SXTDiskRev} for a discussion of the disk instability model in the context of current problems.

The case where the disk temperature varies between being above or below $T_\mathrm{H}$ brings us back to the SXTs. In this case, the outbursts are triggered by the formation of an ionization zone somewhere within the disk, which will be in the hot, viscous, state. This state of heightened energy dissipation favors an increased accretion rate onto the compact object, and thus an increased X-ray luminosity. The increased X-ray luminosity brings more and more of the disk into the hot state, increasing even more the rate of accretion, and so on. The heightened accretion tends to decrease the surface density of the gas, but the central X-ray irradiation stops this from effectively cooling the disk, which must remain trapped in the hot state until the accretion rate itself drops after a considerable accretion onto the compact object. This picture fits well with short-period SXTs, where the radius of the disk is limited, and it can become completely ionized during an outburst. Long-period SXTs, on the other hand, are able to comport much larger disks, a large portion of which can remain cold even during outbursts. The cold disk can act as a mass reservoir, making outbursts much longer than usual, to the point where some known persistent sources may actually be very long transients; a mass reservoir can also cause outbursts to occur in much quicker succession than expected. Disk warps also play a role, as they can affect the efficiency with which the X-ray source heats the full extent of the disk \cite{DiskWarpRev2014,DiskWarpRev2020}. At the same time, a general ``exhaustion'' of the disk's mass during long outbursts points towards very long quiescent periods as well, which implies a population of quiescent transients that have never been observed during an outburst \citep{king2006accretion}.

\begin{figure}[H]
    \includegraphics[width=\textwidth]{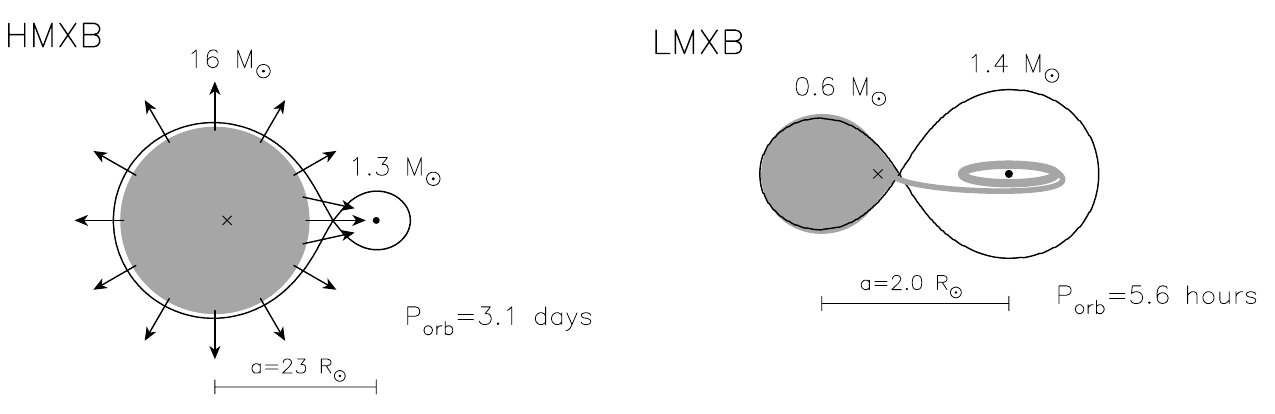}
    \caption{Example of typical LMXB and HMXB. For LMXBs, accretion occurs always through Roche lobe overflow, while, in HMXBs, wind-fed accretion is also possible, sometimes simultaneously to Roche lobe overflow. Note that, while we focus on BH XRBs, the same configurations are found for NS XRBs. {Figure obtained from} \citet{tauris2006} {in Compact Stellar X-ray Sources (Copr. Cambridge University Press 2006), reproduced with permission of Cambridge University Press through PLSclear.}}
    \label{fig:lmxb_hmxb}
\end{figure}

The discussion above leads naturally to the main distinction to be made among XRBs: low-mass X-ray binaries (LMXBs) and high-mass X-ray binaries (HMXBs). LMXBs are XRBs with luminous components with typical masses of $\sim1\Msun$ or less, and so are restricted to RLOF accretion, which means that they virtually always contain an accretion disk (see the right side of {Figure} \ref{fig:lmxb_hmxb}). Most, if not all, LMXBs are SXTs \citep{king2006accretion}. We also see it fit to mention here that the phenomenon of X-ray bursts (and their associated binaries, the X-ray bursters) are entirely distinct from the disk outbursts. Type I X-ray bursts are thermonuclear in nature, similarly to novae, and can only occur in NS-hosting XRBs. For the less well understood Type II X-ray bursts, disk instability models have been proposed, but are not favored (see \cite{TypeIIBurst} for a review of some current proposals).

\begin{figure}[H]
    \includegraphics[width=0.7\textwidth]{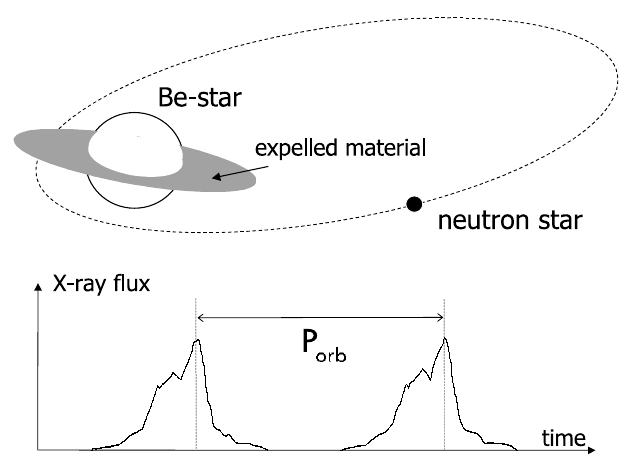}
    \caption{Illustration of a Be X-ray binary. The compact object does not accrete for most of its eccentric orbit as the Be star does not fill its Roche lobe, and the period is too long for wind-fed accretion to take place. At periastron, however, the compact object crosses the circumstellar disk of the Be star and builds up an accretion disk, leading to a burst. Although the example indicates a neutron star, Be star-black hole X-ray binaries can also form. {Figure obtained from} \citet{tauris2006} {in Compact Stellar X-ray Sources (Copr. Cambridge University Press 2006), reproduced with permission of Cambridge University Press through PLSclear.}}
    \label{fig:be_xray}
\end{figure}

HMXBs contain massive companions ($\gtrsim$10 $\Msun$) and generally undergo wind-fed accretion (see the left side of Figure \ref{fig:lmxb_hmxb}), but can also accrete from RLOF, and form a disk. Even when they do have a disk, HMXBs rarely show disk instability outbursts because companion stars of type O or early B are themselves potent enough sources to keep the entire disk ionized ($T>T_\mathrm{H}$). Outbursts can then still occur if the components are distant enough during the orbit, i.e., if the period is longer than $\sim$10 $\text{d}$ \citep{tauris2006}, or for high-eccentricity orbits. These conditions are found in the particular class of Be X-ray binaries \citep{BeXbinary2011}, in which the compact object accretes matter for the Be star's own circumstellar disk; in this case, outbursts are regularly observed and probably caused by a burst of accretion near periastron (Figure \ref{fig:be_xray}). Although for some time only Be-NS binaries could be found, the first Be-BH binary was discovered in 2017 by \citet{first_be_bh}.

Wind-fed HXMBs are generally persistent sources (classical HMXBs), with X-ray luminosities on the order of $\sim$$10^{35}\text{ erg s}^{-1}$ \cite{HMXBLum}. However, a class of Supergiant Fast X-ray Transients (SFXTs) was discovered in 2005 \citep{sguera2005,negueruela2006}, which shows an average X-ray luminosity of $\sim$$10^{34}\text{ erg s}^{-1}$, but bright X-ray flares lasting for a few days, composed of a series of bursts lasting for $\sim$$10^3\text{ s}$ and with $L_X\gtrsim10^{36}\text{ erg s}^{-1}$ \citep{sidoli2017}. 

We can now point toward the sources for the majority of the mass estimates presented here. All but 4 of the 35 systems considered are XRBs; among them, 26 are LMXBs and 5 HMXBs. For all HMXBs and 22 of the LMXBs, we have recalculated the black hole mass $\Mbh$ from known orbital parameters $\Porb$, $\Kcp$, $q$, $i$ and, when available, $e$. Observations of the luminous companion provides a powerful means to constrain the orbital period $\Porb$ and companion velocity semi-ampitude $\Kcp$. These constrain the mass function $f(\Mbh)$, which, together with the companion's spectral type, constrains the mass ratio $q$. For LMXBs, outburst observations are also available, although, for 4 of the LMXBs, only a mass estimate was available, which we adopted as given.

The inclination $i$ is harder to pin down. If an absence of eclipses is confirmed, an upper limit can be established from $q$ as indicated in the next section. If a disk is present, spectral methods for disk reflection can constrain $i$. When jets are present (in this case, sometimes the system is called a \textit{microquasar} \citep{microquasars2011}), their inclination can be measured to a good degree of accuracy; however, it is not a safe assumption to take the jet inclination to be the same as the orbital inclination. In the general case of modeling the companion's motion, detailed models can provide best-fit constraints on the inclination.

Indirect observation of BHs in non-interacting binaries can still occur through the monitoring of the companion's motion, and the masses of three BHs in our sample have been constrained in this manner. Observation of isolated BHs is also possible from the microlensing of background light, but only one such event, which we have also included, has been confirmed so far to have been caused by a BH. Low-frequency Quasi Periodic Oscillations (LF QPOs) in the X-ray spectrum provide yet another way to measure the mass of a black hole in a XRB. QPOs appear in the power spectrum as narrow peaks and evolve through outbursts, with LF QPOs appearing with $\lesssim$$30\text{ Hz}$ (for a review of QPOs and their different classes, see \cite{QPOreview}). XRB spectra can often be fitted by the superposition of a blackbody component and of a power law component \citep{mcclintock_remillard_2006}, and \citet{titarchuk_fiorito_2004} produced a BH mass-dependent model which correlates the fitted photon index $\Gamma$ of the power-law component to the observed LF QPO frequency $\nu$. This model has since also been used to measure the mass of XRBs in which LF QPOs are observed.

\subsection{Mass Computation from Orbital Parameters}
\label{sec:mass_comp}

From all binaries, where available, we collect the orbital period $\Porb$ and the velocity semi-amplitude $\Kcp$ of the companion to the BH. From these quantities, plus the eccentricity $e$, the BH mass function is computed as 

\begin{equation}
    \label{eq:mass_function}
    f\left(\Mbh\right) = \frac{\Porb\Kcp^3}{2\pi G}(1-e^2)^{3/2},
\end{equation}

\noindent when considering a Keplerian orbit \citep{CharlesCoe2006}. The error incurred in the BH mass from assuming a Keplerian orbit is still much smaller than that from observational uncertainties in Keplerian parameters, in particular in the orbital inclination, which might be affected by serious systematic errors not yet taken into account in the present work \citep{kreidberg_mass_2012}. We therefore employ Equation (\ref{eq:mass_function}) as the relation between binary orbital parameters throughout the work.

In the following original sources for the LXMBs, we treat all systems as circularized, although a small but non-zero eccentricity might be allowed for them. For the HMXBs, eccentricities $e\lesssim0.01$ are reported, except for Gaia BH1, which was measured to have a modest $e\approx0.45$. While for Gaia BH1, we consider its eccentricity in Equation (\ref{eq:mass_function}), for all other systems we treat the orbit as circular. Once again, any incurred errors from this approximation are expected to be inferior to those from observational uncertainties and possible systematics not taken into account at present.

For each system, we also collect the binary inclination $i$ and the mass ratio $q=\Mcp/\Mbh$. With these parameters, the BH mass can be determined as

\begin{equation}
    \label{eq:mbh_def}
    \Mbh = f\left(\Mbh\right)\frac{(1+q)^2}{\sin^3i}.
\end{equation}

We employ the above formula as a standardized computation of $\Mbh$ for \textit{all} systems with orbital parameters, even when it results in a difference to the $\Mbh$ reported by the cited sources. In most cases where significant differences arise, they do so only as a broader $1\sigma$ range; and any considerable differences in the central value still keep the original mass within $1\sigma$ of ours. The orbital period $\Porb$ is always an observed quantity, while $\Kcp$ is taken as an observed quantity in all but one case, in which it is computed from $\Porb$ and $f\left(\Mbh\right)$. $q$ is in some cases taken as reported by the respective source, while, in others, it is computed from the source's quoted $\Mbh$ and $\Mcp$; in two cases, we re-estimate it from the companion's spectral type. The inclination $i$ is taken as reported from the sources. In all cases where the sources report a nominal value for a given quantity, that quantity is treated as an asymmetric Gaussian (Section \ref{sec:uncertainties}) with mode at the nominal value; if the upper and lower uncertainties are reported, they are taken as $\sigma_1$ and $\sigma_2$; otherwise, these are set within $10\%$ of the mode. When only upper and lower bounds are reported, the quantity is taken as being distributed uniformly between them.

Whenever the inclination has not receive an upper limit while the mass ratio has been constrained, a lack of eclipses allows the determination of a maximum inclination from

\begin{equation}
    \label{eq:cosi_no_eclipses}
    (\cos i)_\mathrm{min} = 0.462\left(\frac{q}{1+q}\right)^{1/3}.
\end{equation}

Computations of Equations (\ref{eq:mass_function}) and (\ref{eq:mbh_def}) are performed via Monte Carlo, and we fit our results to an asymmetric Gaussian, which in some cases results in a symmetric distribution regardless. The {orbital parameters and resulting mass distributions} are reported in {Tables} \ref{tab:ecc} and \ref{tab:bh_catalog}. Below, we offer a case-by-case, non-exhaustive, discussion of the objects reported in {Table} \ref{tab:bh_catalog} and their references, under the considerations made in Section \ref{sec:uncertainties}. We report the results as given by the references; in some cases, the mass ratio is reported as $Q=\Mbh/\Mcp$, or the masses $\Mbh$, $\Mcp$ themselves are given. In those cases, we use MC to convert their distributions to a distribution of $q$, unless stated otherwise.

\subsection{Collected Systems}
\label{sec:catalog}

\begin{figure}[H]
    \includegraphics[width=\textwidth]{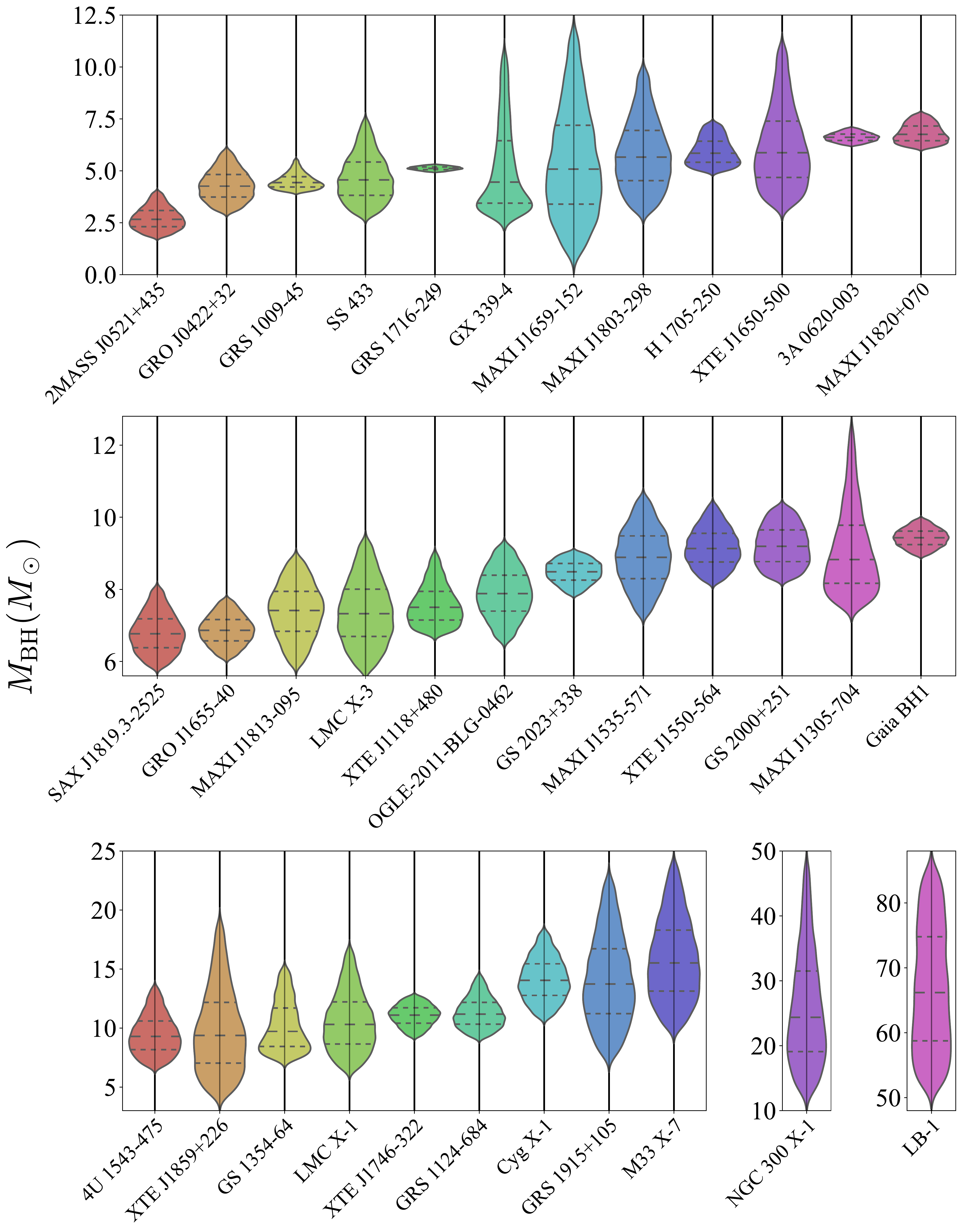}
    \caption{Violin mass plot of all 35 BHs in our sample, ordered by their mean mass (left-right, top-down). The horizontal lines within each of the plots indicate the quartiles of the respective distribution.}
    \label{fig:BHviolin}
\end{figure}

We present below the full catalog of BH mass measurements from 26 LMXBs, 5 HMXBs, 3 non-interacting binaries and 1 isolated BH. We discuss each object in turn, briefly indicating the manner of measurement, the original results we have adopted for our catalog, and conflicts found during their collection. The full sample is displayed in {Table} \ref{tab:bh_catalog} with sources, our standardized mass determination and all orbital parameters except for the eccentricity, which was only explicitly available for six systems; for those we display, the eccentricity in {Table} \ref{tab:ecc}. We show in Figure \ref{fig:BHviolin} the resulting mass distributions for all 35 BHs.

\begin{table}[H] 
\caption{Masses and Binary Parameters, where applicable, for six Black Hole Candidates.\label{tab:ecc}}
\newcolumntype{C}{>{\centering\arraybackslash}X}
	 \begin{tabularx}{\textwidth}{lCr}
		\toprule
            \textbf{Name} & \boldmath$e$ & \textbf{References}\\
            \midrule
            Cyg X-1 & $N(0.018,0.003)$ & \cite{orosz_mass_2011} \\
            LMC X-1 & $N(0.0256,0.0066)$ & \cite{orosz_new_2009} \\
            M33 X-7 & $N(0.0185,0.0077)$ & \cite{orosz_1565-solar-mass_2007} \\
            SS 433 & $N(0.050,0.007)$ & \cite{cherepashchuk_progress_2022} \\
            LB-1 comp. & $N(0.03,0.01)$ & \cite{liu_wide_2019} \\
            2MASS J05215658+4359220 comp. & $N(0.0048,0.0026)$ & \cite{thompson_noninteracting_2019} \\
            Gaia BH1  & $N(0.451,0.005)$ & \cite{el-badry_sun-like_2022} \\ 
        \bottomrule
	\end{tabularx}
		\noindent\footnotesize{{Eccentricities} for the seven objects for which they were given, alongside the reference. $N(\mu,\sigma)$ is a normal distribution with mean $\mu$ and standard deviation $\sigma$, as defined in Section \ref{sec:uncertainties}.}
\end{table}

\subsubsection{LMXBs}

\paragraph{\textbf{Sources with orbital parameters}}

{We} list here the 22 LMXBs for which the necessary orbital parameters for the procedure described in Section \ref{sec:mass_comp} were available. 

\paragraph{4U 1453-475}

\citet{orosz_inventory_2003} compile an inventory of reliable BH-hosting binaries at the time of writing and their relevant parameters. We adopted their reported $\Porb=1.116407(3)\text{ d}$ and $i=20.7\pm1.5^\circ$. We convert their $Q=3.2$--$4.0$ range to a uniform $q$ distribution. We work backwards to compute $\Kcp$ from their $f\left(\Mbh\right)=0.25\pm0.01\Msun$ and $\Porb$.

\paragraph{GRS 1915+105}

\citet{greiner_unusually_2001} use spectroscopic data to measure $\Porb=33.5\pm1.5\text{ d}$ and $\Kcp=140\pm15\text{ km s}^{-1}$. The authors report that the jet angle determination of $70\pm2^\circ$ has been observed to be stable over several years, so that it is reasonable to assume it as the orbital plane inclination $i$. With this inclination and a companion mass estimate of $\Mcp=1.2\pm0.2\Msun$, they find $\Mbh=14\pm4\Msun$. We adopt their $\Porb$, $\Kcp$ and $i$, and compute $q$ from their $M$ estimates.

\paragraph{GS 1354-64}

\citet{casares_refined_2009} present a refined treatment of GS 1354-64 (BW Cir) in relation to the first evidence of the presence of a BH presented in \cite{casares_evidence_2004}, and with spectroscopic and photometric data find $\Porb=2.54451(8)\text{ d}$, $\Kcp=279.0\pm4.7\text{ km s}^{-1}$ and $q=0.12^{+0.03}_{-0.04}$, which we adopt. They report only an upper limit of $i\leq79^\circ$ for the inclination, so we adopt $U(50,80)^\circ$ from \cite{farr_mass_2011}.

\paragraph{GRS 1124-684}

\citet{wu_dynamical_2015} present a study of optical spectroscopic and photometric data of GRS 1124-683 (Nova Muscae 1991), and from the radial velocity of the companion are able to determine $\Porb=0.43260249(9)\text{ d}$, $\Kcp=406.8\pm2.7\text{ km s}^{_1}$ and $q=0.079\pm0.007$, which we adopt. We adopt the inclination from their following study of the object \citep{wu_mass_2016}, in which they constrain it to $i=(43.2^{+2.1}_{-2.7})^\circ$. 

\paragraph{XTE J1118+480}

\citet{gonzalez_hernandez_fast_2014} use NIR spectroscopic data of XTE J1118+480 to measure $\Porb=0.1699337(2)\text{ d}$, yielding values compatible with \cite{torres_mmt_2004, gonzalez_hernandez_fast_2012}; and $\Kcp=710.0\pm2.6\text{ km s}^{-1}$, also compatible with previous work \citep{hernandez_chemical_2008}. We adopt their results, and their quoted $q=0.024\pm0.009$ \citep{calvelo_doppler_2009} and $i=73.5\pm5.5^\circ$ \citep{khargharia_mass_2012} from previous works.

\paragraph{3A 0620-003}

\citet{gonzalez_hernandez_fast_2014} recalculate the orbital period derivative for 3A 0620-003 with new spectroscopic data from \cite{hernandez_doppler_2010} and obtain a more precise value of $\Porb=0.32301415(7)\text{ d}$ for the period, which is compatible with previous work \citep{mcclintock_black_1986}. We adopt their quoted $\Kcp=435.4\pm0.5\text{ km s}^{-1}$, $q=0.060\pm0.004$ \citep{neilsen_eccentric_2008} and $i=51.0\pm0.9^\circ$ \citep{cantrell_inclination_2010}.

\paragraph{GS 2000+251}

\citet{ioannou_mass_2004} use I- and R-band photometric data to constrain the orbital parameters of GS 2000+251, and we adopt their nominal result of $\Porb=0.344086(2)\text{ d}$, $Q=24^{+9}_{-6}$ and the quoted $\Kcp=519.5\pm5.1\text{ km s}^{-1}$ from \cite{Harlaftis1996_ratio}. For the inclination, from their lower and upper limits, we adopt a uniform $U(54,60)^\circ$ distribution.

\paragraph{MAXI J1659-152}

\citet{torres_delimiting_2021} use spectroscopic data of the MAXI J1659-152 X-ray transient's quiescent counterpart to constrain its orbital properties and confirm the compact object's BH nature. We adopt their $P_\mathrm{orb}=2.414\pm0.005\text{ h}$ and $\Kcp=750\pm80\text{ km s}^{-1}$. We adopt their report lower and upper limits for the mass ratio and inclination as distributions $U(0.02,0.07)$ and $U(70,80)^\circ$, respectively.

\paragraph{MAXI J1305-704}

\citet{mata_sanchez_dynamical_2021} present photometric and spectroscopic data of the quiescent state of MAXI J1305-704 to confirm the presence of a BH in the binary. We adopt their determined $\Porb=0.394\pm0.004\text{ d}$ and $\Kcp=554\pm8\text{ km s}^{-1}$. For the mass ratio, the authors adopt a $N(0.05,0.02)$ distribution truncated to $0.01<q<0.07$. We adapt this distribution to an asymmetric normal by keeping the central value at $0.05$ and considering its distance to the truncation limits as $3\sigma$, resulting in a distribution $AN(0.05, 0.007, 0.013)$. We keep the authors' favored inclination model, $i=(72^{+5}_{-8})^\circ$.

\paragraph{GS 2023+338}

\citet{casares_optical_1994} use spectroscopic data to constrain the system parameters of GS 2023+338. We adopt from their Table 2 $\Porb=6.4714(1)\text{ d}$, $\Kcp=208.5\pm0.7\text{ km s}^{-1}$ and $q=0.060^{+0.004}_{-0.005}$. Based on the parameters from \cite{casares_optical_1994,khargharia_near-infrared_2010}, use new NIR spectroscopic data of GS 2023+338 to measure the inclination of the system as $i=(67^{+3}_{-1})^\circ$, which we adopt.

\paragraph{XTE J1650-500}

\citet{orosz_orbital_2004} use R-band photometric data from XTE J1650-500 and confirm an orbital period $\Porb=0.3205(7)\text{ d}$. Reanalysis of archival spectroscopic data yields the velocity semi-amplitude $\Kcp=435\pm30\text{ km s}^{-1}$. From the light curve, the authors find an inclination lower limit of $50\pm3^\circ$ and report an upper lower limit of $\approx80^\circ$ from the absence of eclipses. Although the exact upper limit depends on the mass ratio, we adopt a $U(50,80)^\circ$ distribution for $i$. The mass ratio can be measured from the companion's spectral type, but due to the small amount of template spectra available, cannot be well-constrained. From the sample available, the best match is reported to be K4 V, from which we derive a conservative distribution of $U(0.01,0.5)$ for $q$.

\paragraph{GRO J0422+32}

\citet{webb_tio_2000} use photometric and spectroscopic data of GRO J0422+32 to model the binary. We adopt their measured $\Kcp=378\pm16\text{ km s}^{-1}$ and $\Porb=0.2121600(2)\text{ d}$, and also the derived $Q=9.0^{+2.2}_{-2.7}$. They constrain the mass of the compact object to >$2.2\Msun$, at that point still allowing for a massive NS. Ref. \cite{gelino_gro_2003} studied new optical and IR photometric data of the X-ray transient and were able to show that it should contain a light BH, not an NS. They measure the most likely value for the inclination as $i=45\pm2^\circ$, which corresponds to a most likely $M_\mathrm{BH}=3.97\pm0.95\Msun$. While \cite{webb_tio_2000} find M4 V as the best-matching spectral type for the companion, Ref. \cite{gelino_gro_2003} arrive at M1 V; they argue that an M4 V type cannot be made compatible with the observations unless an extra source of blue luminosity is posited, but that this source is also incompatible with observations.

\paragraph{H 1705-250}

\citet{remillard_dynamical_1996} use photometric and spectroscopic data of H 1705-250 (Nova Ophiuchi 1977) to support the presence of a BH in the system and measure $\Porb=0.5228(44)\text{ d}$ and $\Kcp=441\pm6\text{ km s}^{-1}$. The companion's spectrum is best matched by a type K5, yielding $q=0.014^{+0.019}_{-0.012}$, which we adopt.

\citet{harlaftis_doppler_1997} use spectroscopic data from Keck to model H 1705-250 and fit also a K5 spectrum to the companion. They constrain the inclination to $60^\circ<i<80^\circ$, and we adopt this result as a distribution $U(60,80)^\circ$.

\paragraph{GRO J1655-40}

\citet{hernandez_black_2008} use new optical and UV, as well as archival NIR, spectrographic data to study GRO J1655-40 (Nova Scorpii 1994) with the main goal of performing an abundance analysis of its secondary. They provided updated measurements of the orbital parameters which we adopt: $\Porb=2.62120(14)\text{ d}$, $\Kcp=226.1\pm0.8\text{ km s}^{-1}$ and $q=0.329\pm0.047$. For the inclination, they adopt $i=68.65\pm1.5^\circ$ from \cite{beer_quiescent_2002}, and so we also adopt this result. 

The authors suggest that the $q=0.419\pm0.028$ found by \citet{shahbaz_determining_2003} might be more accurate by virtue of having incorporated the secondary's Roche geometry in their spectral analysis. However, \citet{shahbaz_determining_2003} use $i=70.2\pm1.9^\circ$ from \citet{greene_optical_2001} in their study. We choose to keep both the $q$ and $i$ from \citet{hernandez_black_2008} for consistency. 

\paragraph{XTE J1859+226}

\citet{corral-santana_evidence_2011} use spectroscopic and photometric data of XTE J1859+226 and find from the secondary's motion $\Porb=6.58\pm0.05\text{ h}$ and $\Kcp=541\pm70\text{ km s}^{-1}$, which we adopt. These values imply a mass function $f\left(\Mbh\right)=4.5\pm0.6\Msun$, in considerable excess of the previously reported $f\left(\Mbh\right)=7.4\pm1.1\Msun$ \citep{filippenko_xte_2001}, but we note that the older result has only been presented in an IAU Circular and has been considered unreliable \citep{ozel_black_2010}. The authors also report a best-match K5 V for the companion's spectral type, and we set for $q$ a distribution $U(0.01,0.5)$.

The absence of eclipses imposes an upper limit $i\lesssim 70^\circ$ on the inclination, but the available data are considered not accurate enough to properly estimate the inclination angle. By assuming a secondary similar to that of 3A 0620-00 and the known orbital parameters, the authors arrive at a preferred range $60^\circ\leq i\leq70^\circ$. We thus adopt for the inclination a distribution $U(60,70)^\circ$.

\paragraph{MAXI J1803-298}

\citet{sanchez_hard-state_2022} use optical photometric data taken during the discovery outburst of MAXI J1803-298 to model the binary and provide evidence supporting the compact object's BH nature. Its orbital period has so far only been measured as $\Porb\approx0.29\text{ d}$, so we adopt a $N(0.29,0.03)\text{ d}$ distribution. The authors are able to constrain the companion's velocity semi-amplitude to a $\Kcp\sim$460--570$\text{ km s}^{-1}$ (410--620$\text{ km s}^{-1}$) $68\%$ ($95\%$) confidence interval, so we take $N(515,55)\text{ km s}^{-1}$ as our $\Kcp$ distribution. For the mass ratio, they assume a typical $0.01\leq q\leq0.2$ range, which we take as a $U(0.01,0.2)$ distribution. The authors report a >$65^\circ$ lower inclination limit but no upper limit; as no eclipses are reported, we compute an upper limit of $\sim85^\circ$ and take $U(65,85)^\circ$ as the $i$ distribution.

\paragraph{MAXI J1820+070}

\citet{torres_dynamical_2019} confirm the presence of a BH in the X-ray binary MAXI J1820+070 with spectroscopy from its decline to the quiescent state, and we adopt their $\Porb=0.68549(1)\text{ d}$ and $\Kcp=417.7\pm3.9\text{ km s}^{-1}$. Ref. \cite{torres_binary_2020} use optical spectroscopy of the object to further constrain the mass ratio to $q=0.072\pm0.012$ and the inclination to $66^\circ<i<81^\circ$, both of which we adopt, the inclination as $U(66,81)^\circ$. 

We note also that \citet{atri_radio_2020} provide a tighter constraint for the jet inclination as $63\pm3^\circ$. \citet{torres_binary_2020} consider the case where the orbital inclination is taken to be the same as the jet inclination, obtaining a $8.48^{+0.79}_{-0.72}\Msun$ estimate for $\Mbh$, while their more conservative inclination range yields $5.73<\Mbh/M_\odot<8.34$.

\paragraph{XTE J1550-564} 

\citet{orosz_improved_2011} fit optical and NIR spectroscopic and photometric data from observations of XTE J1550-564 to a set of eight lightcurve models. Although the authors point out that there are conflicts between data taken at different times, we adopt their nominal result $\Porb=1.5420333(24)\text{ d}$, $\Kcp=363.14\pm5.97\text{ km s}^{-1}$ and $i=74.7\pm3.8^\circ$; and, from their $Q\approx30$, we adopt $q=N(0.033,0.003)$. Ref. \cite{connors_evidence_2020} reports an inclination estimate of $i=37\pm4^\circ$ which is inconsistent with the older estimate. They conclude that it may be the case that this object has a warped disk, or that the disk structure may be obscuring blueward line emission, resulting in lower inclination estimates from reflection modeling. 

\paragraph{GX 339-4}

\citet{heida_mass_2017} measure the donor star RV curve with data from VLT/X-shooter, chiefly NIR and optical, from which they obtain $\Porb=1.7587(5)\text{ d}$ and $\Kcp=219\pm3\text{ km s}^{-1}$, which is significantly lower than the previous lower limit of $317\text{ km s}^{-1}$ \citep{hynes_dynamical_2003}, leading also to a lower $f\left(\Mbh\right)$ than previously reported. The authors argue that the scenario in which their $\Kcp$ is underestimated would imply variations of the absorption lines of the donor along the orbit, while their data cover most orbital phases and shows no such variations, which, among other factors, expanded upon in the cited work, support their measured $\Kcp$. 

From their $\Kcp$, \citet{heida_mass_2017} find $q=0.18\pm0.05$ and a lower limit for the inclination of $37^\circ$. We adopt the quoted upper limit of $78^\circ$ derived from the absence of eclipses and treat the inclination distribution as $U(37,78)^\circ$. We adopt their nominal results for $\Porb,\Kcp$ and $q$.

\paragraph{GRS 1009-45}

\citet{filippenko_black_1999} employ optical spectra from GRS 1009-45 (Nova Velorum 1993) during quiescence and measure $\Porb=0.2852060(14)\text{ d}$ and $\Kcp=475.4\pm5.9\text{ km s}^{-1}$ from the companion's RV curve. Approximate radial velocities of the compact primary from the H$\alpha$ emission line allow for a mass ratio determination $q=0.137\pm0.015$ and inclination $\approx$$78^\circ$. We adopt their results for $\Porb$, $\Kcp$ and $q$, and by our standard treatment assume that the inclination is distributed as $N(78,7.8)^\circ$. We highlight that the lack of eclipses imposes a $\lesssim$$80^\circ$ limit on inclination, but because the mass depends on $\sin^3i$ only, which varies minimally for $i>80^\circ$, we keep $N(78,7.8)^\circ$ as our distribution for simplicity. 

\citet{macias_refined_2011} have reported refined measurements of $q\approx0.006$ and $\Mbh=8.5\Msun$ for this system. We, however, do not employ these results here as they are only available in the form of an abstract, and the cited source provides neither $\Kcp$ nor $f\left(\Mbh\right)$.

\paragraph{LMC X-3}

With a companion of about $\approx3.63\Msun$, LMC X-3 does not fit neatly into the LMXB / HMXB dichotomy we have described before; it can be placed instead in the small class of \textit{intermediate-mass} XRBs, which suffer from a negative selection bias (see \cite{tauris2006}). \citet{orosz_mass_2014} study a large set of new and archival spectroscopic and photometric data to model LMC X-3. We employ the ``Adopted value'' for the parameters from the X-ray heating model in their Table 10, $Q=1.93\pm0.20$ and $i=69.24\pm0.72^\circ$. $\Porb=1.7048089(11)\text{ d}$ and $\Kcp=241.1\pm6.2\text{ km s}^{-1}$ are taken as quoted. Their resulting $\Mbh=6.98\pm0.56\Msun$ is considerably different from older estimates but is deemed more reliable as it derives from a much larger body of observations, and was obtained after measuring and taking into account for the first time the rotation of the secondary. The authors include a brief review of previous results and discuss the sources of inconsistencies.

\paragraph{SAX J1819.3-2525}

From spectroscopic observations of SAX J1819.3-2525 (V4641 Sgr), \citet{orosz_black_2001} confirm the BH nature of the system's compact object and measure $\Porb=2.81730(1)\text{ d}$ and $\Kcp=211.0\pm3.1\text{ km s}^{-1}$, which we adopt. \citet{macdonald_black_2014} gather $~10\text{ yr}$ of photometric data from the system and are able to measure $i=72.3\pm4.1^\circ$ and $Q=2.2\pm0.2$, which we adopt. Although their estimated distance ($6.2\text{ kpc}$) is below that of \citet{orosz_black_2001} ($7.40\leq d\leq12.31\text{kpc}$), their $i$ and $q$ were derived by fixing the same orbital period and a compatible $\Kcp$ \citep{lindstrom_new_2005}, and so we consider the parameters to be compatible. With the companion mass estimate as $5.49\leq\Mcp/\Msun\leq8.14$ ($90\%$) {by} \citet{orosz_black_2001}, this system is also an IMXB candidate.

\paragraph{\textbf{Sources with mass estimate only}}

For another four LMXBs, mass estimates were available, but not their orbital parameters. Although this means that, for these objects, we cannot perform the standardized mass computation from Section \ref{sec:mass_comp}, we still include their masses in our sample. We treat them as symmetric or asymmetric Gaussians and their reported uncertainties as $1\sigma$ unless stated otherwise by the sources. The systems here included are: GRS J1716-249, MAXI J1813-095, MAXI J1535-571, and XTE J1746-322.

\paragraph{GRS J1716-249}

\citet{zhang_testing_2022} study X-ray spectroscopic data from the 2016--2017 outburst of binary GRS J1716-249 with continuum-fitting and ironline methods, with the aim of testing the Kerr nature of the system's BH. For their preferred result of assuming an exact Kerr metric, they find $\Mbh=5.12\pm0.11\Msun$, and within the same model constrain the inclination to $47.5\pm0.6^\circ$. We include both results in our sample.

\paragraph{MAXI J1813-095}

\citet{jana_accretion_2021} perform timing and spectral analyses of the 2018 outburst of binary MAXI J1813-095. They consider three sets of observational data and obtain a mass estimate from each one. We include as a mass estimate the average result, $\Mbh=7.41^{+1.47}_{-1.52}\Msun$ ($90\%$). They also place constraints on the inclination, which we adopt as a distribution $U(28,45)^\circ$.

\paragraph{MAXI J1535-571}

\citet{shang_evolution_2019} perform timing and spectral analysis of data from its 2017--2018 outburst. From their spectral analysis, they find the mass range $\Mbh\sim$7.9--9.9$\Msun$, and we include in our sample their suggested mass $\Mbh=8.9\pm1.0\Msun$.

\paragraph{XTE J1746-322}

Spectral studies of the microquasar XTE J1746-322 (H 1743-322) have been performed recently by \citet{molla_estimation_2017} and \citet{tursunov_constraints_2018}. \citet{molla_estimation_2017} perform timing and spectral analyses on observations from two outbursts, in 2010 and 2011, from which they estimate a mass range $\Mbh=9.25$--$12.86\Msun$. They also study observed QPOs during the outburst, from which a second mass estimate is derived, $\Mbh=11.65\pm0.67\Msun$. By combining the two results, they arrive at $\Mbh=11.21^{+1.65}_{-1.96}\Msun$, which we include in our sample. Notably, \citet{tursunov_constraints_2018} employ a different method to also study the QPOs observed in this object and arrive at a mass estimate of $\Mbh=11.2\,\Msun$.

\subsubsection{HMXBs}

{We list here the five HMXBs with mass estimates, all of which include the necessary orbital parameters for the procedure described in Section} \ref{sec:mass_comp}.

\paragraph{Cyg X-1}

\citet{orosz_mass_2011} use an improved distance measurement for Cyg X-1 together with previously published optical data to model the system. The period is fixed to $\Porb=5.599836(24)\text{ d}$ from \cite{brocksopp_improved_1998} and the data are fitted to four models. We adopt the fixed period and the final parameters from Table 2: $i=27.06\pm0.76^\circ$, $e=0.018\pm0.003$, $\Mbh=14.81\pm0.98\Msun$ and $\Mcp=19.19\pm1.90\Msun$. We compute $q$ with MC from $\Mbh$ and $\Mcp$. We adopt their more scattered $\Kcp=75.57\pm0.70\text{ km s}^{-1}$ fitted {to} the data from \cite{gies_wind_2003}.

\paragraph{LMC X-1}

\citet{orosz_new_2009} use optical spectroscopic, and optical and NIR photometric, data to model LMC X-1 and confirm earlier work by \citet{hutchings_optical_1987} with much higher precision. We employ the ``Adopted Value'' from their Table 3 for $\Porb=3.90917(5)\text{ d}$, $\Kcp=71.61\pm1.10\text{ km s}^{-1}$, $\Mbh=10.91\pm1.54\Msun$, $\Mcp=31.79\pm3.67\Msun$ and $i=36.38\pm2.02^\circ$. We also report their resulting $e=0.0256\pm0.0066$ from their "Eccentric Orbit" values in Table 3.

\paragraph{M33 X-7}

\citet{orosz_1565-solar-mass_2007} model the eclipsing binary M33 X-7 with optical spectroscopic and photometric data, fixing the orbital period to the previously determined $\Porb=3.453014(20)\text{ d}$ \citep{pietsch_m33_2006}. We adopt their selected parameters from Table 2: $\Kcp=108.9\pm5.7\text{ km s}^{-1}$, $e=0.0185\pm0.0077$, $i=74.6\pm1.0^\circ$, $\Mbh=15.65\pm1.45\Msun$ and $\Mcp=70.0\pm6.9\Msun$.

\paragraph{NGC 300 X-1}

\citet{crowther_ngc_2010} confirmed the nature of system NGC 300 X-1 as a Wolf--Rayet/black hole binary located in the Sculptor group galaxy NGC 300 ($d=1.88\text{ Mpc}$). They use optical spectroscopic data and determine $\Porb=32.3\pm0.2\text{ h}$ and $\Kcp=267.5\pm7.7\text{ km s}^{-1}$, of which we adopt the latter. For the period, we adopt the more recent $\Porb=32.7921(3)\text{ h}$ from \citet{binder_wolfrayet_2021}, where new X-ray and UV observations are combined with archival X-ray observations to further constrain the binary model. We adopt their mass estimate of $\Mcp=26^{+7}_{-5}\Msun$ for the WR companion and $\Mbh=17\pm4\Msun$ to recover the $q$ distribution; these estimates are consistent with $60^\circ\leq i\leq75^\circ$, so we adopt a $U(60,75)$ distribution for the inclination.

\paragraph{SS 433}

Discovered in the 1970s, SS 433 is the first observed Galactic microquasar, today generally agreed on to be an eclipsing X-ray binary undergoing supercritical accretion onto the compact object. Over the more than 40 years during which this object has been studied, considerably different estimates of the component masses have been put forward. We have taken our parameters for this system from the recent review by \citet{cherepashchuk_progress_2022}, which reports results from {the} over 40 years of observations of SS433 in the optical, radio, and X-rays. We adopt the quoted nominal value for the orbital period from the review and assign it a conservative uncertainty in the last digit, yielding $\Porb=13.082(5)\text{ d}$. The inclination is given as $i=79^\circ$, so we adopt the distribution $N(79,7.9)^\circ$. We also collected the measured eccentricity of $e=0.050\pm0.007$.

For the velocity semi-amplitude, we adopt the reported $\Kcp=58.2\pm3.1\text{ km s}^{-1}$ from \citet{picchi_optical_2020}, and, for consistency, also adopt the mass estimates from that same work, $\Mbh=4.2\pm0.4\Msun$ and $\Mcp=11.3\pm0.6\Msun$, which result in a mass ratio $q\approx2.64$.

A more recent estimate by \citet{cherepashchuk_discovery_2021}, however, has led to a lower limit of $q>0.8$ which implies a relatively massive BH with $\Mbh>8\Msun$. We do not adopt this result presently as it does not provide the parameters necessary for our standardized mass computation.

\subsubsection{{Other Sources}}

In addition to the X-ray binaries, we have collected four recent BH observations with mass and, where applicable, orbital parameter measurements, which we detail here. These are the three non-interacting binaries, LB-1, 2MASS J05215658+4359220 (J0521+435) and Gaia BH1; and the microlensing event OGLE-2011-BLG-0462. 

\paragraph{LB-1 comp.}

\citet{liu_wide_2019} report the results from over two years of radial-velocity {measurements} of the Galactic B-type star LB-1 (LS V +22 25) and find that its motion requires the presence of a $\Mbh=68^{+11}_{-13}\Msun$ ($90\%$) BH in a $\Porb=78.9(3)\text{ d}$ orbit. From the best-fit model for the H$\alpha$ emission line, they also find $\Kcp=52.8\pm0.7\text{ km s}^{-1}$ and $e=0.03\pm0.01$, and, from the companion's spectra, they find $\Mcp=8.2^{+0.9}_{-1.2}\Msun$. We adopt these values for the object.

\paragraph{2MASS J0521+435 comp.}

\citet{thompson_noninteracting_2019} study spectroscopic and photometric data from the giant star J0521+435 and determine that it must be in a $\Porb=83.20(6)\text{ d}$, $\Kcp=44.6\pm0.1\text{ km s}^{-1}$ and $e=0.0048\pm0.0026$ orbit with and an unseen BH companion. By comparing the giant's properties to single-star evolutionary models, they find for the inclination $\sin i=(0.97^{+0.03}_{-0.7})^\circ$ and for the masses $\Mcp=3.2^{+1.0}_{-1.0}\Msun$ $\Mbh=3.3^{2.8}_{0.7}\Msun$ ($2\sigma$). We adopt these nominal results for the object, although the authors also employ other methods that arrive at different but compatible estimates; the unseen companion's mass is consistently estimated to be in the $\sim$2.9--$4.0\Msun$ range, which keeps it a likely low-mass black hole.

\paragraph{Gaia BH1}

\citet{el-badry_sun-like_2022} study a a bright solar-type star in Ophiuchus with astrometric data from the Gaia mission and follow-up spectroscopy, and determine from the Gaia orbital solution and RV measurements that the star must orbiting an unseen companion. We adopted the quoted $\Mcp=0.93\pm0.05\Msun$ mass for the solar-type companion, and their results derived from astrometric and RV data: $\Porb=185.59(5)\text{ d}$, $\Mbh=9.62\pm0.18\Msun$, $i=126.6\pm0.4^\circ$ and $e=0.451\pm0.005$. We adopt the $\Kcp=66.7\pm0.6\text{ km s}^{-1}$ derived from RV data only. This is the only confirmed BH binary with a modest but non-negligible eccentricity, as well as the closest known BH, at the moment of writing, and potentially originated from either a triple system or dynamical assembly.

\startlandscape
\begin{table}[H] 
\tablesize{\scriptsize}
\caption{Masses and Binary Parameters, where applicable, for 35 Black Hole Systems.\label{tab:bh_catalog}}
	\setlength{\cellWidtha}{\columnwidth/8-2\tabcolsep+0.0in}
\setlength{\cellWidthb}{\columnwidth/8-2\tabcolsep+0.0in}
\setlength{\cellWidthc}{\columnwidth/8-2\tabcolsep+0.0in}
\setlength{\cellWidthd}{\columnwidth/8-2\tabcolsep+0.0in}
\setlength{\cellWidthe}{\columnwidth/8-2\tabcolsep+0.0in}
\setlength{\cellWidthf}{\columnwidth/8-2\tabcolsep+0.0in}
\setlength{\cellWidthg}{\columnwidth/8-2\tabcolsep+0.0in}
\setlength{\cellWidthh}{\columnwidth/8-2\tabcolsep+0.0in}
\scalebox{1}[1]{\begin{tabularx}{\columnwidth}{
>{\PreserveBackslash\raggedright}m{\cellWidtha}
>{\PreserveBackslash\centering}m{\cellWidthb}
>{\PreserveBackslash\centering}m{\cellWidthc}
>{\PreserveBackslash\centering}m{\cellWidthd}
>{\PreserveBackslash\centering}m{\cellWidthe}
>{\PreserveBackslash\centering}m{\cellWidthf}
>{\PreserveBackslash\centering}m{\cellWidthg}
>{\PreserveBackslash\raggedleft}m{\cellWidthh}}
		\toprule
\textbf{Name}	& \boldmath$\Porb$	& \boldmath$\Kcp$ & \boldmath$f(M_\mathrm{BH})$ & \boldmath$i$ & \boldmath$q$ & \boldmath$M_\mathrm{BH}$ & \textbf{References}\\
{} & \textbf{day} & \text{\textbf{km s}}\boldmath$^{-1}$ & \boldmath$\Msun$ & \textbf{deg} & {} & $\Msun$ & {}\\

\midrule
4U 1543-475 & $N(1.116407,0.000003)$ & $AN(129.33,1.68,1.76)$ & $N(0.25,0.01)$ & $N(20.7,1.5)$ & $U(0.25,0.31)$ & $AN(8.78,2.45,1.45)$ & \cite{orosz_inventory_2003} \\
GRS 1915+105 & $N(33.5,1.5)$ & $N(140,15)$ & $AN(8.79,3.69,2.35)$ & $N(70,2)$ & $AN(0.08,0.04,0.02)$ & $AN(12.92,5.39,3.57)$ & \cite{greiner_unusually_2001}\\
GS 1354-64 & $N(2.54451,0.00008)$ & $N(279.0,4.7)$ & $AN(5.72,0.30,0.28)$ & $U(50,80)$ & $AN(0.12,0.03,0.04)$ & $AN(8.23,3.76,0.84)$ & \cite{casares_refined_2009,farr_mass_2011}\\
GRS 1124-684 & $N(0.43260249,0.00000009)$ & $N(406.8,2.7)$ & $N(3.018,0.06)$ & $AN(43.2,2.1,2.7)$ & $N(0.079,0.007)$ & $AN(10.70,1.98,1.06)$ & \cite{wu_dynamical_2015,wu_mass_2016}\\
XTE J1118+480 & $N(0.1699337, 0.0000002)$ & $N(710.0, 2.6)$ & $AN(6.304, 0.069, 0.07)$ & $N(73.5,5.5)$& $N(0.024,0.009)$ & $AN(7.23, 0.88, 0.37)$ & \cite{gonzalez_hernandez_fast_2014} \\
3A 0620-003 & $N(0.32301415,0.00000007)$ & $N(435.4,0.5)$ & $AN(2.762,0.009,0.01)$ & $N(51.0,0.9)$ & $N(0.060,0.004)$ & $AN(6.60,0.27,0.24)$ & \cite{gonzalez_hernandez_fast_2014} \\
GS 2000+251 & $N(0.344086, 0.000002)$ & $N(519.5, 5.1)$ & $AN(5.00,0.15,0.14)$ & $U(54,60)$ & $AN(0.034, 0.015,0.006)$ & $AN(8.97,0.84,0.51)$ & \cite{ioannou_mass_2004} \\
MAXI J1659-152 & $N(0.1006,0.0002)$ & $N(750,80)$ & $AN(3.58,3.15,1.91)$ & $U(70,80)$ & $U(0.02,0.7)$ & $AN(4.25,3.88,2.28)$ & \cite{torres_delimiting_2021} \\
MAXI J1305-704 & $N(0.394,0.004)$ & $N(554,8)$ & $AN(6.92,0.32,0.29)$ & $AN(72,5,8)$ & $AN(0.050,0.007,0.013)$ & $AN(8.18,1.87,0.57)$ & \cite{mata_sanchez_dynamical_2021} \\
GS 2023+338 & $N(6.4714,0.0001)$ & $N(208.5,0.7)$ & $N(6.078,0.061)$ & $AN(67,3,1)$ & $AN(0.060,0.004,0.005)$ & $AN(8.58,0.31,0.45)$ & \cite{casares_optical_1994,khargharia_near-infrared_2010}\\
XTE J1650-500 & $N(0.3205,0.0007)$ & $N(435,30)$ & $AN(2.64,0.64,0.48)$ & $U(50,80)$ & $U(0.01,0.5)$ & $AN(5.03,3.04,1.36)$ & \cite{orosz_orbital_2004}  \\
GRO J0422+32 & $N(0.2121600, 0.0000002)$ & $N(378,16)$ & $AN(1.19,0.16,0.14)$ & $N(45,2)$ & $AN(0.116,0.079,0.071)$ & $AN(4.06,1.08,0.71)$ & \cite{webb_tio_2000,gelino_gro_2003} \\
H 1705-250 & $N(0.5228,0.0044)$ & $N(441,6)$ & $AN(4.64,0.20,0.19)$ & $U(60,80)$ & $AN(0.014,0.019,0.012)$ & $AN(5.41,1.15,0.36)$ & \cite{harlaftis_doppler_1997,remillard_dynamical_1996}\\
GRO J1655-40 & $N(2.62120,0.00014)$ & $N(226.1,0.8)$ & $N(3.139,0.033)$ & $N(68.65,1.5)$ & $N(0.329,0.047)$ & $AN(6.86,0.52,0.49)$ & \cite{hernandez_black_2008,beer_quiescent_2002} \\
XTE J1859+226 & $N(0.274,0.002)$ & $N(541,70)$ & $AN(4.08,2.12,1.29)$ & $U(60,70)$ & $U(0.01,0.5)$ & $AN(7.79,5.70,2.64)$ & \cite{corral-santana_evidence_2011,kreidberg_mass_2012} \\
MAXI J1803-298 & $N(0.29,0.03)$ & $N(515,55)$ & $AN(3.77,1.66,1.04)$ & $U(65,85)$ & $U(0.01,0.2)$ & $AN(5.06,2.51,1.45)$ & \cite{mata_sanchez_dynamical_2021} \\
MAXI J1820+070 & $N(0.68549, 0.00001)$ & $N(417.7,3.9)$ & $AN(5.18,0.15,0.14)$ & $U(66,81)$ & $N(0.072,0.012)$ & $AN(6.47,0.80,0.30)$ & \cite{torres_dynamical_2019,torres_binary_2020} \\
XTE J1550-564 & $N(1.5420333, 0.0000024)$ & $N(363.14, 5.97)$ & $AN(7.64, 0.39, 0.36)$ & $N(74.7,3.8)$ & $N(0.033,0.003)$ & $AN(9.00,0.78,0.56)$ & \cite{orosz_improved_2011} \\
GX 339-4 & $N(1.7587,0.0005)$ & $N(219,3)$ & $AN(1.913,0.08,0.007)$ & $U(37,78)$ & $N(0.18,0.05)$ & $AN(3.21,3.50,0.44)$ &  \cite{heida_mass_2017} \\
GRS 1009-45 & $N(0.2852060,0.0000014)$ & $N(475.4,5.9)$ & $N(3.17,0.12)$ & $N(78.0,7.8)$ & $N(0.137,0.015)$ & $AN(4.28,0.57,0.24)$ &  \cite{filippenko_black_1999} \\
LMC X-3 & $N(1.7048089,0.0000011)$ & $N(241.1,6.2)$ & $AN(2.47,0.20,0.18)$ & $N(69.24,0.72)$ & $AN(0.512,0.099,0.084)$ & $AN(7.20,1.23,0.94)$ & \cite{orosz_mass_2014} \\
SAX J1819.3-2525 & $N(2.81730, 0.00001)$ & $N(211.0, 3.1)$ & $N(2.74, 0.12)$ & $N(72.3,4.1)$ & $AN(0.447, 0.048,0.034)$ & $AN(6.64,0.79,0.54)$ & \cite{macdonald_black_2014,orosz_black_2001}\\
GRS 1716-249 & - & - & - & $N(47.5,0.37)$ & - & $N(5.12,0.11)$ & \cite{zhang_testing_2022} \\
MAXI J1813-095 & - & - & - & $U(28,45)$ & - & $AN(7.41,0.90,0.93)$ & \cite{jana_accretion_2021}\\
MAXI J1535-571 & - & - & - & - & - & $N(8.9,1)$ & \cite{shang_evolution_2019} \\
XTE J1746-322 & - & - & - & - & - & $AN(11.21,1.65,1.96)$ & \cite{molla_estimation_2017} \\

\midrule
Cyg X-1 & $N(5.599836,0.000024)$ & $N(75.57,0.70)$ & $N(0.250,0.007)$ & $N(27.06,0.76)$ & $AN(1.29,0.17,0.14)$ & $AN(13.72,2.55,1.88)$ & \cite{brocksopp_improved_1998,orosz_mass_2011} \\
LMC X-1 & $N(3.90917,0.00005)$ & $N(71.61,1.10)$ & $AN(0.143,0.005,0.004)$ & $N(36.38,2.02)$ & $AN(2.81,0.62,0.41)$ & $AN(9.27,3.85,2.03)$ & \cite{orosz_new_2009} \\
M33 X-7 & $N(3.453014,0.000020)$ & $N(108.9,5.7)$ & $AN(0.449,0.082,0.062)$ & $N(74.6,1.0)$ & $AN(4.39,0.68,0.53)$ & $AN(14.39,5.12,3.13)$ & \cite{pietsch_m33_2006,orosz_1565-solar-mass_2007} \\
NGC 300 X-1 & $N(1.3663375,0.0000125)$ & $N(267.5,7.7)$ & $AN(2.68,0.26,0.23)$ & $U(60,75)$ & $AN(1.42,0.70,0.33)$ & $AN(19.79,14.01,5.38)$ & \cite{binder_wolfrayet_2021,crowther_ngc_2010} \\
SS 433 & $N(13.082,0.005)$ & $N(58.2,3.1)$ & $AN(0.263,0.047,0.038)$ & $N(79.0,7.9)$ & $AN(2.64,0.34,0.24)$ & $AN(4.23,1.70,0.98)$ & \cite{cherepashchuk_progress_2022,picchi_optical_2020}\\

\midrule 
LB-1 cp. & $N(78.9,0.3)$ & $N(52.8,0.7)$  & $AN(1.203,0.049,0.047)$ & $U(15,18)$ & $AN(0.117,0.022,0.019)$ & $AN(56.86,18.59,4.75)$ & \cite{liu_wide_2019} \\
2MASS J0521+435 cp.  & $N(83.2,0.06)$ & $N(44.6,0.1)$ & $N(0.765,0.005)$ & $AN(75.43,5.16,2.35)$ & $AN(0.74,0.26,0.17)$ & $AN(2.45,0.84,0.45)$ & \cite{thompson_noninteracting_2019} \\
Gaia BH1 & $N(185.59,0.05)$ & $N(66.7,0.6)$ & $N(4.08,0.08)$ & $N(126.6,0.4)$ & $N(0.097,0.005)$ & $AN(9.42,0.33,0.31)$ & \cite{el-badry_sun-like_2022} \\
OGLE-2011-BLG-0462 & - & - & - & - & - & $N(7.88,0.82)$ & \cite{mroz_systematic_2022} \\
\bottomrule
\end{tabularx}}
\begin{adjustwidth}{}{0cm}
\noindent\footnotesize{{In} each line, we identify the name of the source containing the black hole, as well the BH mass and available orbital parameters; when these are available, the indicated BH mass is the result of our standardized calculation (Section \ref{sec:mass_comp}); otherwise, it corresponds to the nominal result from the references. For the systems with more than one source, which parameters where taken from which source are specified in Section \ref{sec:catalog}. The distributions are given in terms of the notation established in Section \ref{sec:uncertainties}: $N(\mu,\sigma)$ indicates a normal distribution with mean $\mu$ and standard deviation $\sigma$, while $AN(m, \sigma_1, \sigma_2)$ an asymmetrical normal distribution with mode $m$, standard deviation $\sigma_1$ above $m$ and standard deviation $\sigma_2$ below. $U(x_1,x_2)$ indicates a uniform distribution between $x_1$ and $x_2$.}
\end{adjustwidth}
\end{table}
\finishlandscape

\paragraph{OGLE-2011-BLG-0462}

The microlensing event OGLE-2011-BLG-046S (OB110462) marks the first discovery an isolated BH. It was made independently by \citet{lam_isolated_2022} with data from the Optical Gravitational Lensing Experiment (OGLE), and by \citet{sahu_isolated_2022} with data from the Microlensing Observations in Astrophysics (MOA) survey, with initially conflicting mass estimates of 1.6--$4.4\Msun$ and $7.1\pm1.3\Msun$, respectively. A simultaneous re-analysis of all available data from OB110462 by \citet{mroz_systematic_2022} later confirmed a mass of $\Mbh=7.88\pm0.8\Msun$ for the object, which we include in our sample.

\subsection{The Full Mass Distribution}
\label{sec:full_mass_distr}

One of the great interests in building an updated and standardized catalog of BH masses is to verify the shape of the full Galactic BH mass distribution and to match it to analytical models, which may be made available for use in further work concerning BH populations. \citet{ozel_black_2010} and \citet{farr_mass_2011} performed the last such modeling more than 10 years ago, with samples of 16 and 20 BHs, respectively. At that time, through Bayesian methods, they found that, for a simple Gaussian, the preferred distribution had a mean $\mu\sim$$7.8\Msun$ and standard deviation $\sigma\sim$$1.2\Msun$. On the other hand, \citet{farr_mass_2011} also considered a double Gaussian model, and found for it a Bayesian evidence $\sim30$ greater than that for the simple Gaussian. We present here a simple goodness-of-fit test for an asymmetric Gaussian model to our full sample and compare it to the previous results. A full investigation of the mass distribution with more robust methods and including further corrections for systematics and biases \citep{kreidberg_mass_2012,jonker2021} will be developed in a future work (Bernardo et al., in preparation).

\begin{figure}[H]
    \includegraphics[width=0.75\textwidth]{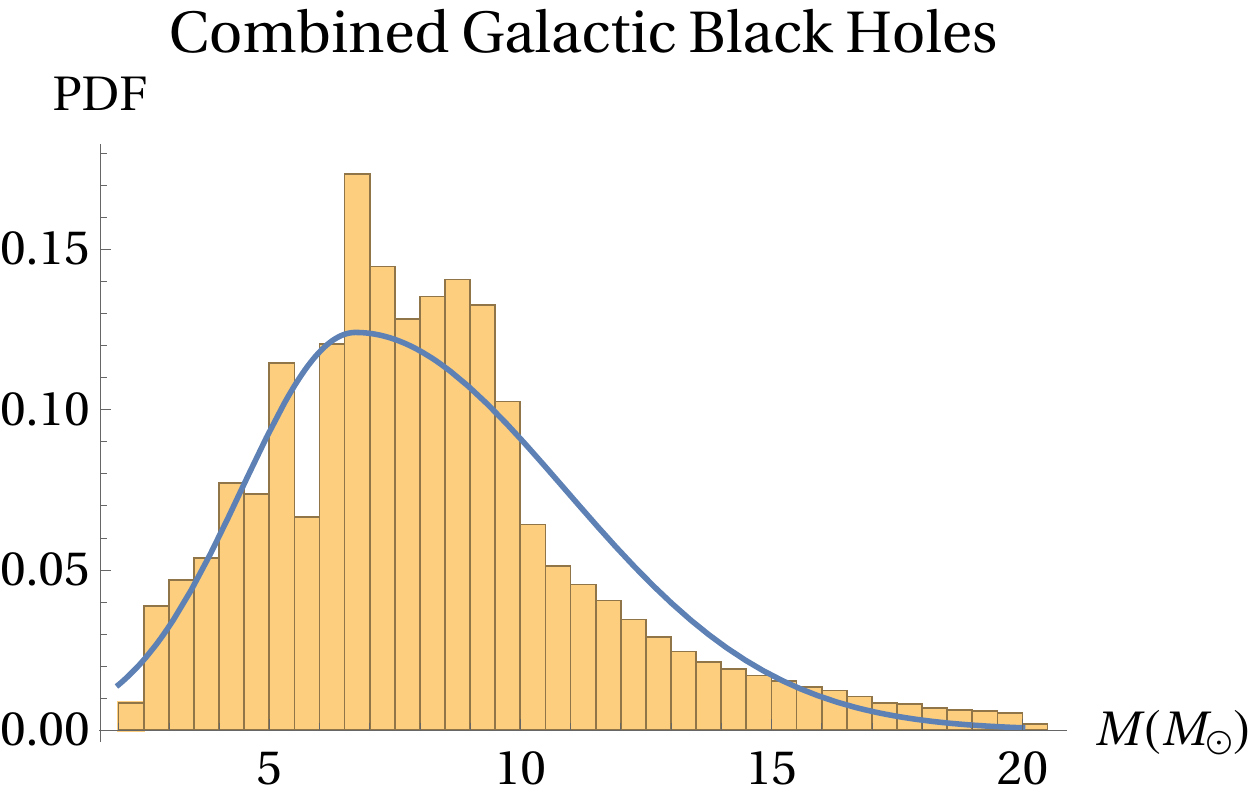}
    \caption{Mixed sample resulting from a draw of $10^6$ masses from each object in Table \ref{tab:bh_catalog} (yellow histogram) and the resulting best asymmetric Gaussian fit as $AN(6.72, 4.17, 2.27)\Msun$ (blue line).}
    \label{fig:gal_gaussian}
\end{figure}

From each of the 35 mass distributions in Table \ref{tab:bh_catalog}, we draw $10^6$ masses, and fit an asymmetric Gaussian to the mixture of these draws, resulting in $m=6.72\Msun$, $\sigma_1=4.17\Msun$ and $\sigma_2=2.27\Msun$. The sample histogram is displayed along with the asymmetric Gaussian fit in Figure \ref{fig:gal_gaussian}. The mode of the distribution (\textit{m}) results in a value not too distinct to the result from \cite{ozel_black_2010} for a sample two times smaller, indicating the robustness of a peak in this region, while allowing for asymmetry, leads to quite different $\sigma$.

While a full analysis of the currently known extragalactic BH masses measured by LVK is out of the scope of this work, some first considerations and a simple comparison stand to be made. It must first be highlighted that there is no simple way to derive a single distribution from both Galactic and extragalactic samples simultaneously, as not only are they affected by different observational biases, but also the underlying physical distribution itself is not necessarily the same. Different environmental conditions at the higher redshifts probed by GW observations could change the shape of the distribution by affecting the process of stellar and binary evolution. Lower metallicites, for example, make wind mass loss from giant stars less efficient (e.g., \cite{WindPols}), while both low metallicities and high star formation rates have been linked to a top-heavy initial mass function (see the reviews by \cite{Kroupa2013,Hopkins2018}), which could be expected to lead to a BH mass distribution shifted to higher masses.

\begin{figure}[H]
    \includegraphics[width=0.75\textwidth]{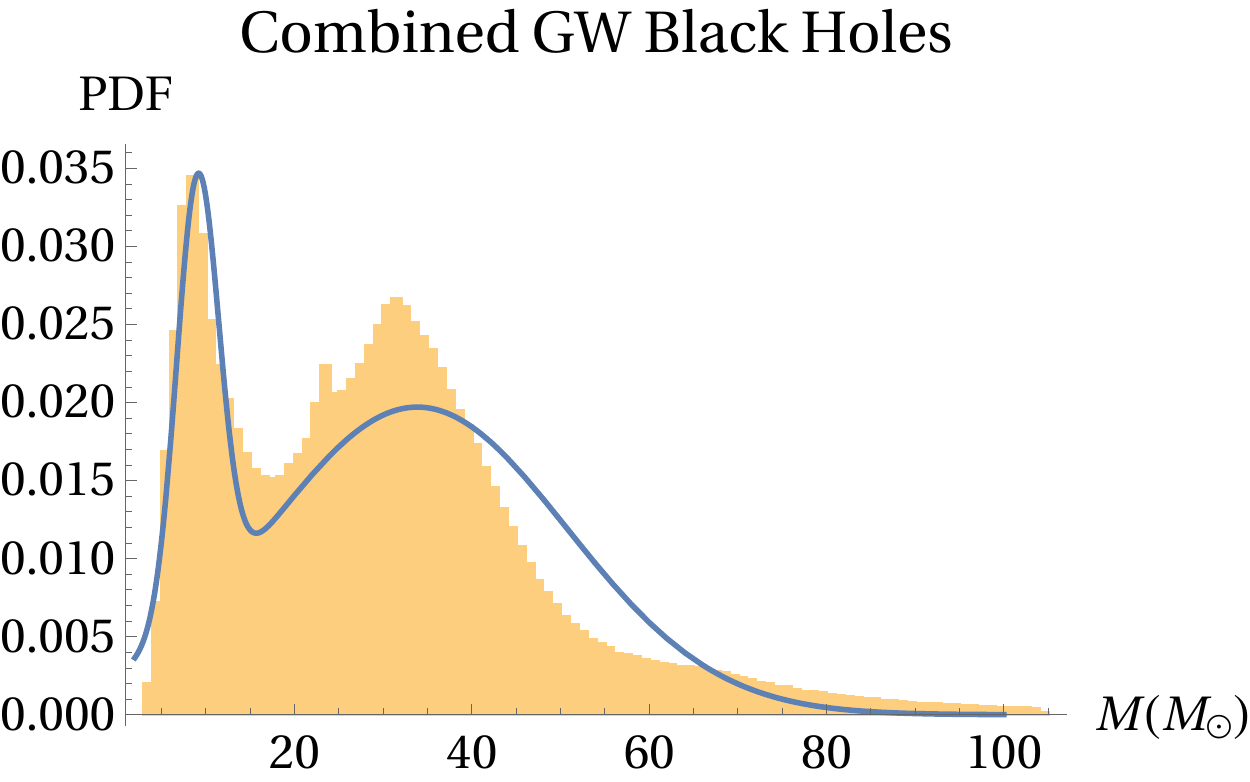}
    \caption{Mixed sample resulting from a draw of $10^6$ masses from each BH component from GWTC-3 (yellow histogram) and the resulting best double Gaussian fit (blue line), with parameters given in the text.}
    \label{fig:combgw}
\end{figure}

\begin{figure}[H]
    \includegraphics[width=0.8\textwidth]{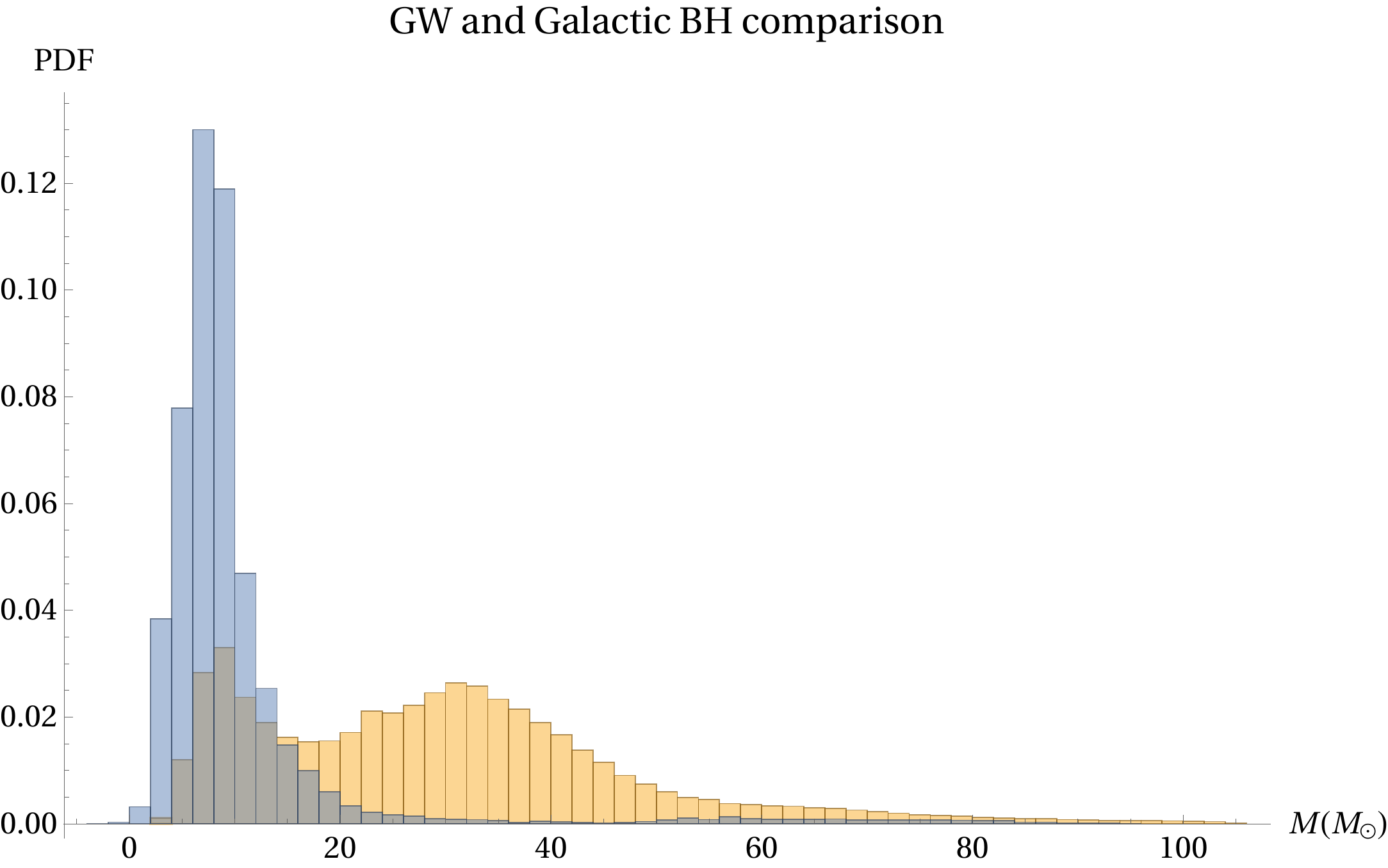}
    \caption{Comparison between the same mass draw from Galactic BHs shown in Figure \ref{fig:gal_gaussian} (blue histogram) and a mass draw from the extragalactic BHs from GWTC-3 (yellow histogram). We note the alignment between the Galactic peak and the lower extragalactic peak.}
    \label{fig:comparison}
\end{figure}

With that in mind, we have collected the raw BH mass distributions made available by LVK (Monte Carlo samples of $8,000$ masses for each object) as part of GWTC-3 \cite{gwtc3} and produced their mixture distribution with no further processing of the data. Unlike the Galactic distribution, the extragalactic distribution shows two peaks, which we fitted to a double Gaussian in order to determine the location and breadth of each one. Figure \ref{fig:combgw} shows the resulting fit, with a lower peak $\mu_1=9.07\Msun$, $\sigma_1=2.39\Msun$, with weight $r_1=0.17$; and an upper peak $\mu_2=33.86\Msun$, $\sigma_2=16.85\Msun$. Even though a direct comparison cannot be directly made, as discussed, it is still noteworthy that a lower peak from the GW sources seems to align with the single Galactic peak as shown in Figure \ref{fig:comparison}. This lower peak is mainly caused by the new additions to GWTC-3 in relation to GWTC-2 \cite{gwtc3pop}. It may then be connected to the improvements in sensitivity made to the observatories between each new run, which allow for the detection of lower-mass sources than in previous runs. Although a safe conclusion may only be derived after a careful modeling and consideration of observational biases for both samples, this might lend credence to the idea that a Galactic-like population of BHs exists in galaxies at higher redshift ($z$$\sim$1).

\begin{figure}[H]
    \includegraphics[width=0.8\textwidth]{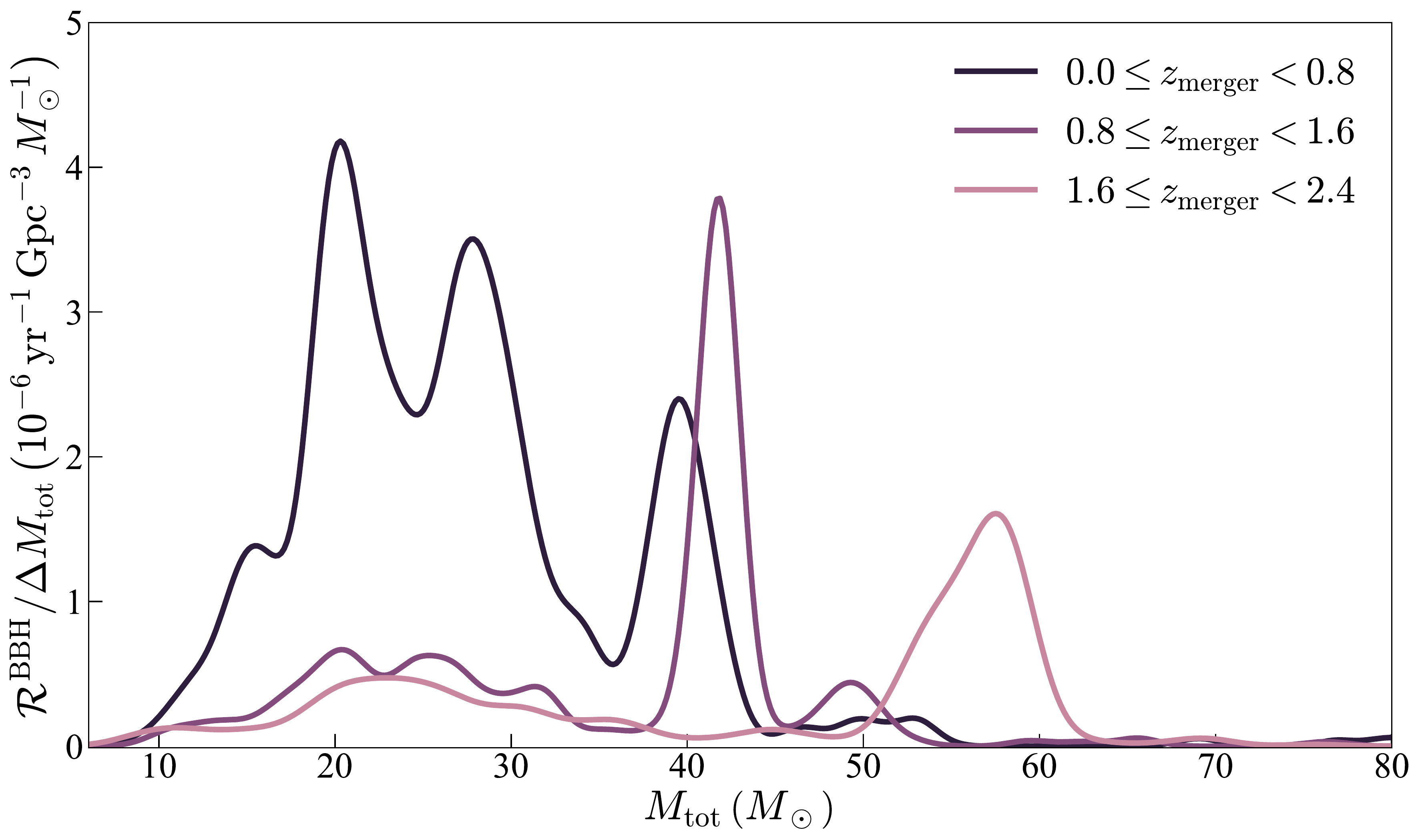}
    \caption{{BH-BH} merger rate ($\mathcal{R}^\mathrm{BBH}$) per total binary mass ($M_\mathrm{tot}$) bin in three different ranges of redshift at merger $z_\mathrm{merger}$ from \cite{desa_synthesis}, resulting from binary population synthesis with an IMF that becomes top-heavy at high $z$.}
    \label{fig:synthesis_redshift}
\end{figure}

On the other hand, the GW mass distribution shows a high-mass peak which is entirely absent from the Galactic distribution. Although it is hard to treat the existing Galactic BH sample as reflective of the real mass distribution with its current size, this also might indicate that high-redshift populations allow for the production of more massive BHs than local populations. If mass-redshift correlations are present in the GW sample, then they might support the aforementioned idea that low-metallicity and high-star formation rate conditions at high redshift lead to the birth of more massive stars, \cite{Chruslinska2020} which in turn allow for an increased production of more massive BHs, and consequently of more massive BH mergers, as has been indicated by population synthesis studies that consider environment-dependent initial mass functions and star formation rates (\citep{Neijssel2019,desa_synthesis}, {see also Figure} \ref{fig:synthesis_redshift}) and tentatively supported by GW observations \cite{Fishbach2021When}. A full consideration will, however, still require accounting for observational biases and for a study of the redshift distribution in GWTC-3 and future catalogs. 

\subsection{The Lower Mass Gap}
\label{sec:gap}

We note from Figures \ref{fig:gal_gaussian} and \ref{fig:combgw} that both the Galactic and extragalactic mass distributions smoothly extend into the <$5\Msun$ region to a significant level; in fact, six of our objects --- MAXI J1659$-$152, GRO J0422+32, GX 339$-$4, GRS 1009$-$45, SS 433 and 2MASS J0521+435 cp. --- all have mass distributions with central values below $5\Msun$. This makes them candidates for being within the $2$--$5\Msun$ \textit{lower mass gap} in the BH mass distribution, an apparently empty mass range above the maximum NS mass first indicated by \cite{bailyn1998} from a sample of seven LMXBs. With their expanded samples of $~20$ XRBs, \citet{ozel_black_2010} in 2010 and \citet{farr_mass_2011} in 2011 still found evidence for a gap in the BH mass distribution below $\sim$$5\Msun$, and so its existence has remained an open question for over 20 years. The gap has been connected to the characteristic convection timescale in core-collapse supernovae \citep{fryer2012,belczynski2012}, but even in this case a partial filling has not been ruled out. Ongoing work on this issue \citep{Fryer2022,Olejak2022} can now be taken both ways, as while in one hand an appropriate convection timescale can help explain the existence of the gap, a measure of the gap depth can help constrain the convection timescale. Other mechanisms for compact object production, such as AIC in WD binaries and accretion in "spider" binaries, have also been suggested to fill the gap to some extent \cite{WSChapter}, but the hypothesis of a partially filled gap has remained mostly speculative until recently.

Objects such as the six we mention from our sample, as well as light BHs detected by LVK have in the past few years started to offer observational support for the partial gap hypothesis. \citet{farah2022} performed a Bayesian study of the full sample of BH and NS masses measured by LVK and considered a single, overall compact object distribution in the form of a broken power law with a parametric dip representing the gap. Although they cannot conclusively confirm the existence or absence of the gap, their results show a preference for a partially depleted gap with a preferred lower edge at $2.2^{+0.7}_{-0.5}\Msun$, not incompatible with $\sim$$2.5\Msun$ maximum NS mass estimates. The broken power law showed a preference for a break at $2.4^{+0.5}_{-0.5}\Msun$, which could point towards a high NS $M_\mathrm{max}$.

\begin{figure}[H]
    \includegraphics[width=0.9\textwidth]{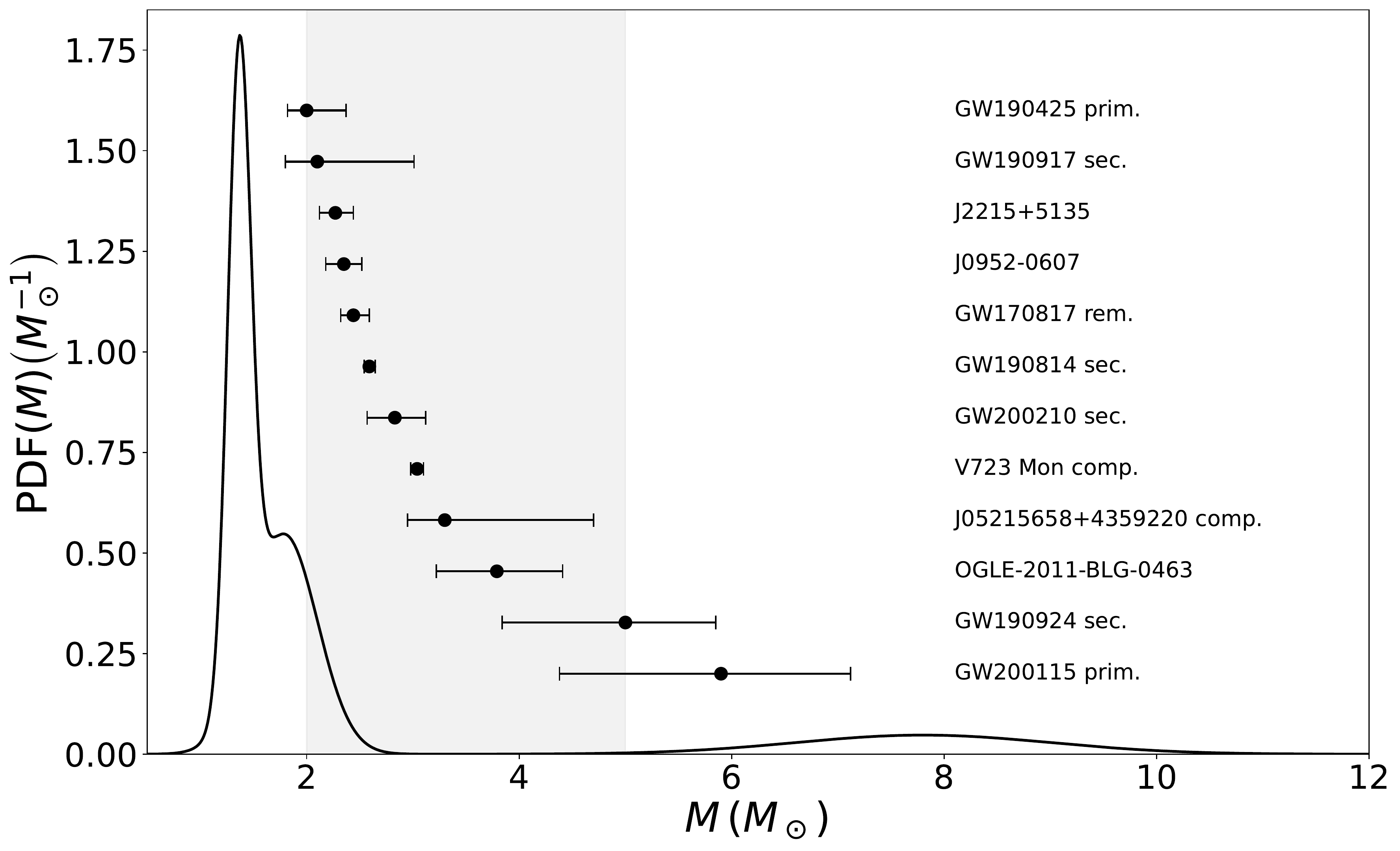}
    \caption{{Mass} distribution and gap object candidates from \cite{desa_gap}. The solid line shows the mixture of the BH mass distribution from \cite{ozel_black_2010} and the NS mass distribution from \cite{rocha2021}. In addition, shown are the mass measurements, with their $1\sigma$ confidence intervals, of the 12 gap candidates the authors consider, a mix of Galactic and extragalactic sources.}
    \label{fig:gapcandidates}
\end{figure}

\citet{desa_gap} later performed a frequentist analysis of 12 current gap candidates, both Galactic and extragalactic, in relation to the then most recent BH and NS mass distributions --- which show a gap between $\sim$$2.2$--$5\Msun$ --- displayed in {Figure} \ref{fig:gapcandidates}. They calculate both the likelihood of individual objects falling in the gap, and by a likelihood ratio test between the existing distributions and a simple partially depleted gap model are able to show that current detections place the probability of a desert gap below $\sim$$3\Msun$ as being at most $\approx$$10\%$, and that a mass distribution with a partially depleted gap is strongly preferred over one with a desert gap, with a likelihood ratio of $\sim$$10^3$. It seems thus clear at this point that, if there is a gap, it should not start below $\sim$$2.5\Msun$, and that the investigation of alternative formation channels for compact objects is needed to further constrain the degree to which the $2.5$--$5\Msun$ range is populated.

\section{Conclusions}
\label{sec:conclusions}

The size and density of the available compact object sample has greatly grown since their existence was first confirmed in the 1960s, with the sample of NSs surpassing 100 and that of Galactic BHs reaching 35 objects, while the GW observations from LVK have already detected more than 150 BHs and around 10 NSs at high redshift. Slowly, this growing sample has allowed us to start to see the shape of the underlying compact object mass distribution, provided biases do not dominate the samples. Once thought to be clustered around a canonical mass of $\sim$$1.4\Msun$, today NSs have been shown beyond doubt to have masses $\geq$$2.2\Msun$, with at least two peaks in the mass distribution. More recently, growing evidence suggests that their maximum mass might be closer to $\sim$$2.5\Msun$. This slow upward shift of the maximum NS mass value serves also as a cautionary tale with regard to the BH mass distribution, which is still based on only a few tens of measurements. As the canonical model for NS masses changed considerably while the sample was below $\sim$$100$ objects, the BH sample may still need an accumulation of reliable data points before it can properly constrain the shape of the physical mass distribution. It is expected that $\geq$$10^7$ BHs and a higher number of NSs should exist in the Galaxy, originating from a great diversity of evolutionary paths which we are only starting to understand. Progress in this front will also be needed before we can properly evaluate the degree of completeness of the sample due to observational biases. 


\vspace{6pt}
\authorcontributions{{Conceptualization, L.M.S., A.B., R.R.A.B., L.S.R., P.H.R.S.M. and J.E.H.; methodology, L.M.S., A.B., R.R.A.B., L.S.R., P.H.R.S.M. and J.E.H; formal analysis,, L.M.S., A.B., R.R.A.B., L.S.R., P.H.R.S.M. and J.E.H; investigation,, L.M.S., A.B., R.R.A.B., L.S.R., P.H.R.S.M. and J.E.H; data curation, , L.M.S., A.B., R.R.A.B., L.S.R., P.H.R.S.M. and J.E.H; writing-original draft preparation, , L.M.S., A.B., R.R.A.B., L.S.R., P.H.R.S.M. and J.E.H; writing-review and editing, , L.M.S., A.B., R.R.A.B., L.S.R., P.H.R.S.M. and J.E.H. All authors have read and agreed to the published version of the manuscript.}}

\funding{This research was funded by the Fapesp Agency through the grant 13/26258-4 and 2020/08518-2 and the CNPq (Federal Government) for the award of a Research Fellowship to J.E.H. L.M.S. acknowledges CNPq for financial support. The CAPES Agency (Federal Government) is acknowledged for financial support in the form of scholarships.}

\dataavailability{{All observational data reported in this work is available within the quoted references. All data compiled and calculated in Tables 12 and 13 is publicly available in} \url{https://doi.org/10.5281/zenodo.7508626}.} 

\acknowledgments{We would like to thank the anonymous referees for their careful reports on the manuscript which helped lead to a more complete final version. This work has also made extensive use of NASA\textquotesingle s Astrophysics Data System (ADS).}

\conflictsofinterest{{The authors declare no conflict of interest.}}
\begin{adjustwidth}{-\extralength}{0cm}
\reftitle{References}

\PublishersNote{}
\end{adjustwidth}

\begin{thebibliography}{999}

\bibitem[{Walter} {et~al.}(1996){Walter}, {Wolk}, and
  {Neuh{\"a}user}]{Walter1996}
{Walter}, F.M.; {Wolk}, S.J.; {Neuh{\"a}user}, R.
\newblock {Discovery of a nearby isolated neutron star}.
\newblock {\em Nature} {\bf 1996}, {\em 379},~233--235.
\newblock {{https://doi.org/10.1038/379233a0}}.

\bibitem[{Zampieri} {et~al.}(2001){Zampieri}, {Campana}, {Turolla},
  {Chieregato}, {Falomo}, {Fugazza}, {Moretti}, and {Treves}]{Zampieri2001}
{Zampieri}, L.; {Campana}, S.; {Turolla}, R.; {Chieregato}, M.; {Falomo}, R.;
  {Fugazza}, D.; {Moretti}, A.; {Treves}, A.
\newblock {1RXS J214303.7+065419/RBS 1774: A new Isolated Neutron Star
  candidate}.
\newblock {\em Astron. Astrophys.} {\bf 2001}, {\em 378},~L5--L9.
\newblock {{https://doi.org/10.1051/0004-6361:20011151}}.

\bibitem[{Turolla}(2009)]{Turolla2009}
{Turolla}, R.
\newblock {Isolated Neutron Stars: The Challenge of Simplicity}.
\newblock In \emph{Proceedings of the Astrophysics and Space Science Library};
  {Becker}, W., Ed.; {Springer}: {Berlin, Germany} 2009; Volume 357, p. 141.
\newblock {{https://doi.org/10.1007/978-3-540-76965-1\_7}}.

\bibitem[{Ertan} {et~al.}(2014){Ertan}, {{\c{C}}al{\i}{\c{s}}kan}, {Benli},
  and {Alpar}]{Ertan2014}
{Ertan}, {\"U}.; {{\c{C}}al{\i}{\c{s}}kan}, {\c{S}}.; {Benli}, O.; {Alpar},
  M.A.
\newblock {Long-term evolution of dim isolated neutron stars}.
\newblock {\em Mon. Not. R. Astron. Soc.} {\bf 2014}, {\em 444},~1559--1565.
\newblock {{https://doi.org/10.1093/mnras/stu1523}}.

\bibitem[{Esposito} {et~al.}(2021){Esposito}, {Rea}, and
  {Israel}]{MagnetarReview}
{Esposito}, P.; {Rea}, N.; {Israel}, G.L.
\newblock {Magnetars: A Short Review and Some Sparse Considerations}.
\newblock In \emph{Proceedings of the Astrophysics and Space Science Library};
  {Belloni}, T.M.; {M{\'e}ndez}, M.; {Zhang}, C., Eds.;  Astrophysics and Space Science Library: New York, NY, USA,  2021, Volume 461, pp. 97--142,
\newblock {{https://doi.org/10.1007/978-3-662-62110-3\_3}}.

\bibitem[{De Luca}(2017)]{CCODeLuca}
{De Luca}, A.
\newblock {Central compact objects in supernova remnants}.
\newblock In \emph{Proceedings of the Journal of Physics Conference Series};{ IOP Publishing: Bristol, UK}, 2017,
  Volume 932, p. 012006,
\newblock {{https://doi.org/10.1088/1742-6596/932/1/012006}}.

\bibitem[{Mayer} and {Becker}(2021)]{CCOMayer}
{Mayer}, M.G.F.; {Becker}, W.
\newblock {A kinematic study of central compact objects and their host
  supernova remnants}.
\newblock {\em Astron. Astrophys.} {\bf 2021}, {\em 651},~A40.
\newblock {{https://doi.org/10.1051/0004-6361/202141119}}.

\bibitem[{Ho}(2013)]{CCOho2013}
{Ho}, W.C.G.
\newblock {Central compact objects and their magnetic fields}.
\newblock In \emph{Proceedings of the Neutron Stars and Pulsars: Challenges and
  Opportunities after 80 Years}; {van Leeuwen}, J., Ed.;  {2013}; Volume 291, pp.
  101--106. Publisher: Cambridge University Press (Cambridge) 
\newblock {{https://doi.org/10.1017/S1743921312023289}}.

\bibitem[{Ho} {et~al.}(2021){Ho}, {Zhao}, {Heinke}, {Kaplan}, {Shternin},
  and {Wijngaarden}]{CCOho2021}
{Ho}, W.C.G.; {Zhao}, Y.; {Heinke}, C.O.; {Kaplan}, D.L.; {Shternin}, P.S.;
  {Wijngaarden}, M.J.P.
\newblock {X-ray bounds on cooling, composition, and magnetic field of the
  Cassiopeia A neutron star and young central compact objects}.
\newblock {\em Mon. Not. R. Astron. Soc.} {\bf 2021}, {\em 506},~5015--5029.
\newblock {{https://doi.org/10.1093/mnras/stab2081}}.

\bibitem[{Abhishek} {et~al.}(2022){Abhishek}, {Tanushree}, {Hegde}, and
  {Konar}]{RRATabhishek}
{Abhishek}, Malusare, N.; {Tanushree}, N.; {Hegde}, G.; {Konar}, S.
\newblock {Radio pulsar sub-populations (II): The mysterious RRATs}.
\newblock {\em J. Astrophys. Astron.} {\bf 2022}, {\em 43},~75.
\newblock {{https://doi.org/10.1007/s12036-022-09862-3}}.

\bibitem[{Keane} and {McLaughlin}(2011)]{RRATkeane}
{Keane}, E.F.; {McLaughlin}, M.A.
\newblock {Rotating radio transients}.
\newblock {\em Bull. Astron. Soc. India} {\bf 2011}, {\em
 39},~333--352

\bibitem[Hurley-Walker {et~al.}(2022)Hurley-Walker, Zhang, Bahramian,
  McSweeney, O’Doherty, Hancock, Morgan, Anderson, Heald, and Galvin]{GLEAM}
Hurley-Walker, N.; Zhang, X.; Bahramian, A.; McSweeney, S.J.; O’Doherty,
  T.N.; Hancock, P.J.; Morgan, J.S.; Anderson, G.E.; Heald, G.H.; Galvin, T.J.
\newblock A radio transient with unusually slow periodic emission.
\newblock {\em Nature} {\bf 2022}, {\em 601},{~526--530}.
\newblock  {{https://doi.org/10.1038/s41586-021-04272-x}}.

\bibitem[{Corral-Santana} {et~al.}(2016){Corral-Santana}, {Casares},
  {Mu{\~n}oz-Darias}, {Bauer}, {Mart{\'\i}nez-Pais}, and {Russell}]{blackcat}
{Corral-Santana}, J.M.; {Casares}, J.; {Mu{\~n}oz-Darias}, T.; {Bauer}, F.E.;
  {Mart{\'\i}nez-Pais}, I.G.; {Russell}, D.M.
\newblock {BlackCAT: A catalogue of stellar-mass black holes in X-ray
  transients}.
\newblock {\em Astron. Astrophys.} {\bf 2016}, {\em 587},~A61.
\newblock {{https://doi.org/10.1051/0004-6361/201527130}}.

\bibitem[{Tetarenko} {et~al.}(2016){Tetarenko}, {Sivakoff}, {Heinke}, and
  {Gladstone}]{watchdog}
{Tetarenko}, B.E.; {Sivakoff}, G.R.; {Heinke}, C.O.; {Gladstone}, J.C.
\newblock {WATCHDOG: A Comprehensive All-sky Database of Galactic Black Hole
  X-ray Binaries}.
\newblock {\em Astrophys. J., Suppl. Ser.} {\bf 2016}, {\em 222},~15.
\newblock {{https://doi.org/10.3847/0067-0049/222/2/15}}.

\bibitem[{Ingram} and {Motta}(2019)]{QPOreview}
{Ingram}, A.R.; {Motta}, S.E.
\newblock {A review of quasi-periodic oscillations from black hole X-ray
  binaries: Observation and theory}.
\newblock {\em New Astron. Rev.} {\bf 2019}, {\em 85},~101524.
\newblock {{https://doi.org/10.1016/j.newar.2020.101524}}.

\bibitem[Lam {et~al.}(2022)Lam, Lu, Udalski, Bond, Bennett, Skowron, Mróz,
  Poleski, Sumi, Szymański, Kozłowski, Pietrukowicz, Soszyński, Ulaczyk,
  Wyrzykowski, Miyazaki, Suzuki, Koshimoto, Rattenbury, Hosek, Abe, Barry,
  Bhattacharya, Fukui, Fujii, Hirao, Itow, Kirikawa, Kondo, Matsubara,
  Matsumoto, Muraki, Olmschenk, Ranc, Okamura, Satoh, Silva, Toda, Tristram,
  Vandorou, Yama, Abrams, Agarwal, Rose, and Terry]{lam_isolated_2022}
Lam, C.Y.; Lu, J.R.; Udalski, A.; Bond, I.; Bennett, D.P.; Skowron, J.; Mróz,
  P.; Poleski, R.; Sumi, T.; Szymański, M.K.;  et~al.
\newblock An {Isolated} {Mass}-gap {Black} {Hole} or {Neutron} {Star}
  {Detected} with {Astrometric} {Microlensing}.
\newblock {\em  Astrophys. J.} {\bf 2022}, {\em 933},{~L23.}
\newblock {{https://doi.org/10.3847/2041-8213/ac7442}}. 

\bibitem[Sahu {et~al.}(2022)Sahu, Anderson, Casertano, Bond, Udalski,
  Dominik, Calamida, Bellini, Brown, Rejkuba, Bajaj, Kains, Ferguson, Fryer,
  Yock, Mróz, Kozłowski, Pietrukowicz, Poleski, Skowron, Soszyński,
  Szymański, Ulaczyk, Wyrzykowski, Collaboration), Barry, Bennett, Bond,
  Hirao, Silva, Kondo, Koshimoto, Ranc, Rattenbury, Sumi, Suzuki, Tristram,
  Vandorou, Collaboration), Beaulieu, Marquette, Cole, Fouqué, Hill, Dieters,
  Coutures, Dominis-Prester, Bennett, Bachelet, Menzies, Albrow, Pollard,
  Collaboration), Gould, Yee, Allen, Almeida, Christie, Drummond, Gal-Yam,
  Gorbikov, Jablonski, Lee, Maoz, Manulis, McCormick, Natusch, Pogge,
  Shvartzvald, Collaboration), Jørgensen, Alsubai, Andersen, Bozza, Novati,
  Burgdorf, Hinse, Hundertmark, Husser, Kerins, Longa-Peña, Mancini, Penny,
  Rahvar, Ricci, Sajadian, Skottfelt, Snodgrass, Southworth, Tregloan-Reed,
  Wambsganss, Wertz, Consortium), Tsapras, Street, Bramich, Horne, Steele, and
  Collaboration)]{sahu_isolated_2022}
Sahu, K.C.; Anderson, J.; Casertano, S.; Bond, H.E.; Udalski, A.; Dominik, M.;
  Calamida, A.; Bellini, A.; Brown, T.M.; Rejkuba, M.;  et~al.
\newblock An {Isolated} {Stellar}-mass {Black} {Hole} {Detected} through
  {Astrometric} {Microlensing}.
\newblock {\em Astrophys. J.} {\bf 2022}, {\em 933},{83.}
\newblock {{https://doi.org/10.3847/1538-4357/ac739e}}.

\bibitem[Burrows and Vartanyan(2021)]{Adam}
Burrows, A.; Vartanyan, D.
\newblock Core-collapse supernova explosion theory.
\newblock {\em Nature} {\bf 2021}, {\em 589},{~29--39.}
\newblock  {{https://doi.org/10.1038/s41586-020-03059-w}}. 

\bibitem[Ertl {et~al.}(2020)Ertl, Woosley, Sukhbold, and Janka]{Ertl}
Ertl, T.; Woosley, S.E.; Sukhbold, T.; Janka, H.T.
\newblock The {Explosion} of {Helium} {Stars} {Evolved} with {Mass} {Loss}.
\newblock {\em  Astrophys. J.} {\bf 2020}, {\em 890},{~51.}
\newblock  {{https://doi.org/10.3847/1538-4357/ab6458}}. 

\bibitem[{Tauris} {et~al.}(2013){Tauris}, {Sanyal}, {Yoon}, and
  {Langer}]{AICtauris}
{Tauris}, T.M.; {Sanyal}, D.; {Yoon}, S.C.; {Langer}, N.
\newblock {Evolution towards and beyond accretion-induced collapse of massive
  white dwarfs and formation of millisecond pulsars}.
\newblock {\em Astron. Astrophys.} {\bf 2013}, {\em 558},~A39.
\newblock {{https://doi.org/10.1051/0004-6361/201321662}}.

\bibitem[{Wang} and {Liu}(2020)]{AICWang2020}
{Wang}, B.; {Liu}, D.
\newblock {The formation of neutron star systems through accretion-induced
  collapse in white-dwarf binaries}.
\newblock {\em Res. Astron. Astrophys.} {\bf 2020}, {\em
  20},~135
\newblock {{https://doi.org/10.1088/1674-4527/20/9/135}}.

\bibitem[{Wang} {et~al.}(2022){Wang}, {Liu}, and {Chen}]{AICWang2022}
{Wang}, B.; {Liu}, D.; {Chen}, H.
\newblock {Formation of millisecond pulsars with long orbital periods by
  accretion-induced collapse of white dwarfs}.
\newblock {\em Mon. Not. R. Astron. Soc.} {\bf 2022}, {\em 510},~6011--6021.
\newblock {{https://doi.org/10.1093/mnras/stac114}}.

\bibitem[{Bailyn} {et~al.}(1998){Bailyn}, {Jain}, {Coppi}, and
  {Orosz}]{bailyn1998}
{Bailyn}, C.D.; {Jain}, R.K.; {Coppi}, P.; {Orosz}, J.A.
\newblock {The Mass Distribution of Stellar Black Holes}.
\newblock {\em Astrophys. J.} {\bf 1998}, {\em 499},~367--374.
\newblock {{https://doi.org/10.1086/305614}}.

\bibitem[Özel {et~al.}(2010)Özel, Psaltis, Narayan, and
  McClintock]{ozel_black_2010}
Özel, F.; Psaltis, D.; Narayan, R.; McClintock, J.E.
\newblock {THE} {BLACK} {HOLE} {MASS} {DISTRIBUTION} {IN} {THE} {GALAXY}.
\newblock {\em  Astrophys. J.} {\bf 2010}, {\em 725},{~1918.}
\newblock {{https://doi.org/10.1088/0004-637X/725/2/1918}}. 

\bibitem[Farr {et~al.}(2011)Farr, Sravan, Cantrell, Kreidberg, Bailyn,
  Mandel, and Kalogera]{farr_mass_2011}
Farr, W.M.; Sravan, N.; Cantrell, A.; Kreidberg, L.; Bailyn, C.D.; Mandel, I.;
  Kalogera, V.
\newblock {THE} {MASS} {DISTRIBUTION} {OF} {STELLAR}-{MASS} {BLACK} {HOLES}.
\newblock {\em Astrophys. J.} {\bf 2011}, {\em 741},{~103.}
\newblock   {{https://doi.org/10.1088/0004-637X/741/2/103}}. 

\bibitem[{Fryer} {et~al.}(2012){Fryer}, {Belczynski}, {Wiktorowicz},
  {Dominik}, {Kalogera}, and {Holz}]{fryer2012}
{Fryer}, C.L.; {Belczynski}, K.; {Wiktorowicz}, G.; {Dominik}, M.; {Kalogera},
  V.; {Holz}, D.E.
\newblock {Compact Remnant Mass Function: Dependence on the Explosion Mechanism
  and Metallicity}.
\newblock {\em Astrophys. J.} {\bf 2012}, {\em 749},~91.
\newblock {{https://doi.org/10.1088/0004-637X/749/1/91}}.

\bibitem[{Belczynski} {et~al.}(2012){Belczynski}, {Wiktorowicz}, {Fryer},
  {Holz}, and {Kalogera}]{belczynski2012}
{Belczynski}, K.; {Wiktorowicz}, G.; {Fryer}, C.L.; {Holz}, D.E.; {Kalogera},
  V.
\newblock {Missing Black Holes Unveil the Supernova Explosion Mechanism}.
\newblock {\em Astrophys. J.} {\bf 2012}, {\em 757},~91.
\newblock {{https://doi.org/10.1088/0004-637X/757/1/91}}.

\bibitem[{Farah} {et~al.}(2022){Farah}, {Fishbach}, {Essick}, {Holz}, and
  {Galaudage}]{farah2022}
{Farah}, A.; {Fishbach}, M.; {Essick}, R.; {Holz}, D.E.; {Galaudage}, S.
\newblock {Bridging the Gap: Categorizing Gravitational-wave Events at the
  Transition between Neutron Stars and Black Holes}.
\newblock {\em Astrophys. J.} {\bf 2022}, {\em 931},~108.
\newblock {{https://doi.org/10.3847/1538-4357/ac5f03}}.

\bibitem[de~Sá {et~al.}(2022)de~Sá, Bernardo, Bachega, Horvath, Rocha, and
  Moraes]{desa_gap}
de~Sá, L.M.; Bernardo, A.; Bachega, R.R.A.; Horvath, J.E.; Rocha, L.S.;
  Moraes, P.H.R.S.
\newblock Quantifying the Evidence Against a Mass Gap between Black Holes and
  Neutron Stars.
\newblock {\em  Astrophys. J.} {\bf 2022}, {\em 941},~130.
\newblock {{https://doi.org/10.3847/1538-4357/aca076}}.

\bibitem[Ye and Fishbach(2022)]{Ye2022}
Ye, C.; Fishbach, M.
\newblock Inferring the Neutron Star Maximum Mass and Lower Mass Gap in Neutron
  Star–Black Hole Systems with Spin.
\newblock {\em Astrophys. J.} {\bf 2022}, {\em 937},~73.
\newblock {{https://doi.org/10.3847/1538-4357/ac7f99}}.

\bibitem[Landau(1932)]{Landau}
Landau, L.
\newblock On the {Theory} of {Stars}.
\newblock {\em Phys. Z. Sowjetunion} {\bf 1932}, {\em
  1},~285--288.

\bibitem[Baade and Zwicky(1934)]{BZ}
Baade, W.; Zwicky, F.
\newblock On {Super}-{Novae}.
\newblock {\em Proc. Natl. Acad. Sci. USA} {\bf 1934},
  {\em 20},{~254--259.}
\newblock {{https://doi.org/10.1073/pnas.20.5.254}}. 

\bibitem[Hewish {et~al.}(1968)Hewish, Bell, Pilkington, Scott, and
  Collins]{Jocelyn}
Hewish, A.; Bell, S.J.; Pilkington, J.D.H.; Scott, P.F.; Collins, R.A.
\newblock Observation of a {Rapidly} {Pulsating} {Radio} {Source}.
\newblock {\em Nature} {\bf 1968}, {\em 217},~709--713.
\newblock Number: 5130 Publisher: Nature Publishing Group,
  {{https://doi.org/10.1038/217709a0}}.

\bibitem[Pacini(1968)]{Franco}
Pacini, F.
\newblock Rotating {Neutron} {Stars}, {Pulsars} and {Supernova} {Remnants}.
\newblock {\em Nature} {\bf 1968}, {\em 219},{~145--146.}
\newblock {{https://doi.org/10.1038/219145a0}}.

\bibitem[Gold(1968)]{Gold}
Gold, T.
\newblock Rotating {Neutron} {Stars} as the {Origin} of the {Pulsating} {Radio}
  {Sources}.
\newblock {\em Nature} {\bf 1968}, {\em 218},{~731--732.}
\newblock {{https://doi.org/10.1038/218731a0}}. 

\bibitem[{Bisnovatyi-Kogan}(1992)]{BisnovatyiKogan1992}
{Bisnovatyi-Kogan}, G.S.
\newblock {The Neutron Star Population in the Galaxy}.
\newblock In {\em Proceedings of IAU Symposium \#149}; {Barbuy},
  B.; {Renzini}, A., Eds.;  {Spinger: Dordrecht, The Netherlands 1992}; Volume 149, p. 379. 

\bibitem[Lyne {et~al.}(1985)Lyne, Manchester, and Taylor]{LMT}
Lyne, A.G.; Manchester, R.N.; Taylor, J.H.
\newblock The galactic population of pulsars.
\newblock {\em Mon. Not. R. Astron. Soc.} {\bf 1985}, {\em 213},~613--639.
\newblock {{https://doi.org/10.1093/mnras/213.3.613}}.

\bibitem[Dirson {et~al.}(2022)Dirson, Pétri, and Mitra]{dirson}
Dirson, L.; Pétri, J.; Mitra, D.
\newblock The {Galactic} population of canonical pulsars.
\newblock {\em Astron. Astrophys.} {\bf 2022}, {\em 667},{~A82.}
\newblock {{https://doi.org/10.1051/0004-6361/202243305}}. 

\bibitem[Kaspi(2010)]{Kaspi}
Kaspi, V.M.
\newblock Grand unification of neutron stars.
\newblock {\em Proc. Natl. Acad. Sci. USA} {\bf 2010},
  {\em 107},{~7147--7152.}
\newblock {{https://doi.org/10.1073/pnas.1000812107}}. 

\bibitem[{Rutledge} {et~al.}(2008){Rutledge}, {Fox}, and
  {Shevchuk}]{CalveraDiscovery}
{Rutledge}, R.E.; {Fox}, D.B.; {Shevchuk}, A.H.
\newblock {Discovery of an Isolated Compact Object at High Galactic Latitude}.
\newblock {\em Astrophys. J.} {\bf 2008}, {\em 672},~1137--1143.
\newblock {{https://doi.org/10.1086/522667}}.

\bibitem[{Potekhin} {et~al.}(2015){Potekhin}, {De Luca}, and
  {Pons}]{CoolingNS}
{Potekhin}, A.Y.; {De Luca}, A.; {Pons}, J.A.
\newblock {Neutron Stars{\textemdash}Thermal Emitters}.
\newblock {\em Space Sci. Rev.} {\bf 2015}, {\em 191},~171--206,
\newblock {{https://doi.org/10.1007/s11214-014-0102-2}}.

\bibitem[Agar {et~al.}(2021)Agar, Weltevrede, Bondonneau, Grießmeier,
  Hessels, Huang, Karastergiou, Keith, Kondratiev, Künsemöller, Li, Peng,
  Sobey, Stappers, Tan, Theureau, Wang, Zhang, Cecconi, Girard, Loh, and
  Zarka]{Agar}
Agar, C.H.; Weltevrede, P.; Bondonneau, L.; Grießmeier, J.M.; Hessels, J.W.T.;
  Huang, W.J.; Karastergiou, A.; Keith, M.J.; Kondratiev, V.I.; Künsemöller,
  J.;  et~al.
\newblock A broad-band radio study of {PSR} {J0250}+5854: the slowest spinning
  radio pulsar known.
\newblock {\em Mon. Not. R. Astron. Soc.} {\bf 2021}, {\em 508},~1102--1114.
\newblock {{https://doi.org/10.1093/mnras/stab2496}}.

\bibitem[Morello {et~al.}(2020)Morello, Keane, Enoto, Guillot, Ho, Jameson,
  Kramer, Stappers, Bailes, Barr, Bhandari, Caleb, Flynn, Jankowski, Johnston,
  van Straten, Arzoumanian, Bogdanov, Gendreau, Malacaria, Ray, and
  Remillard]{Morello}
Morello, V.; Keane, E.F.; Enoto, T.; Guillot, S.; Ho, W.C.G.; Jameson, A.;
  Kramer, M.; Stappers, B.W.; Bailes, M.; Barr, E.D.;  et~al.
\newblock The {SUrvey} for {Pulsars} and {Extragalactic} {Radio} {Bursts} –
  {IV}. {Discovery} and polarimetry of a 12.1-s radio pulsar.
\newblock {\em Mon. Not. R. Astron. Soc.} {\bf 2020}, {\em 493},~1165--1177.
\newblock {{https://doi.org/10.1093/mnras/staa321}}.

\bibitem[Young {et~al.}(1999)Young, Manchester, and Johnston]{Young}
Young, M.D.; Manchester, R.N.; Johnston, S.
\newblock A radio pulsar with an 8.5-second period that challenges emission
  models.
\newblock {\em Nature} {\bf 1999}, {\em 400},{~848--849.}
\newblock {{https://doi.org/10.1038/23650}}. 

\bibitem[Turolla and Esposito(2013)]{Roberto}
Turolla, R.; Esposito, P.
\newblock Low-magnetic-field magnetars.
\newblock {\em Int. J. Mod. Phys. D} {\bf 2013}, {\em 22},{~1330024.}
\newblock  {{https://doi.org/10.1142/S0218271813300243}}. 

\bibitem[{Caleb} {et~al.}(2022){Caleb}, {Heywood}, {Rajwade}, {Malenta},
  {Stappers}, {Barr}, {Chen}, {Morello}, {Sanidas}, {van den Eijnden},
  {Kramer}, {Buckley}, {Brink}, {Motta}, {Woudt}, {Weltevrede}, {Jankowski},
  {Surnis}, {Buchner}, {Bezuidenhout}, {Driessen}, and
  {Fender}]{UltraLongPeriodPSR}
{Caleb}, M.; {Heywood}, I.; {Rajwade}, K.; {Malenta}, M.; {Stappers}, B.W.;
  {Barr}, E.; {Chen}, W.; {Morello}, V.; {Sanidas}, S.; {van den Eijnden}, J.;
  et~al.
\newblock {{Discovery of a radio-emitting neutron star with an ultra-long spin
  period of 76 s}}. 
\newblock {\em Nat. Astron.} {\bf 2022}, {\em 6},~828--836.
\newblock {{https://doi.org/10.1038/s41550-022-01688-x}}.

\bibitem[{Buckley} {et~al.}(2017){Buckley}, {Meintjes}, {Potter}, {Marsh},
  and {G{\"a}nsicke}]{WDPulsar}
{Buckley}, D.A.H.; {Meintjes}, P.J.; {Potter}, S.B.; {Marsh}, T.R.;
  {G{\"a}nsicke}, B.T.
\newblock {{Polarimetric evidence of a white dwarf pulsar in the binary system
  AR Scorpii}}. 
\newblock {\em Nat. Astron.} {\bf 2017}, {\em 1},~29.
\newblock {{https://doi.org/10.1038/s41550-016-0029}}.

\bibitem[{Vigan{\`o}} {et~al.}(2013){Vigan{\`o}}, {Rea}, {Pons}, {Perna},
  {Aguilera}, and {Miralles}]{MagneticFieldDecay}
{Vigan{\`o}}, D.; {Rea}, N.; {Pons}, J.A.; {Perna}, R.; {Aguilera}, D.N.;
  {Miralles}, J.A.
\newblock Unifying the observational diversity of isolated neutron stars via
  magneto-thermal evolution models.
\newblock {\em Mon. Not. R. Astron. Soc.} {\bf 2013}, {\em 434},~123--141.
\newblock {{https://doi.org/10.1093/mnras/stt1008}}.

\bibitem[{Igoshev} {et~al.}(2021){Igoshev}, {Popov}, and
  {Hollerbach}]{Igoshev2021}
\textls[-65]{{Igoshev}, A.P.; {Popov}, S.B.; {Hollerbach}, R.
 {Evolution of Neutron Star Magnetic Fields}.
 {\em Universe} {\bf 2021}, {\em 7},~351.
 {{https://doi.org/10.3390/universe7090351}}.}

\bibitem[Benvenuto {et~al.}(2014)Benvenuto, De~Vito, and Horvath]{BDH}
Benvenuto, O.G.; De~Vito, M.A.; Horvath, J.E.
\newblock Understanding the {Evolution} of {Close} {Binary} {Systems} with
  {Radio} {Pulsars}.
\newblock {\em  Astrophys. J.} {\bf 2014}, {\em 786},{~L7.}
\newblock  {{https://doi.org/10.1088/2041-8205/786/1/L7}}. 

\bibitem[Mendes {et~al.}(2018)Mendes, de~Avellar, Horvath, Souza, Benvenuto,
  and De~Vito]{nos}
Mendes, C.; de~Avellar, M.G.B.; Horvath, J.E.; Souza, R.A.d.; Benvenuto, O.G.;
  De~Vito, M.A.
\newblock Magnetic field decay in black widow pulsars.
\newblock {\em Mon. Not. R. Astron. Soc.} {\bf 2018}, {\em 475},{~2178--2184.}
\newblock {{https://doi.org/10.1093/mnras/stx3319}}. 

\bibitem[Igoshev and Popov(2020)]{braking}
Igoshev, A.P.; Popov, S.B.
\newblock Braking indices of young radio pulsars: theoretical perspective.
\newblock {\em Mon. Not. R. Astron. Soc.} {\bf 2020}, {\em 499},~2826--2835.
\newblock {{https://doi.org/10.1093/mnras/staa3070}}.

\bibitem[Allen and Horvath(1997)]{AllenHorvath}
Allen, M.P.; Horvath, J.E.
\newblock Glitches, torque evolution and the dynamics of young pulsars.
\newblock {\em Mon. Not. R. Astron. Soc.} {\bf 1997}, {\em 287},~615--621.
\newblock {{https://doi.org/10.1093/mnras/287.3.615}}.

\bibitem[Yoneyama {et~al.}(2019)Yoneyama, Hayashida, Nakajima, and
  Matsumoto]{Yone}
Yoneyama, T.; Hayashida, K.; Nakajima, H.; Matsumoto, H.
\newblock Unification of strongly magnetized neutron stars with regard to
  {X}-ray emission from hot spots.
\newblock {\em Astron. Nachrichten} {\bf 2019}, {\em 340},~221--225.
\newblock 
 Available online:  \url{https://onlinelibrary.wiley.com/doi/pdf/10.1002/asna.201913593} ({accessed on 21 Dec 2022})
  {{https://doi.org/10.1002/asna.201913593}}.

\bibitem[{Hurley}(2010)]{SGRrev}
{Hurley}, K.
\newblock {Soft gamma repeaters}.
\newblock {\em Mem. Soc. Astron. Ital.} {\bf 2010}, {\em 81},~432.

\bibitem[{Kaspi}(2007)]{AXPreview}
{Kaspi}, V.M.
\newblock {Recent progress on anomalous X-ray pulsars}.
\newblock {\em Astrophys. Space Sci.} {\bf 2007}, {\em 308},~1--11.
\newblock {{https://doi.org/10.1007/s10509-007-9309-y}}.

\bibitem[Keane(2016)]{keane}
Keane, E.F.
\newblock Classifying {RRATs} and {FRBs}.
\newblock {\em Mon. Not. R. Astron. Soc.} {\bf 2016}, {\em 459},~1360--1362.
\newblock {{https://doi.org/10.1093/mnras/stw767}}.

\bibitem[{Kaspi} and {Kramer}(2016)]{KaspiKramer2016}
{Kaspi}, V.M.; {Kramer}, M.
\newblock {Radio Pulsars: The Neutron Star Population \& Fundamental Physics}.
\newblock {\em arXiv} {\bf 2016}, arXiv:1602.07738,

\bibitem[Beniamini {et~al.}(2022)Beniamini, Wadiasingh, Hare, Rajwade,
  Younes, and van~der Horst]{Benia}
Beniamini, P.; Wadiasingh, Z.; Hare, J.; Rajwade, K.; Younes, G.; van~der
  Horst, A.J.
\newblock Evidence for an abundant old population of {Galactic} {Ultra} long
  period magnetars and implications for fast radio bursts. \emph{arXiv} \textbf{2022}, {arXiv:2210.09323.}
\newblock {{https://doi.org/10.48550/arxiv.2210.09323}}. 

\bibitem[Rocha {et~al.}(2021)Rocha, Bachega, Horvath, and Moraes]{rocha2021}
Rocha, L.S.; Bachega, R.R.A.; Horvath, J.E.; Moraes, P.H.R.S.
\newblock The maximum mass of neutron stars may be higher than expected: an
  inference from binary systems. \emph{arXiv} \textbf{2021}, {arXiv:2107.08822.}
\newblock {{https://doi.org/10.48550/ARXIV.2107.08822}}. 

\bibitem[{Postnov} and {Prokhorov}(2001)]{PostnovProkhorov2001}
{Postnov}, K.A.; {Prokhorov}, M.E.
\newblock {The Relation between the Observed Mass Distribution for Compact
  Stars and the Mechanism for Supernova Explosions}.
\newblock {\em Astron. Rep.} {\bf 2001}, {\em 45},~899--907.
\newblock {{https://doi.org/10.1134/1.1416279}}.

\bibitem[Valentim {et~al.}(2011)Valentim, Rangel, and
  Horvath]{valentim2011mass}
Valentim, R.; Rangel, E.; Horvath, J.
\newblock On the mass distribution of neutron stars.
\newblock {\em Mon. Not. R. Astron. Soc.} {\bf 2011}, {\em 414},~1427--1431.

\bibitem[{\"O}zel and Freire(2016)]{ozel2016masses}
{\"O}zel, F.; Freire, P.
\newblock Masses, radii, and the equation of state of neutron stars.
\newblock {\em Annu. Rev. Astron. Astrophys.} {\bf 2016}, {\em
  54},~401--440.

\bibitem[Kiziltan {et~al.}(2013)Kiziltan, Kottas, De~Yoreo, and
  Thorsett]{kiziltan2013neutron}
Kiziltan, B.; Kottas, A.; De~Yoreo, M.; Thorsett, S.E.
\newblock The neutron star mass distribution.
\newblock {\em  Astrophys. J.} {\bf 2013}, {\em 778},~66.

\bibitem[Zhang {et~al.}(1997)Zhang, Strohmayer, and Swank]{zhang1997neutron}
Zhang, W.; Strohmayer, T.; Swank, J.
\newblock Neutron star masses and radii as inferred from kilohertz
  quasi-periodic oscillations.
\newblock {\em  Astrophys. J.} {\bf 1997}, {\em 482},~L167.

\bibitem[Alsing {et~al.}(2018)Alsing, Silva, and Berti]{alsing2018evidence}
Alsing, J.; Silva, H.O.; Berti, E.
\newblock Evidence for a maximum mass cut-off in the neutron star mass
  distribution and constraints on the equation of state.
\newblock {\em Mon. Not. R. Astron. Soc.} {\bf 2018}, {\em 478},~1377--1391.

\bibitem[Horvath {et~al.}(2022)Horvath, Rocha, Bernardo, de~Avellar, and
  Valentim]{WSChapter}
Horvath, J.; Rocha, L.; Bernardo, A.C.; de~Avellar, M.; Valentim, R.
\newblock {Birth events, masses and the maximum mass of Compact Stars}. In {\em
  Astrophysics in the XXI Century with Compact Stars}; Vasconcellos, C.Z., Weber, F., Eds.;  {World Scientific Publishing: Singapore, Singapore 2022}; pp. 1--51.

\bibitem[{Rhoades} and {Ruffini}(1974)]{RhoadesRuffini}
\textls[-35]{{Rhoades}, C.E.; {Ruffini}, R.}
\newblock \textls[-35]{{Maximum Mass of a Neutron Star}.}
\newblock \textls[-35]{{\em Phys. Rev. Lett.} {\bf 1974}, {\em 32},~324--327.}
\newblock \textls[-35]{{{https://doi.org/10.1103/PhysRevLett.32.324}}.}

\bibitem[{Margalit} and {Metzger}(2017)]{margalit2017}
{Margalit}, B.; {Metzger}, B.D.
\newblock {Constraining the Maximum Mass of Neutron Stars from Multi-messenger
  Observations of GW170817}.
\newblock {\em Astrophys. J. Lett.} {\bf 2017}, {\em 850},~L19.
\newblock {{https://doi.org/10.3847/2041-8213/aa991c}}.

\bibitem[{Rezzolla} {et~al.}(2018){Rezzolla}, {Most}, and
  {Weih}]{Rezzolla2018}
{Rezzolla}, L.; {Most}, E.R.; {Weih}, L.R.
\newblock {Using Gravitational-wave Observations and Quasi-universal Relations
  to Constrain the Maximum Mass of Neutron Stars}.
\newblock {\em Astrophys. J. Lett.} {\bf 2018}, {\em 852},~L25.
\newblock {{https://doi.org/10.3847/2041-8213/aaa401}}.

\bibitem[{Ruiz} {et~al.}(2018){Ruiz}, {Shapiro}, and {Tsokaros}]{Ruiz2018}
{Ruiz}, M.; {Shapiro}, S.L.; {Tsokaros}, A.
\newblock {GW170817, general relativistic magnetohydrodynamic simulations, and
  the neutron star maximum mass}.
\newblock {\em Phys. Rev. D} {\bf 2018}, {\em 97},~021501,
\newblock {{https://doi.org/10.1103/PhysRevD.97.021501}}.

\bibitem[{Shibata} {et~al.}(2019){Shibata}, {Zhou}, {Kiuchi}, and
  {Fujibayashi}]{Shibata2019}
{Shibata}, M.; {Zhou}, E.; {Kiuchi}, K.; {Fujibayashi}, S.
\newblock {Constraint on the maximum mass of neutron stars using GW170817
  event}.
\newblock {\em Phys. Rev. D} {\bf 2019}, {\em 100},~023015.
\newblock {{https://doi.org/10.1103/PhysRevD.100.023015}}.

\bibitem[{Ai} {et~al.}(2020){Ai}, {Gao}, and {Zhang}]{Ai2020}
{Ai}, S.; {Gao}, H.; {Zhang}, B.
\newblock {What Constraints on the Neutron Star Maximum Mass Can One Pose from
  GW170817 Observations?}
\newblock {\em Astrophys. J.} {\bf 2020}, {\em 893},~146.
\newblock {{https://doi.org/10.3847/1538-4357/ab80bd}}.

\bibitem[Romani {et~al.}(2021)Romani, Kandel, Filippenko, Brink, and
  Zheng]{Romani}
Romani, R.; Kandel, D.; Filippenko, A.; Brink, T.; Zheng, W.
\newblock {{PSR J0952-0607: The Fastest and Heaviest Known Galactic Neutron Star} }.
\newblock {\em Astrophys. J. Lett.} {\bf 2022}, {\em 934}, ~L17  .

\bibitem[{Ecker} and {Rezzolla}(2022)]{RezzollaHighMMax}
{Ecker}, C.; {Rezzolla}, L.
\newblock {Impact of large-mass constraints on the properties of neutron
  stars}.
\newblock {\em arXiv} {\bf 2022}, arXiv:2209.08101,

\bibitem[{The LIGO Scientific Collaboration} {et~al.}(2021){The LIGO
  Scientific Collaboration}, {The Virgo Collaboration}, {The KAGRA
  Collaboration}, et~al.]{gwtc3}
Abbott, R.; Abbott, T.D.; Acernese, F.; Ackley, K.; Adams, C.; Adhikari, N.; Adhikari, R.X.; Adya, V.B.; Affeldt, C.; Agarwal, D.; et al.
\newblock GWTC-3: Compact Binary Coalescences Observed by LIGO and Virgo During
  the Second Part of the Third Observing Run. \emph{arXiv} \textbf{2021},{ arXiv:2111.03606.}
\newblock {{https://doi.org/10.48550/ARXIV.2111.03606}}. 

\bibitem[{Doroshenko} {et~al.}(2022){Doroshenko}, {Suleimanov},
  {P{\"u}hlhofer}, and {Santangelo}]{LightNSDoroshenko}
{Doroshenko}, V.; {Suleimanov}, V.; {P{\"u}hlhofer}, G.; {Santangelo}, A.
\newblock {A strangely light neutron star within a supernova remnant}.
\newblock {\em Nat. Astron.} {\bf 2022}, {\em 6},~1444--1451.
\newblock {{https://doi.org/10.1038/s41550-022-01800-1}}.

\bibitem[{Di Clemente} {et~al.}(2022){Di Clemente}, {Drago}, and
  {Pagliara}]{DragoSS}
{Di Clemente}, F.; {Drago}, A.; {Pagliara}, G.
\newblock {Is the compact object associated with HESS J1731-347 a strange quark
  star?}
\newblock {\em arXiv} {\bf 2022}, arXiv:2211.07485.

\bibitem[Weber(2005)]{reviewSS}
Weber, F.
\newblock Strange quark matter and compact stars.
\newblock {\em Prog. Part. Nucl. Phys.} {\bf 2005}, {\em
  54},~193--288.

\bibitem[Martinez {et~al.}(2015)Martinez, Stovall, Freire, Deneva, Jenet,
  McLaughlin, Bagchi, Bates, and Ridolfi]{Martinez}
Martinez, J.G.; Stovall, K.; Freire, P.C.C.; Deneva, J.S.; Jenet, F.A.;
  McLaughlin, M.A.; Bagchi, M.; Bates, S.D.; Ridolfi, A.
\newblock {PULSAR} J0453+1559: A {DOUBLE} {NEUTRON} {STAR} {SYSTEM} {WITH} A
  {LARGE} {MASS} {ASYMMETRY}.
\newblock {\em  Astrophys. J.} {\bf 2015}, {\em 812},~143.
\newblock {{https://doi.org/10.1088/0004-637x/812/2/143}}.

\bibitem[{Fonseca} {et~al.}(2016){Fonseca}, {Pennucci}, {Ellis}, {Stairs},
  {Nice}, {Ransom}, {Demorest}, {Arzoumanian}, {Crowter}, {Dolch}, {Ferdman},
  {Gonzalez}, {Jones}, {Jones}, {Lam}, {Levin}, {McLaughlin}, {Stovall},
  {Swiggum}, and {Zhu}]{Fonsece2016}
{Fonseca}, E.; {Pennucci}, T.T.; {Ellis}, J.A.; {Stairs}, I.H.; {Nice}, D.J.;
  {Ransom}, S.M.; {Demorest}, P.B.; {Arzoumanian}, Z.; {Crowter}, K.; {Dolch},
  T.;  et~al.
\newblock {The NANOGrav Nine-year Data Set: Mass and Geometric Measurements of
  Binary Millisecond Pulsars}.
\newblock {\em Astrophys. J.} {\bf 2016}, {\em 832},~167.
\newblock {{https://doi.org/10.3847/0004-637X/832/2/167}}.

\bibitem[{Lattimer}(2019)]{LattimerRev}
{Lattimer}, J.M.
\newblock {Neutron Star Mass and Radius Measurements}.
\newblock {\em Universe} {\bf 2019}, {\em 5},~159.
\newblock {{https://doi.org/10.3390/universe5070159}}.

\bibitem[Riley and {et al.}(2021)]{NICER1}
Riley, T.; {et al.}.
\newblock {A NICER View of the Massive Pulsar PSR J0740+6620 Informed by Radio
  Timing and XMM-Newton Spectroscopy}.
\newblock {\em Astrophys. J.} {\bf 2021}, {\em 918},~L27.

\bibitem[Riley and {et al.}(2019)]{NICER2}
Riley, T.; {et al.}.
\newblock {A NICER View of PSR J0030+0451: Millisecond Pulsar Parameter
  Estimation}.
\newblock {\em Astrophys. J.} {\bf 2019}, {\em 887},~L21.

\bibitem[Horvath and Moraes(2021)]{JP}
Horvath, J.; Moraes, P. H. R. S.
\newblock {{Modeling a 2.5Mo compact star with quark matter }}.
\newblock {\em Int. Jour. Mod. Phys. D} {\bf 2021}, {\em 30},~2150016. 

\bibitem[{Einstein}(1915)]{Einstein1915}
{Einstein}, A.
\newblock {\emph{Die Feldgleichungen der Gravitation}};
\newblock {Sitzungsberichte der Preussischen Akademie
der Wissenschaften zu Berlin}: Berlin, Germany, {1915}; pp. 844--847.

\bibitem[{Schwarzschild}(1916)]{Schwarzschild1916}
{Schwarzschild}, K.
\newblock {On the Gravitational Field of a Mass Point According to Einstein's
  Theory}.
\newblock {\em Abh. Konigl. Preuss. Akad.} {\bf 1916},{~189--196}.

\bibitem[{Kerr}(1963)]{Kerr1963}
{Kerr}, R.P.
\newblock {Gravitational Field of a Spinning Mass as an Example of
  Algebraically Special Metrics}.
\newblock {\em Phys. Rev. Lett.} {\bf 1963}, {\em 11},~237--238.
\newblock {{https://doi.org/10.1103/PhysRevLett.11.237}}.

\bibitem[{Newman} {et~al.}(1965){Newman}, {Couch}, {Chinnapared}, {Exton},
  {Prakash}, and {Torrence}]{Newman1965}
{Newman}, E.T.; {Couch}, E.; {Chinnapared}, K.; {Exton}, A.; {Prakash}, A.;
  {Torrence}, R.
\newblock {Metric of a Rotating, Charged Mass}.
\newblock {\em J. Math. Phys.} {\bf 1965}, {\em 6},~918--919.
\newblock {{https://doi.org/10.1063/1.1704351}}.

\bibitem['t~Hooft(1991)]{hooft1991}
't~Hooft, G.
\newblock {The Black hole horizon as a quantum surface}.
\newblock {\em Phys. Scr.} {\bf 1991}, {\em 36},~247--252.
\newblock {{https://doi.org/10.1088/0031-8949/1991/T36/026}}.

\bibitem[Hawking(2015)]{hawking2015information}
Hawking, S.W.
\newblock The information paradox for black holes.
\newblock {\em arXiv} {\bf 2015}, arXiv:1509.01147.

\bibitem[Susskind(2006)]{susskind2006paradox}
Susskind, L.
\newblock The paradox of quantum black holes.
\newblock {\em Nature Physics} {\bf 2006}, {\em 2},~665--677.

\bibitem[{Oppenheimer} and {Volkoff}(1939)]{OppenheimerVolkoff1939}
{Oppenheimer}, J.R.; {Volkoff}, G.M.
\newblock {On Massive Neutron Cores}.
\newblock {\em Phys. Rev.} {\bf 1939}, {\em 55},~374--381.
\newblock {{https://doi.org/10.1103/PhysRev.55.374}}.

\bibitem[{Tolman}(1939)]{Tolman1939}
{Tolman}, R.C.
\newblock {Static Solutions of Einstein's Field Equations for Spheres of
  Fluid}.
\newblock {\em Phys. Rev.} {\bf 1939}, {\em 55},~364--373.
\newblock {{https://doi.org/10.1103/PhysRev.55.364}}.

\bibitem[Zel'dovich(1964)]{Zeldovich1964}
Zel'dovich, Y.B.
\newblock The fate of a star, and the liberation of gravitation energy in
  accretion.
\newblock {\em Dokl. Akad. Nauk SSSR} {\bf 1964}, {\em 155}, {~67--69}. 

\bibitem[{Salpeter}(1964)]{Salpeter1964}
{Salpeter}, E.E.
\newblock {Accretion of Interstellar Matter by Massive Objects.}
\newblock {\em Astrophys. J.} {\bf 1964}, {\em 140},~796--800.
\newblock {{https://doi.org/10.1086/147973}}.

\bibitem[{Bolton}(1972)]{BoltonCyg}
{Bolton}, C.T.
\newblock {Identification of Cygnus X-1 with HDE 226868}.
\newblock {\em Nature} {\bf 1972}, {\em 235},~271--273.
\newblock {{https://doi.org/10.1038/235271b0}}.

\bibitem[{Webster} and {Murdin}(1972)]{LouisePaulCyg}
{Webster}, B.L.; {Murdin}, P.
\newblock {Cygnus X-1-a Spectroscopic Binary with a Heavy Companion ?}
\newblock {\em Nature} {\bf 1972}, {\em 235},~37--38.
\newblock {{https://doi.org/10.1038/235037a0}}.

\bibitem[Gelino and Harrison(2003)]{gelino_gro_2003}
Gelino, D.M.; Harrison, T.E.
\newblock {GRO} {J0422}+32: {The} {Lowest} {Mass} {Black} {Hole}?
\newblock {\em  Astrophys. J.} {\bf 2003}, {\em 599},{~1254--1259.}
\newblock {{https://doi.org/10.1086/379311}}.

\bibitem[Liu {et~al.}(2019)Liu, Zhang, Howard, Bai, Lu, Soria, Justham, Li,
  Zheng, Wang, Belczynski, Casares, Zhang, Yuan, Dong, Lei, Isaacson, Wang,
  Bai, Shao, Gao, Wang, Niu, Cui, Zheng, Mu, Zhang, Wang, Heger, Qi, Liao,
  Lattanzi, Gu, Wang, Wu, Shao, Shen, Wang, Bregman, Di~Stefano, Liu, Han,
  Zhang, Wang, Ren, Zhang, Zhang, Wang, Cabrera-Lavers, Corradi, Rebolo, Zhao,
  Zhao, Chu, and Cui]{liu_wide_2019}
Liu, J.; Zhang, H.; Howard, A.W.; Bai, Z.; Lu, Y.; Soria, R.; Justham, S.; Li,
  X.; Zheng, Z.; Wang, T.;  et~al.
\newblock A wide star–black-hole binary system from radial-velocity
  measurements.
\newblock {\em Nature} {\bf 2019}, {\em 575},{~618--621.}
\newblock{{https://doi.org/10.1038/s41586-019-1766-2}}.  

\bibitem[Thompson {et~al.}(2019)Thompson, Kochanek, Stanek, Badenes, Post,
  Jayasinghe, Latham, Bieryla, Esquerdo, Berlind, Calkins, Tayar, Lindegren,
  Johnson, Holoien, Auchettl, and Covey]{thompson_noninteracting_2019}
Thompson, T.A.; Kochanek, C.S.; Stanek, K.Z.; Badenes, C.; Post, R.S.;
  Jayasinghe, T.; Latham, D.W.; Bieryla, A.; Esquerdo, G.A.; Berlind, P.;
  et~al.
\newblock A noninteracting low-mass black hole-giant star binary system.
\newblock {\em Science} {\bf 2019}, {\em 366},{~637--640.}
\newblock {{https://doi.org/10.1126/science.aau4005}}. 

\bibitem[El-Badry {et~al.}(2022)El-Badry, Rix, Quataert, Howard, Isaacson,
  Fuller, Hawkins, Breivik, Wong, Rodriguez, Conroy, Shahaf, Mazeh, Arenou,
  Burdge, Bashi, Faigler, Weisz, Seeburger, Monter, and
  Wojno]{el-badry_sun-like_2022}
El-Badry, K.; Rix, H.W.; Quataert, E.; Howard, A.W.; Isaacson, H.; Fuller, J.;
  Hawkins, K.; Breivik, K.; Wong, K.W.K.; Rodriguez, A.C.;  et~al.
\newblock A {Sun}-like star orbiting a black hole.
\newblock {\em Mon. Not. R. Astron. Soc.} {\bf 2022}, \emph{518}, 1057--1085.
\newblock {{https://doi.org/10.1093/mnras/stac3140}}.

\bibitem[{King}(2006)]{king2006accretion}
{King}, A.R.
\newblock {Accretion in compact binaries}. In {\em Compact Stellar X-ray
  Sources};  {Cambridge University Press: Cambridge, UK}, {2006}; Volume~39, pp. 507--546. 
  
\bibitem[{Shakura} and {Sunyaev}(1973)]{ShakuraDIsk}
{Shakura}, N.I.; {Sunyaev}, R.A.
\newblock {Black holes in binary systems. Observational appearance.}
\newblock {\em Astron. Atrophys.} {\bf 1973}, {\em 24},~337--355. 

\bibitem[{Shapiro} {et~al.}(1976){Shapiro}, {Lightman}, and
  {Eardley}]{ShapiroDisk}
{Shapiro}, S.L.; {Lightman}, A.P.; {Eardley}, D.M.
\newblock {A two-temperature accretion disk model for Cygnus X-1: structure and
  spectrum.}
\newblock {\em Astrophys. J.} {\bf 1976}, {\em 204},~187--199.
\newblock {{https://doi.org/10.1086/154162}}.

\bibitem[{Frank} {et~al.}(2002){Frank}, {King}, and {Raine}]{Frank2002}
{Frank}, J.; {King}, A.; {Raine}, D.J.
\newblock {\em {Accretion Power in Astrophysics,}} 3rd ed.;  {Cambridge University Press: Cambridge, UK,} 2002. 

\bibitem[{Liao} {et~al.}(2020){Liao}, {Liu}, {Zheng}, and
  {Gou}]{VelaX1WindDisk}
{Liao}, Z.; {Liu}, J.; {Zheng}, X.; {Gou}, L.
\newblock {Spectral evidence of an accretion disc in wind-fed X-ray pulsar Vela
  X-1 during an unusual spin-up period}.
\newblock {\em Mon. Not. R. Astron. Soc.} {\bf 2020}, {\em 492},~5922--5929.
\newblock {{https://doi.org/10.1093/mnras/staa162}}.

\bibitem[{Tauris} and {van den Heuvel}(2006)]{tauris2006}
{Tauris}, T.M.; {van den Heuvel}, E.P.J.
\newblock {Formation and evolution of compact stellar X-ray sources}. In {\em
  Compact Stellar X-ray Sources}; {Cambridge University Press: Cambridge, UK,} 2006; Volume~39, pp. 623--665. 

\bibitem[Hameury(2020)]{SXTDiskRev}
Hameury, J.
\newblock A review of the disc instability model for dwarf novae, soft X-ray
  transients and related objects.
\newblock {\em Adv. Space Res.} {\bf 2020}, {\em 66},{~1004--1024.}
\newblock {{https://doi.org/https://doi.org/10.1016/j.asr.2019.10.022}}. 

\bibitem[Tremaine and Davis(2014)]{DiskWarpRev2014}
Tremaine, S.; Davis, S.W.
\newblock {Dynamics of warped accretion discs}.
\newblock {\em Mon. Not. R. Astron. Soc.} {\bf 2014}, {\em 441},~1408--1434.
  Available online:  \url{https://academic.oup.com/mnras/article-pdf/441/2/1408/3724550/stu663.pdf} {(accessed on 21 Dec 2022)}.
\newblock {{https://doi.org/10.1093/mnras/stu663}}. 

\bibitem[{Abarr} and {Krawczynski}(2020)]{DiskWarpRev2020}
{Abarr}, Q.; {Krawczynski}, H.
\newblock {Exploring the Physics of Warped Accretion Disks with the Imaging
  X-ray Polarimetry Explorer}.
\newblock In Proceedings of the American Astronomical Society Meeting Abstracts
  \#235. {\em Am. Astron. Soc. Meet. Abstr.} \textbf{2020}, \emph{235}, 346.02.

\bibitem[{Bagnoli} {et~al.}(2015){Bagnoli}, {in't Zand}, {D'Angelo}, and
  {Galloway}]{TypeIIBurst}
{Bagnoli}, T.; {in't Zand}, J.J.M.; {D'Angelo}, C.R.; {Galloway}, D.K.
\newblock {A population study of type II bursts in the Rapid Burster}.
\newblock {\em Mon. Not. R. Astron. Soc.} {\bf 2015}, {\em 449},~268--287.
\newblock {{https://doi.org/10.1093/mnras/stv330}}.

\bibitem[{Reig}(2011)]{BeXbinary2011}
{Reig}, P.
\newblock {Be/X-ray binaries}.
\newblock {\em Astrophys. Space Sci.} {\bf 2011}, {\em 332},~1--29.
\newblock {{https://doi.org/10.1007/s10509-010-0575-8}}.

\bibitem[{Rib{\'o}} {et~al.}(2017){Rib{\'o}}, {Munar-Adrover}, {Paredes},
  {Marcote}, {Iwasawa}, {Mold{\'o}n}, {Casares}, {Migliari}, and
  {Paredes-Fortuny}]{first_be_bh}
{Rib{\'o}}, M.; {Munar-Adrover}, P.; {Paredes}, J.M.; {Marcote}, B.; {Iwasawa},
  K.; {Mold{\'o}n}, J.; {Casares}, J.; {Migliari}, S.; {Paredes-Fortuny}, X.
\newblock {The First Simultaneous X-Ray/Radio Detection of the First Be/BH
  System MWC 656}.
\newblock {\em Astrophys. J. Lett.} {\bf 2017}, {\em 835},~L33.
\newblock {{https://doi.org/10.3847/2041-8213/835/2/L33}}.

\bibitem[{Lutovinov} {et~al.}(2013){Lutovinov}, {Revnivtsev}, {Tsygankov},
  and {Krivonos}]{HMXBLum}
{Lutovinov}, A.A.; {Revnivtsev}, M.G.; {Tsygankov}, S.S.; {Krivonos}, R.A.
\newblock {Population of persistent high-mass X-ray binaries in the Milky Way}.
\newblock {\em Mon. Not. R. Astron. Soc.} {\bf 2013}, {\em 431},~327--341.
\newblock {{https://doi.org/10.1093/mnras/stt168}}.

\bibitem[{Sguera} {et~al.}(2005){Sguera}, {Barlow}, {Bird}, {Clark}, {Dean},
  {Hill}, {Moran}, {Shaw}, {Willis}, {Bazzano}, {Ubertini}, and
  {Malizia}]{sguera2005}
{Sguera}, V.; {Barlow}, E.J.; {Bird}, A.J.; {Clark}, D.J.; {Dean}, A.J.;
  {Hill}, A.B.; {Moran}, L.; {Shaw}, S.E.; {Willis}, D.R.; {Bazzano}, A.;
  et~al.
\newblock {INTEGRAL observations of recurrent fast X-ray transient sources}.
\newblock {\em Astron. Astrophys.} {\bf 2005}, {\em 444},~221--231.
\newblock {{https://doi.org/10.1051/0004-6361:20053103}}.

\bibitem[{Negueruela} {et~al.}(2006){Negueruela}, {Smith}, {Reig}, {Chaty},
  and {Torrej{\'o}n}]{negueruela2006}
{Negueruela}, I.; {Smith}, D.M.; {Reig}, P.; {Chaty}, S.; {Torrej{\'o}n}, J.M.
\newblock {Supergiant Fast X-ray Transients: A New Class of High Mass X-ray
  Binaries Unveiled by INTEGRAL}.
\newblock In \emph{Proceedings of the The X-ray Universe 2005}; {Wilson}, A., Ed.; {ESA Publications Division: Noordwijk, The Netherlands}
  2006; Volume 604,p. 165.

\bibitem[Sidoli(2017)]{sidoli2017}
Sidoli, L.
\newblock \textls[-25]{Supergiant Fast X-ray Transients-A short review. \emph{arXiv} \textbf{2017}, {arXiv:1710.03943. }}
\newblock {{https://doi.org/10.48550/ARXIV.1710.03943}}. 

\bibitem[{Corbel}(2011)]{microquasars2011}
{Corbel}, S.
\newblock {Microquasars: an observational review.}
\newblock In \emph{Proceedings of IAU Symposium \#275}; {Romero}, G.E., {Sunyaev},
  R.A., {Belloni}, T., Eds.;  {Cambridge University Press: Cambridge, UK 2011}; pp. 205--214.
\newblock {{https://doi.org/10.1017/S1743921310016054}}.

\bibitem[McClintock and Remillard(2006)]{mcclintock_remillard_2006}
McClintock, J.E.; Remillard, R.A., Black hole binaries.
\newblock In {\em Compact Stellar X-ray Sources}; Lewin, W., van~der Klis, M.,
  Eds.; Cambridge Astrophysics, Cambridge University Press: Cambridge, UK, 2006; p.
  157–214.
\newblock {{https://doi.org/10.1017/CBO9780511536281.005}}.

\bibitem[{Titarchuk} and {Fiorito}(2004)]{titarchuk_fiorito_2004}
{Titarchuk}, L.; {Fiorito}, R.
\newblock {Spectral Index and Quasi-Periodic Oscillation Frequency Correlation
  in Black Hole Sources: Observational Evidence of Two Phases and Phase
  Transition in Black Holes}.
\newblock {\em Astrophys. J.} {\bf 2004}, {\em 612},~988--999.
\newblock {{https://doi.org/10.1086/422573}}.

\bibitem[{Charles} and {Coe}(2006)]{CharlesCoe2006}
{Charles}, P.A.; {Coe}, M.J.
\newblock {Optical, ultraviolet and infrared observations of X-ray binaries}.
  In {\em Compact Stellar X-ray Sources};  {Cambridge University Press: Cambridge, UK}, 2006; Volume 39, pp. 215--265. 

\bibitem[Kreidberg {et~al.}(2012)Kreidberg, Bailyn, Farr, and
  Kalogera]{kreidberg_mass_2012}
Kreidberg, L.; Bailyn, C.D.; Farr, W.M.; Kalogera, V.
\newblock Mass {Measurements} of {Black} {Holes} in {X}-{Ray} {Transients}:
  {Is} {There} a {Mass} {Gap}?
\newblock {\em  Astrophys. J.} {\bf 2012}, {\em 757},{~36.}
\newblock  {{https://doi.org/10.1088/0004-637X/757/1/36}}. 

\bibitem[Orosz {et~al.}(2009)Orosz, Steeghs, McClintock, Torres, Bochkov,
  Gou, Narayan, Blaschak, Levine, Remillard, Bailyn, Dwyer, and
  Buxton]{orosz_new_2009}
Orosz, J.A.; Steeghs, D.; McClintock, J.E.; Torres, M.A.P.; Bochkov, I.; Gou,
  L.; Narayan, R.; Blaschak, M.; Levine, A.M.; Remillard, R.A.;  et~al.
\newblock A {NEW} {DYNAMICAL} {MODEL} {FOR} {THE} {BLACK} {HOLE} {BINARY} {LMC}
  {X}-1.
\newblock {\em Astrophys. J.} {\bf 2009}, {\em 697},{~573.}
\newblock {{https://doi.org/10.1088/0004-637X/697/1/573}}. 

\bibitem[Orosz {et~al.}(2011)Orosz, McClintock, Aufdenberg, Remillard, Reid,
  Narayan, and Gou]{orosz_mass_2011}
Orosz, J.A.; McClintock, J.E.; Aufdenberg, J.P.; Remillard, R.A.; Reid, M.J.;
  Narayan, R.; Gou, L.
\newblock {THE} {MASS} {OF} {THE} {BLACK} {HOLE} {IN} {CYGNUS} {X}-1.
\newblock {\em Astrophys. J.} {\bf 2011}, {\em 742},{~84.}
\newblock  {{https://doi.org/10.1088/0004-637X/742/2/84}}. 

\bibitem[Orosz {et~al.}(2007)Orosz, McClintock, Narayan, Bailyn, Hartman,
  Macri, Liu, Pietsch, Remillard, Shporer, and
  Mazeh]{orosz_1565-solar-mass_2007}
Orosz, J.A.; McClintock, J.E.; Narayan, R.; Bailyn, C.D.; Hartman, J.D.; Macri,
  L.; Liu, J.; Pietsch, W.; Remillard, R.A.; Shporer, A.;  et~al.
\newblock A 15.65-solar-mass black hole in an eclipsing binary in the nearby
  spiral galaxy {M} 33.
\newblock {\em Nature} {\bf 2007}, {\em 449},{~872--875.}
\newblock {{https://doi.org/10.1038/nature06218}}. 

\bibitem[Cherepashchuk(2022)]{cherepashchuk_progress_2022}
Cherepashchuk, A.
\newblock Progress in {Understanding} the {Nature} of {SS433}.
\newblock {\em Universe} {\bf 2022}, {\em 8},{~13.}
\newblock {{https://doi.org/10.3390/universe8010013}}. 

\bibitem[Orosz(2003)]{orosz_inventory_2003}
Orosz, J.A.
\newblock Inventory of black hole binaries. In \emph{Proceedings of IAU Symposium \#212}; {Cambridge University Press: Cambridge, UK}, {\bf 2003}.
\newblock {Volume 212}, ~365. 

\bibitem[Greiner {et~al.}(2001)Greiner, Cuby, and
  McCaughrean]{greiner_unusually_2001}
Greiner, J.; Cuby, J.G.; McCaughrean, M.J.
\newblock An unusually massive stellar black hole in the {Galaxy}.
\newblock {\em Nature} {\bf 2001}, {\em 414},{~522--525.}
\newblock {{https://doi.org/10.1038/35107019}}. 

\bibitem[Casares {et~al.}(2009)Casares, Orosz, Zurita, Shahbaz,
  Corral-Santana, McClintock, Garcia, Martínez-Pais, Charles, Fender, and
  Remillard]{casares_refined_2009}
Casares, J.; Orosz, J.A.; Zurita, C.; Shahbaz, T.; Corral-Santana, J.M.;
  McClintock, J.E.; Garcia, M.R.; Martínez-Pais, I.G.; Charles, P.A.; Fender,
  R.P.;  et~al.
\newblock Refined {Orbital} {Solution} and {Quiescent} {Variability} in the
  {Black} {Hole} {Transient} {GS} 1354-64 (= {BW} {Cir}).
\newblock {\em  Astrophys. J. Suppl. Ser.} {\bf 2009}, {\em
  181},~238--243.
\newblock {{https://doi.org/10.1088/0067-0049/181/1/238}}.

\bibitem[Casares {et~al.}(2004)Casares, Zurita, Shahbaz, Charles, and
  Fender]{casares_evidence_2004}
Casares, J.; Zurita, C.; Shahbaz, T.; Charles, P.A.; Fender, R.P.
\newblock Evidence of a {Black} {Hole} in the {X}-{Ray} {Transient} {GS}
  1354-64 (={BW} {Circini}).
\newblock {\em  Astrophys. J.} {\bf 2004}, {\em 613},{~L133--L136.}
\newblock {{https://doi.org/10.1086/425145}}. 

\bibitem[Wu {et~al.}(2015)Wu, Orosz, McClintock, Steeghs, Longa-Peña,
  Callanan, Gou, Ho, Jonker, Reynolds, and Torres]{wu_dynamical_2015}
Wu, J.; Orosz, J.A.; McClintock, J.E.; Steeghs, D.; Longa-Peña, P.; Callanan,
  P.J.; Gou, L.; Ho, L.C.; Jonker, P.G.; Reynolds, M.T.;  et~al.
\newblock A {Dynamical} {Study} of the {Black} {Hole} {X}-{Ray} {Binary} {Nova}
  {Muscae} 1991.
\newblock {\em  Astrophys. J.} {\bf 2015}, {\em 806},{~92.}
\newblock 
  {{https://doi.org/10.1088/0004-637X/806/1/92}}. 

\bibitem[Wu {et~al.}(2016)Wu, Orosz, McClintock, Hasan, Bailyn, Gou, and
  Chen]{wu_mass_2016}
Wu, J.; Orosz, J.A.; McClintock, J.E.; Hasan, I.; Bailyn, C.D.; Gou, L.; Chen,
  Z.
\newblock The {Mass} of the {Black} {Hole} in the {X}-ray {Binary} {Nova}
  {Muscae} 1991.
\newblock {\em  Astrophys. J.} {\bf 2016}, {\em 825},{~46.}
\newblock  {{https://doi.org/10.3847/0004-637X/825/1/46}}. 

\bibitem[González~Hernández {et~al.}(2014)González~Hernández, Rebolo,
  and Casares]{gonzalez_hernandez_fast_2014}
González~Hernández, J.I.; Rebolo, R.; Casares, J.
\newblock Fast orbital decays of black hole {X}-ray binaries: {XTE} {J1118}+480
  and {A0620}-00.
\newblock {\em Mon. Not. R. Astron. Soc.} {\bf 2014}, {\em 438},{~L21--L25.}
\newblock {{https://doi.org/10.1093/mnrasl/slt150}}. 

\bibitem[Torres {et~al.}(2004)Torres, Callanan, Garcia, Zhao, Laycock, and
  Kong]{torres_mmt_2004}
Torres, M.A.P.; Callanan, P.J.; Garcia, M.R.; Zhao, P.; Laycock, S.; Kong,
  A.K.H.
\newblock {MMT} {Observations} of the {Black} {Hole} {Candidate} {XTE}
  {J1118}+480 near and in {Quiescence}.
\newblock {\em  Astrophys. J.} {\bf 2004}, {\em 612},{~1026--1033.}
\newblock  {{https://doi.org/10.1086/422740}}. 

\bibitem[González~Hernández {et~al.}(2012)González~Hernández, Rebolo,
  and Casares]{gonzalez_hernandez_fast_2012}
González~Hernández, J.I.; Rebolo, R.; Casares, J.
\newblock The {Fast} {Spiral}-in of the {Companion} {Star} to the {Black}
  {Hole} {XTE} {J1118}+480.
\newblock {\em Astrophys. J.} {\bf 2012}, {\em 744},{~L25.}
\newblock  {{https://doi.org/10.1088/2041-8205/744/2/L25}}. 

\bibitem[Hernández {et~al.}(2008)Hernández, Rebolo, Israelian, Filippenko,
  Chornock, Tominaga, Umeda, and Nomoto]{hernandez_chemical_2008}
Hernández, J.I.G.; Rebolo, R.; Israelian, G.; Filippenko, A.V.; Chornock, R.;
  Tominaga, N.; Umeda, H.; Nomoto, K.
\newblock Chemical {Abundances} of the {Secondary} {Star} in the {Black} {Hole}
  {X}-{Ray} {Binary} {XTE} {J1118}+480.
\newblock {\em Astrophys. J.} {\bf 2008}, {\em 679},{~732.}
\newblock {{https://doi.org/10.1086/586888}}. 

\bibitem[Calvelo {et~al.}(2009)Calvelo, Vrtilek, Steeghs, Torres, Neilsen,
  Filippenko, and González~Hernández]{calvelo_doppler_2009}
Calvelo, D.E.; Vrtilek, S.D.; Steeghs, D.; Torres, M.A.P.; Neilsen, J.;
  Filippenko, A.V.; González~Hernández, J.I.
\newblock Doppler and modulation tomography of {XTEJ1118}+480 in quiescence.
\newblock {\em Mon. Not. R. Astron. Soc.} {\bf 2009}, {\em 399},{~539--549.}
\newblock  {{https://doi.org/10.1111/j.1365-2966.2009.15304.x}}. 

\bibitem[Khargharia {et~al.}(2012)Khargharia, Froning, Robinson, and
  Gelino]{khargharia_mass_2012}
Khargharia, J.; Froning, C.S.; Robinson, E.L.; Gelino, D.M.
\newblock {THE} {MASS} {OF} {THE} {BLACK} {HOLE} {IN} {XTE} {J1118}+480.
\newblock {\em AJ} {\bf 2012}, {\em 145},{~21.}
\newblock  {{https://doi.org/10.1088/0004-6256/145/1/21}}. 

\bibitem[Hernández and Casares(2010)]{hernandez_doppler_2010}
Hernández, J.I.G.; Casares, J.
\newblock Doppler tomography of the black hole binary {A0620}-00 and the origin
  of chromospheric emission in quiescent {X}-ray binaries.
\newblock {\em Astron. Astrophys.} {\bf 2010}, {\em 516},{~A58.}
\newblock {{https://doi.org/10.1051/0004-6361/201014088}}. 

\bibitem[McClintock and Remillard(1986)]{mcclintock_black_1986}
McClintock, J.E.; Remillard, R.A.
\newblock The {Black} {Hole} {Binary} {A0620}-00.
\newblock {\em  Astrophys. J.} {\bf 1986}, {\em 308},{~110.}
\newblock  {{https://doi.org/10.1086/164482}}. 

\bibitem[Neilsen {et~al.}(2008)Neilsen, Steeghs, and
  Vrtilek]{neilsen_eccentric_2008}
Neilsen, J.; Steeghs, D.; Vrtilek, S.D.
\newblock The eccentric accretion disc of the black hole {A0620}-00.
\newblock {\em Mon. Not. R. Astron. Soc.} {\bf 2008}, {\em 384},~849--862.
\newblock {{https://doi.org/10.1111/j.1365-2966.2007.12599.x}}.

\bibitem[Cantrell {et~al.}(2010)Cantrell, Bailyn, Orosz, McClintock,
  Remillard, Froning, Neilsen, Gelino, and Gou]{cantrell_inclination_2010}
Cantrell, A.G.; Bailyn, C.D.; Orosz, J.A.; McClintock, J.E.; Remillard, R.A.;
  Froning, C.S.; Neilsen, J.; Gelino, D.M.; Gou, L.
\newblock The {Inclination} of the {Soft} {X}-{Ray} {Transient} {A0620}-00 and
  the {Mass} of its {Black} {Hole}.
\newblock {\em  Astrophys. J.} {\bf 2010}, {\em 710},{~1127--1141.}
\newblock {{https://doi.org/10.1088/0004-637X/710/2/1127}}.

\bibitem[Ioannou {et~al.}(2004)Ioannou, Robinson, Welsh, and
  Haswell]{ioannou_mass_2004}
Ioannou, Z.; Robinson, E.L.; Welsh, W.F.; Haswell, C.A.
\newblock The {Mass} of the {Black} {Hole} in {GS} 2000+25.
\newblock {\em  Astron. J.} {\bf 2004}, {\em 127},{~481--488.}
\newblock  {{https://doi.org/10.1086/380215}}. 

\bibitem[{Harlaftis} {et~al.}(1996){Harlaftis}, {Horne}, and
  {Filippenko}]{Harlaftis1996_ratio}
{Harlaftis}, E.T.; {Horne}, K.; {Filippenko}, A.V.
\newblock {The Mass Ratio and the Disk Image of the X-Ray Nova GS 2000+25}.
\newblock {\em Publ. Astron. Soc. Pac.} {\bf 1996}, {\em 108},~762.
\newblock {{https://doi.org/10.1086/133799}}.

\bibitem[Torres {et~al.}(2021)Torres, Jonker, Casares, Miller-Jones, and
  Steeghs]{torres_delimiting_2021}
Torres, M.A.P.; Jonker, P.G.; Casares, J.; Miller-Jones, J.C.A.; Steeghs, D.
\newblock Delimiting the black hole mass in the {X}-ray transient {MAXI}
  {J1659}-152 with {H$\alpha$} spectroscopy.
\newblock {\em Mon. Not. R. Astron. Soc.} {\bf 2021}, {\em 501},~2174--2181.
\newblock {{https://doi.org/10.1093/mnras/staa3786}}.

\bibitem[Mata~Sánchez {et~al.}(2021)Mata~Sánchez, Rau, Álvarez
  Hernández, van Grunsven, Torres, and Jonker]{mata_sanchez_dynamical_2021}
Mata~Sánchez, D.; Rau, A.; Álvarez Hernández, A.; van Grunsven, T.F.J.;
  Torres, M.A.P.; Jonker, P.G.
\newblock Dynamical confirmation of a stellar mass black hole in the transient
  {X}-ray dipping binary {MAXI} {J1305}-704.
\newblock {\em Mon. Not. R. Astron. Soc.} {\bf 2021}, {\em 506},{~581--594.}
\newblock{{https://doi.org/10.1093/mnras/stab1714}}. 

\bibitem[Casares and Charles(1994)]{casares_optical_1994}
Casares, J.; Charles, P.A.
\newblock Optical studies of {V404} {Cyg}, the {X}-ray transient {GS} 2023+338.
  {IV}. {The} rotation speed of the companion star.
\newblock {\em Mon. Not. R. Astron. Soc.} {\bf 1994}, {\em 271},{~L5--L9.}
\newblock {{https://doi.org/10.1093/mnras/271.1.L5}}. 

\bibitem[Khargharia {et~al.}(2010)Khargharia, Froning, and
  Robinson]{khargharia_near-infrared_2010}
Khargharia, J.; Froning, C.S.; Robinson, E.L.
\newblock Near-infrared {Spectroscopy} of {Low}-mass {X}-ray {Binaries}:
  {Accretion} {Disk} {Contamination} and {Compact} {Object} {Mass}
  {Determination} in {V404} {Cyg} and {Cen} {X}-4.
\newblock {\em  Astrophys. J.} {\bf 2010}, {\em 716},{~1105--1117.}
\newblock   {{https://doi.org/10.1088/0004-637X/716/2/1105}}. 

\bibitem[Orosz {et~al.}(2004)Orosz, McClintock, Remillard, and
  Corbel]{orosz_orbital_2004}
Orosz, J.A.; McClintock, J.E.; Remillard, R.A.; Corbel, S.
\newblock Orbital {Parameters} for the {Black} {Hole} {Binary} {XTE}
  {J1650}-500.
\newblock {\em  Astrophys. J.} {\bf 2004}, {\em 616},{~376--382.}
\newblock{{https://doi.org/10.1086/424892}}. 

\bibitem[Webb {et~al.}(2000)Webb, Naylor, Ioannou, Charles, and
  Shahbaz]{webb_tio_2000}
Webb, N.A.; Naylor, T.; Ioannou, Z.; Charles, P.A.; Shahbaz, T.
\newblock A {TiO} study of the black hole binary {GRO} {J0422}+32 in a very low
  state.
\newblock {\em Mon. Not. R. Astron. Soc.} {\bf 2000}, {\em 317},{~528--534.}
\newblock {{https://doi.org/10.1046/j.1365-8711.2000.03608.x}}. 

\bibitem[Remillard {et~al.}(1996)Remillard, Orosz, McClintock, and
  Bailyn]{remillard_dynamical_1996}
Remillard, R.A.; Orosz, J.A.; McClintock, J.E.; Bailyn, C.D.
\newblock Dynamical {Evidence} for a {Black} {Hole} in {X}-{Ray} {Nova}
  {Ophiuchi} 1977.
\newblock {\em  Astrophys. J.} {\bf 1996}, {\em 459},{~226.}
\newblock {{https://doi.org/10.1086/176885}}. 

\bibitem[Harlaftis {et~al.}(1997)Harlaftis, Steeghs, Horne, and
  Filippenko]{harlaftis_doppler_1997}
Harlaftis, E.T.; Steeghs, D.; Horne, K.; Filippenko, A.V.
\newblock A doppler map and mass-ration constraint for the black-hole X-ray Nova Ophiuchi 1977.
\newblock {\em  Astron. J.} {\bf 1997}, {\em 114},{~1170--1175.}
\newblock {{https://doi.org/10.1086/118548}}. 

\bibitem[Hernández {et~al.}(2008)Hernández, Rebolo, and
  Israelian]{hernandez_black_2008}
Hernández, J.I.G.; Rebolo, R.; Israelian, G.
\newblock The black hole binary nova {Scorpii} 1994 ({GRO} {J1655}-40): an
  improved chemical analysis.
\newblock {\em Astron. Astrophys.} {\bf 2008}, {\em 478},{~203--217.}
\newblock {{https://doi.org/10.1051/0004-6361:20077141}}. 

\bibitem[Beer and Podsiadlowski(2002)]{beer_quiescent_2002}
Beer, M.E.; Podsiadlowski, P.
\newblock The quiescent light curve and the evolutionary state of {GRO}
  {J1655}-40.
\newblock {\em Mon. Not. R. Astron. Soc.} {\bf 2002}, {\em 331},{~351--360.}
\newblock   {{https://doi.org/10.1046/j.1365-8711.2002.05189.x}}. 

\bibitem[Shahbaz(2003)]{shahbaz_determining_2003}
Shahbaz, T.
\newblock Determining the spectroscopic mass ratio in interacting binaries:
  application to {X}-{Ray} {Nova} {Sco} 1994.
\newblock {\em Mon. Not. R. Astron. Soc.} {\bf 2003}, {\em 339},{~1031--1040.}
\newblock{{https://doi.org/10.1046/j.1365- 8711.2003.06258.x}}. 

\bibitem[Greene {et~al.}(2001)Greene, Bailyn, and
  Orosz]{greene_optical_2001}
Greene, J.; Bailyn, C.D.; Orosz, J.A.
\newblock Optical and {Infrared} {Photometry} of the {Microquasar} {GRO}
  {J1655}–40 in {Quiescence}.
\newblock {\em Astrophys. J.} {\bf 2001}, {\em 554},{~1290.}
\newblock {{https://doi.org/10.1086/321411}}. 

\bibitem[Corral-Santana {et~al.}(2011)Corral-Santana, Casares, Shahbaz,
  Zurita, Martínez-Pais, and Rodríguez-Gil]{corral-santana_evidence_2011}
Corral-Santana, J.M.; Casares, J.; Shahbaz, T.; Zurita, C.; Martínez-Pais,
  I.G.; Rodríguez-Gil, P.
\newblock Evidence for a black hole in the {X}-ray transient {XTE} {J1859}+226.
\newblock {\em Mon. Not. R. Astron. Soc.} {\bf 2011}, {\em 413},{~L15--L19.}
\newblock {{https://doi.org/10.1111/j.1745-3933.2011.01022.x}}.  

\bibitem[Filippenko and Chornock(2001)]{filippenko_xte_2001}
Filippenko, A.V.; Chornock, R.
\newblock {XTE} {J1859}+226.
\newblock {\em Int. Astron. Union Circ.} {\bf 2001}, {\em
  7644},{~2}. 

\bibitem[Sánchez {et~al.}(2022)Sánchez, Muñoz-Darias, Cúneo, Padilla,
  Sánchez-Sierras, Panizo-Espinar, Casares, Corral-Santana, and
  Torres]{sanchez_hard-state_2022}
Sánchez, D.M.; Muñoz-Darias, T.; Cúneo, V.A.; Padilla, M.A.;
  Sánchez-Sierras, J.; Panizo-Espinar, G.; Casares, J.; Corral-Santana, J.M.;
  Torres, M.A.P.
\newblock Hard-state {Optical} {Wind} during the {Discovery} {Outburst} of the
  {Black} {Hole} {X}-{Ray} {Dipper} {MAXI} {J1803}–298.
\newblock {\em Astrophys. J. Lett.} {\bf 2022}, {\em 926},{~L10.}
\newblock {{https://doi.org/10.3847/2041-8213/ac502f}}. 

\bibitem[Torres {et~al.}(2019)Torres, Casares, Jiménez-Ibarra,
  Muñoz-Darias, Armas~Padilla, Jonker, and Heida]{torres_dynamical_2019}
Torres, M.A.P.; Casares, J.; Jiménez-Ibarra, F.; Muñoz-Darias, T.;
  Armas~Padilla, M.; Jonker, P.G.; Heida, M.
\newblock Dynamical {Confirmation} of a {Black} {Hole} in {MAXI} {J1820}+070.
\newblock {\em  Astrophys. J.} {\bf 2019}, {\em 882},{~L21.}
\newblock {{https://doi.org/10.3847/2041-8213/ab39df}}. 

\bibitem[Torres {et~al.}(2020)Torres, Casares, Jiménez-Ibarra, Álvarez
  Hernández, Muñoz-Darias, Armas~Padilla, Jonker, and
  Heida]{torres_binary_2020}
Torres, M.A.P.; Casares, J.; Jiménez-Ibarra, F.; Álvarez Hernández, A.;
  Muñoz-Darias, T.; Armas~Padilla, M.; Jonker, P.G.; Heida, M.
\newblock The {Binary} {Mass} {Ratio} in the {Black} {Hole} {Transient} {MAXI}
  {J1820}+070.
\newblock {\em  Astrophys. J.} {\bf 2020}, {\em 893},{~L37.}
\newblock {{https://doi.org/10.3847/2041-8213/ab863a}}.  

\bibitem[Atri {et~al.}(2020)Atri, Miller-Jones, Bahramian, Plotkin, Deller,
  Jonker, Maccarone, Sivakoff, Soria, Altamirano, Belloni, Fender, Koerding,
  Maitra, Markoff, Migliari, Russell, Russell, Sarazin, Tetarenko, and
  Tudose]{atri_radio_2020}
Atri, P.; Miller-Jones, J.C.A.; Bahramian, A.; Plotkin, R.M.; Deller, A.T.;
  Jonker, P.G.; Maccarone, T.J.; Sivakoff, G.R.; Soria, R.; Altamirano, D.;
  et~al.
\newblock A radio parallax to the black hole {X}-ray binary {MAXI} {J1820}+070.
\newblock {\em Mon. Not. R. Astron. Soc.} {\bf 2020}, {\em 493},{~L81--L86.}
\newblock{{https://doi.org/10.1093/mnrasl/slaa010}}. 

\bibitem[Orosz {et~al.}(2011)Orosz, Steiner, McClintock, Torres, Remillard,
  Bailyn, and Miller]{orosz_improved_2011}
Orosz, J.A.; Steiner, J.F.; McClintock, J.E.; Torres, M.A.P.; Remillard, R.A.;
  Bailyn, C.D.; Miller, J.M.
\newblock An {Improved} {Dynamical} {Model} for the {Microquasar} {XTE}
  {J1550}-564.
\newblock {\em  Astrophys. J.} {\bf 2011}, {\em 730},{~75.}
\newblock {{https://doi.org/10.1088/0004-637X/730/2/75}}. 

\bibitem[Connors {et~al.}(2020)Connors, García, Dauser, Grinberg, Steiner,
  Sridhar, Wilms, Tomsick, Harrison, and Licklederer]{connors_evidence_2020}
Connors, R.M.T.; García, J.A.; Dauser, T.; Grinberg, V.; Steiner, J.F.;
  Sridhar, N.; Wilms, J.; Tomsick, J.; Harrison, F.; Licklederer, S.
\newblock Evidence for {Returning} {Disk} {Radiation} in the {Black} {Hole}
  {X}-{Ray} {Binary} {XTE} {J1550}–564.
\newblock {\em Astrophys. J.} {\bf 2020}, {\em 892},{~47.}
\newblock {{https://doi.org/10.3847/1538-4357/ab7afc}}. 

\bibitem[Heida {et~al.}(2017)Heida, Jonker, Torres, and
  Chiavassa]{heida_mass_2017}
Heida, M.; Jonker, P.G.; Torres, M.A.P.; Chiavassa, A.
\newblock The {Mass} {Function} of {GX} 339–4 from {Spectroscopic}
  {Observations} of {Its} {Donor} {Star}.
\newblock {\em Astrophys. J.} {\bf 2017}, {\em 846},{~132.}
\newblock  {{https://doi.org/10.3847/1538-4357/aa85df}}. 

\bibitem[Hynes {et~al.}(2003)Hynes, Steeghs, Casares, Charles, and
  O'Brien]{hynes_dynamical_2003}
Hynes, R.I.; Steeghs, D.; Casares, J.; Charles, P.A.; O'Brien, K.
\newblock Dynamical {Evidence} for a {Black} {Hole} in {GX} 339-4.
\newblock {\em  Astrophys. J.} {\bf 2003}, {\em 583},{~L95--L98.}
\newblock{{https://doi.org/10.1086/368108}}. 

\bibitem[Filippenko {et~al.}(1999)Filippenko, Leonard, Matheson, Li, Moran,
  and Riess]{filippenko_black_1999}
Filippenko, A.V.; Leonard, D.C.; Matheson, T.; Li, W.; Moran, E.C.; Riess, A.G.
\newblock A {Black} {Hole} in the {X}-{Ray} {Nova} {Velorum} 1993.
\newblock {\em Publ. Astron. Soc. Pac.} {\bf
  1999}, {\em 111},{~969--979.}
\newblock  {{https://doi.org/10.1086/316413}}. 

\bibitem[Macias {et~al.}(2011)Macias, Orosz, Bailyn, Buxton, Schechter,
  Remillard, McClintock, and Steiner]{macias_refined_2011}
Macias, P.; Orosz, J.A.; Bailyn, C.D.; Buxton, M.M.; Schechter, P.L.;
  Remillard, R.A.; McClintock, J.E.; Steiner, J.F.
\newblock A {Refined} {Black} {Hole} {Mass} for the {X}-ray {Transient} {GRS}
  1009-45. \emph{Am. Astron. Soc. Meet. Abstr.} {\bf 2011}.
\newblock {\em 217},{~143.04.} 

\bibitem[Orosz {et~al.}(2014)Orosz, Steiner, McClintock, Buxton, Bailyn,
  Steeghs, Guberman, and Torres]{orosz_mass_2014}
Orosz, J.A.; Steiner, J.F.; McClintock, J.E.; Buxton, M.M.; Bailyn, C.D.;
  Steeghs, D.; Guberman, A.; Torres, M.A.P.
\newblock {THE} {MASS} {OF} {THE} {BLACK} {HOLE} {IN} {LMC} {X}-3.
\newblock {\em Astrophys. J.} {\bf 2014}, {\em 794},{~154.}
\newblock{{https://doi.org/10.1088/0004-637X/794/2/154}}. 

\bibitem[Orosz {et~al.}(2001)Orosz, Kuulkers, van~der Klis, McClintock,
  Garcia, Callanan, Bailyn, Jain, and Remillard]{orosz_black_2001}
Orosz, J.A.; Kuulkers, E.; van~der Klis, M.; McClintock, J.E.; Garcia, M.R.;
  Callanan, P.J.; Bailyn, C.D.; Jain, R.K.; Remillard, R.A.
\newblock A {Black} {Hole} in the {Superluminal} {Source} {SAX} {J1819}.3-2525
  ({V4641} {Sgr}).
\newblock {\em  Astrophys. J.} {\bf 2001}, {\em 555},{~489--503.}
\newblock {{https://doi.org/10.1086/321442}}. 

\bibitem[MacDonald {et~al.}(2014)MacDonald, Bailyn, Buxton, Cantrell,
  Chatterjee, Kennedy-Shaffer, Orosz, Markwardt, and
  Swank]{macdonald_black_2014}
MacDonald, R.K.D.; Bailyn, C.D.; Buxton, M.; Cantrell, A.G.; Chatterjee, R.;
  Kennedy-Shaffer, R.; Orosz, J.A.; Markwardt, C.B.; Swank, J.H.
\newblock The {Black} {Hole} {Binary} {V4641} {Sagitarii}: {Activity} in
  {Quiescence} and {Improved} {Mass} {Determinations}.
\newblock {\em  Astrophys. J.} {\bf 2014}, {\em 784},{~2.}
\newblock {{https://doi.org/10.1088/0004-637X/784/1/2}}. 

\bibitem[Lindstrøm {et~al.}(2005)Lindstrøm, Griffin, Kiss, Uemura,
  Derekas, Mészáros, and Székely]{lindstrom_new_2005}
Lindstrøm, C.; Griffin, J.; Kiss, L.L.; Uemura, M.; Derekas, A.; Mészáros,
  S.; Székely, P.
\newblock New clues on outburst mechanisms and improved spectroscopic elements
  of the black hole binary {V4641} {Sagittarii}.
\newblock {\em Mon. Not. R. Astron. Soc.} {\bf 2005}, {\em 363},{~882--890.}
\newblock {{https://doi.org/10.1111/j.1365-2966.2005.09483.x}}.

\bibitem[Zhang {et~al.}(2022)Zhang, Liu, Abdikamalov, Ayzenberg, Bambi, and
  Zhou]{zhang_testing_2022}
Zhang, Z.; Liu, H.; Abdikamalov, A.B.; Ayzenberg, D.; Bambi, C.; Zhou, M.
\newblock Testing the {Kerr} {Black} {Hole} {Hypothesis} with {GRS} 1716-249 by
  {Combining} the {Continuum} {Fitting} and the {Iron}-line {Methods}.
\newblock {\em Astrophys. J.} {\bf 2022}, {\em 924},{~72.}
\newblock {{https://doi.org/10.3847/1538-4357/ac350e}}. 

\bibitem[Jana {et~al.}(2021)Jana, Jaisawal, Naik, Kumari, Chatterjee,
  Chatterjee, Bhowmick, Chakrabarti, Chang, and Debnath]{jana_accretion_2021}
Jana, A.; Jaisawal, G.K.; Naik, S.; Kumari, N.; Chatterjee, D.; Chatterjee, K.;
  Bhowmick, R.; Chakrabarti, S.K.; Chang, H.K.; Debnath, D.
\newblock Accretion properties of {MAXI} {J1813}-095 during its failed outburst
  in 2018.
\newblock {\em Res. Astron. Astrophys.} {\bf 2021}, {\em 21},{~125.}
\newblock {{https://doi.org/10.1088/1674-4527/21/5/125}}. 

\bibitem[Shang {et~al.}(2019)Shang, Debnath, Chatterjee, Jana, Chakrabarti,
  Chang, Yap, and Chiu]{shang_evolution_2019}
Shang, J.R.; Debnath, D.; Chatterjee, D.; Jana, A.; Chakrabarti, S.K.; Chang,
  H.K.; Yap, Y.X.; Chiu, C.L.
\newblock Evolution of {X}-{Ray} {Properties} of {MAXI} {J1535}-571: {Analysis}
  with the {TCAF} {Solution}.
\newblock {\em Astrophys. J.} {\bf 2019}, {\em 875},{~4.}
\newblock{{https://doi.org/10.3847/1538-4357/ab0c1e}}. 

\bibitem[Molla {et~al.}(2017)Molla, Chakrabarti, Debnath, and
  Mondal]{molla_estimation_2017}
Molla, A.A.; Chakrabarti, S.K.; Debnath, D.; Mondal, S.
\newblock {Estimation} {of} {Mass} {of} {Compact} {Object} {in} {H} 1743-322
  {From} 2010 {and} 2011 {Outbursts} {Using} {Tcaf} {Solution} {and} {Spectral}
  {Index}–{qpo} {Frequency} {Correlation}.
\newblock {\em Astrophys. J.} {\bf 2017}, {\em 834},~88.
\newblock {Publisher: The American Astronomical Society,}
  {{https://doi.org/10.3847/1538-4357/834/1/88}}.

\bibitem[Tursunov and Kološ(2018)]{tursunov_constraints_2018}
Tursunov, A.A.; Kološ, M.
\newblock Constraints on {Mass}, {Spin} and {Magnetic} {Field} of {Microquasar}
  {H} 1743-322 from {Observations} of {QPOs}.
\newblock {\em Phys. Atom. Nuclei} {\bf 2018}, {\em 81},~279--282.
\newblock {{https://doi.org/10.1134/S1063778818020187}}.

\bibitem[{Brocksopp} {et~al.}(1999){Brocksopp}, {Tarasov}, {Lyuty}, and
  {Roche}]{brocksopp_improved_1998}
{Brocksopp}, C.; {Tarasov}, A.E.; {Lyuty}, V.M.; {Roche}, P.
\newblock {An improved orbital ephemeris for Cygnus X-1}.
\newblock {\em Astron. Astrophys.} {\bf 1999}, {\em 343},~861--864,

\bibitem[Gies {et~al.}(2003)Gies, Bolton, Thomson, Huang, McSwain, Riddle,
  Wang, Wiita, Wingert, Csák, and Kiss]{gies_wind_2003}
Gies, D.R.; Bolton, C.T.; Thomson, J.R.; Huang, W.; McSwain, M.V.; Riddle,
  R.L.; Wang, Z.; Wiita, P.J.; Wingert, D.W.; Csák, B.;  et~al.
\newblock Wind {Accretion} and {State} {Transitions} in {Cygnus} {X}-1.
\newblock {\em Astrophys. J.} {\bf 2003}, {\em 583},{~424.}
\newblock
{{https://doi.org/10.1086/345345}}. 

\bibitem[Hutchings {et~al.}(1987)Hutchings, Crampton, Cowley, Bianchi, and
  Thompson]{hutchings_optical_1987}
Hutchings, J.B.; Crampton, D.; Cowley, A.P.; Bianchi, L.; Thompson, I.B.
\newblock Optical and {UV} {Spectroscopy} of the {Black} {Hole} {Binary}
  {Candidate} {LMC} {X}-1.
\newblock {\em  Astron. J.} {\bf 1987}, {\em 94},{~340.}
\newblock {{https://doi.org/10.1086/114475}}. 

\bibitem[Pietsch {et~al.}(2006)Pietsch, Haberl, Sasaki, Gaetz, Plucinsky,
  Ghavamian, Long, and Pannuti]{pietsch_m33_2006}
Pietsch, W.; Haberl, F.; Sasaki, M.; Gaetz, T.J.; Plucinsky, P.P.; Ghavamian,
  P.; Long, K.S.; Pannuti, T.G.
\newblock M33 {X}-7: {ChASeM33} {Reveals} the {First} {Eclipsing} {Black}
  {Hole} {X}-{Ray} {Binary}.
\newblock {\em Astrophys. J.} {\bf 2006}, {\em 646},~420.
\newblock {{https://doi.org/10.1086/504704}}.

\bibitem[Crowther {et~al.}(2010)Crowther, Barnard, Carpano, Clark, Dhillon,
  and Pollock]{crowther_ngc_2010}
Crowther, P.A.; Barnard, R.; Carpano, S.; Clark, J.S.; Dhillon, V.S.; Pollock,
  A.M.T.
\newblock {NGC} 300 {X}-1 is a {Wolf}-{Rayet}/black hole binary.
\newblock {\em Mon. Not. R. Astron. Soc. Lett.} {\bf 2010}, {\em
  403},~L41--L45.
\newblock {{https://doi.org/10.1111/j.1745-3933.2010.00811.x}}.

\bibitem[Binder {et~al.}(2021)Binder, Sy, Eracleous, Christodoulou,
  Bhattacharya, Cappallo, Laycock, Plucinsky, and
  Williams]{binder_wolfrayet_2021}
Binder, B.A.; Sy, J.M.; Eracleous, M.; Christodoulou, D.M.; Bhattacharya, S.;
  Cappallo, R.; Laycock, S.; Plucinsky, P.P.; Williams, B.F.
\newblock The {Wolf}–{Rayet} + {Black} {Hole} {Binary} {NGC} 300 {X}-1:
  {What} is the {Mass} of the {Black} {Hole}?
\newblock {\em Astrophys. J.} {\bf 2021}, {\em 910},{~74.}
\newblock {{https://doi.org/10.3847/1538-4357/abe6a9}}. 

\bibitem[Picchi {et~al.}(2020)Picchi, Shore, Harvey, and
  Berdyugin]{picchi_optical_2020}
Picchi, P.; Shore, S.N.; Harvey, E.J.; Berdyugin, A.
\newblock An optical spectroscopic and polarimetric study of the microquasar
  binary system {SS} 433.
\newblock {\em Astron. Astrophys.} {\bf 2020}, {\em 640},{~A96.}
\newblock {{https://doi.org/10.1051/0004-6361/202037960}}. 

\bibitem[Cherepashchuk {et~al.}(2021)Cherepashchuk, Belinski, Dodin, and
  Postnov]{cherepashchuk_discovery_2021}
Cherepashchuk, A.M.; Belinski, A.A.; Dodin, A.V.; Postnov, K.A.
\newblock Discovery of orbital eccentricity and evidence for orbital period
  increase of {SS433}.
\newblock {\em Mon. Not. R. Astron. Soc. Lett.} {\bf 2021}, {\em
  507},~L19--L23.
\newblock {{https://doi.org/10.1093/mnrasl/slab083}}.

\bibitem[Mróz {et~al.}(2022)Mróz, Udalski, and
  Gould]{mroz_systematic_2022}
Mróz, P.; Udalski, A.; Gould, A.
\newblock Systematic {Errors} as a {Source} of {Mass} {Discrepancy} in {Black}
  {Hole} {Microlensing} {Event} {OGLE}-2011-{BLG}-0462.
\newblock {\em Astrophys. J. Lett.} {\bf 2022}, {\em 937},{~L24.}
\newblock {{https://doi.org/10.3847/2041-8213/ac90bb}}. 

\bibitem[{Jonker} {et~al.}(2021){Jonker}, {Kaur}, {Stone}, and
  {Torres}]{jonker2021}
{Jonker}, P.G.; {Kaur}, K.; {Stone}, N.; {Torres}, M.A.P.
\newblock {The Observed Mass Distribution of Galactic Black Hole LMXBs Is
  Biased against Massive Black Holes}.
\newblock {\em Astrophys. J.} {\bf 2021}, {\em 921},~131.
\newblock {{https://doi.org/10.3847/1538-4357/ac2839}}.

\bibitem[{Pols} {et~al.}(1998){Pols}, {Schr{\"o}der}, {Hurley}, {Tout}, and
  {Eggleton}]{WindPols}
{Pols}, O.R.; {Schr{\"o}der}, K.P.; {Hurley}, J.R.; {Tout}, C.A.; {Eggleton},
  P.P.
\newblock {Stellar evolution models for Z = 0.0001 to 0.03}.
\newblock {\em Mon. Not. R. Astron. Soc.} {\bf 1998}, {\em 298},~525--536.
\newblock {{https://doi.org/10.1046/j.1365-8711.1998.01658.x}}.

\bibitem[Kroupa {et~al.}(2013)Kroupa, Weidner, Pflamm-Altenburg, Thies,
  Dabringhausen, Marks, and Maschberger]{Kroupa2013}
Kroupa, P.; Weidner, C.; Pflamm-Altenburg, J.; Thies, I.; Dabringhausen, J.;
  Marks, M.; Maschberger, T.
\newblock The Stellar and Sub-Stellar Initial Mass Function of Simple and
  Composite Populations. In {\em Planets, Stars and Stellar Systems}; Springer: Amsterdam, 
  The Netherlands,  2013; pp. 115--242.
\newblock {{https://doi.org/10.1007/978-94-007-5612-0\_4}}.

\bibitem[{Hopkins}(2018)]{Hopkins2018}
{Hopkins}, A.M.
\newblock {The Dawes Review 8: Measuring the Stellar Initial Mass Function}.
\newblock {\em Publ. Astron. Soc. Pac.} {\bf 2018}, {\em 35},~e039.
\newblock {{https://doi.org/10.1017/pasa.2018.29}}.

\bibitem[{The LIGO Scientific Collaboration} {et~al.}(2021){The LIGO
  Scientific Collaboration}, {the Virgo Collaboration}, {the KAGRA
  Collaboration}, et~al.]{gwtc3pop}
{The LIGO Scientific Collaboration}.; {the Virgo Collaboration}.; {the KAGRA
  Collaboration}.;  et~al.
\newblock {The population of merging compact binaries inferred using
  gravitational waves through GWTC-3}.
\newblock {\em arXiv} {\bf 2021},  arXiv:2111.03634.

\bibitem[de~Sá {et~al.}()de~Sá, Bernardo, Bachega, Rocha, and
  Horvath]{desa_synthesis}
de~Sá, L.M.; Bernardo, A.; Bachega, R.R.A.; Rocha, L.S.; Horvath, J.E.
\newblock Effects of a Non-Universal IMF and Binary Parameter Correlations on
  Compact Binary Mergers.
  Available online:  \url{http://xxx.lanl.gov/abs/https://onlinelibrary.wiley.com/doi/pdf/10.1002/asna.20220089} {(accessed on 21 Dec 2022)}.
\newblock {{https://doi.org/https://doi.org/10.1002/asna.20220089}}. 

\bibitem[{Chru{\'s}li{\'n}ska} {et~al.}(2020){Chru{\'s}li{\'n}ska},
  {Je{\v{r}}{\'a}bkov{\'a}}, {Nelemans}, and {Yan}]{Chruslinska2020}
{Chru{\'s}li{\'n}ska}, M.; {Je{\v{r}}{\'a}bkov{\'a}}, T.; {Nelemans}, G.;
  {Yan}, Z.
\newblock {The effect of the environment-dependent IMF on the formation and
  metallicities of stars over the cosmic history}.
\newblock {\em Astron. Astrophys.} {\bf 2020}, {\em 636},~A10.
\newblock {{https://doi.org/10.1051/0004-6361/202037688}}.

\bibitem[{Neijssel} {et~al.}(2019){Neijssel}, {Vigna-G{\'o}mez},
  {Stevenson}, {Barrett}, {Gaebel}, {Broekgaarden}, {de Mink}, {Sz{\'e}csi},
  {Vinciguerra}, and {Mandel}]{Neijssel2019}
{Neijssel}, C.J.; {Vigna-G{\'o}mez}, A.; {Stevenson}, S.; {Barrett}, J.W.;
  {Gaebel}, S.M.; {Broekgaarden}, F.S.; {de Mink}, S.E.; {Sz{\'e}csi}, D.;
  {Vinciguerra}, S.; {Mandel}, I.
\newblock {The effect of the metallicity-specific star formation history on double compact object mergers}.
\newblock {\em Mon. Not. R. Astron. Soc.} {\bf 2019}, {\em 490},~3740--3759.
\newblock {{https://doi.org/10.1093/mnras/stz2840}}.

\bibitem[{Fishbach} {et~al.}(2021){Fishbach}, {Doctor}, {Callister},
  {Edelman}, {Ye}, {Essick}, {Farr}, {Farr}, and {Holz}]{Fishbach2021When}
{Fishbach}, M.; {Doctor}, Z.; {Callister}, T.; {Edelman}, B.; {Ye}, J.;
  {Essick}, R.; {Farr}, W.M.; {Farr}, B.; {Holz}, D.E.
\newblock {When Are LIGO/Virgo's Big Black Hole Mergers?}
\newblock {\em Astrophys. J.} {\bf 2021}, {\em 912},~98.
\newblock {{https://doi.org/10.3847/1538-4357/abee11}}.

\bibitem[Fryer {et~al.}(2022)Fryer, Olejak, and Belczynski]{Fryer2022}
Fryer, C.L.; Olejak, A.; Belczynski, K.
\newblock The Effect of Supernova Convection On Neutron Star and Black Hole
  Masses.
\newblock {\em  Astrophys. J.} {\bf 2022}, {\em 931},~94.
\newblock {{https://doi.org/10.3847/1538-4357/ac6ac9}}.

\bibitem[{Olejak} {et~al.}(2022){Olejak}, {Fryer}, {Belczynski}, and
  {Baibhav}]{Olejak2022}
{Olejak}, A.; {Fryer}, C.L.; {Belczynski}, K.; {Baibhav}, V.
\newblock {The role of supernova convection for the lower mass gap in the
  isolated binary formation of gravitational wave sources}.
\newblock {\em Mon. Not. R. Astron. Soc.} {\bf 2022}, {\em 516},~2252--2271,
\newblock {\url{https://doi.org/10.1093/mnras/stac2359}}.

\end{thebibliography}
\end{document}